\documentclass[aps,groupedaddress,superscriptaddress,twocolumn,amsmath,amssymb,pra,longbibliography]{revtex4-2}
\usepackage{dcolumn}
\usepackage{bm}
\usepackage{xcolor}
\usepackage[hidelinks]{hyperref}
\usepackage{float}
\usepackage{graphicx}
\usepackage{amsfonts}
\usepackage{verbatim}
\usepackage{bbm}
\usepackage{braket}
\usepackage{qcircuit}
\usepackage[normalem]{ulem}
\usepackage[T1]{fontenc} 
\usepackage{lmodern} % In your preamble
\definecolor{darkgreen}{RGB}{0,100,0}
\hypersetup{
    colorlinks=true,
   linkcolor=blue,
   citecolor=magenta,      
    urlcolor=purple,
}
\usepackage{multirow}
\usepackage{array}
\usepackage{tabularx}

\usepackage{graphicx}% Include figure files
\usepackage{dcolumn}% Align table columns on decimal point

\usepackage[hidelinks]{hyperref}
\usepackage{float}
\usepackage{cancel}

\usepackage{multirow}
\usepackage{array}
\usepackage{tabularx}
\def\sgn{\mathrm{sgn}}

\def\RWA{_\mathrm{RWA}}
\def\1{^{(1)}}
\def\2{^{(2)}}
\def\ep{\varepsilon}

\def\kb {\mathbf{k}}
\def\0{^{(0)}}

\def\llangle{\langle\!\langle}
\def\rrangle{\rangle\!\rangle}

\begin{document}

\title{Anomalous parametric resonance in a spin-1/2 chain: dynamical effects of nontrivial topology}

\author{Mahmoud T. Elewa}
\author{M. I. Dykman}
\affiliation{ Department of Physics and Astronomy, Michigan State University, East Lansing, Michigan 48824}
\date{\today}

\begin{abstract}
Resonant parametric modulation is a major tool of studying magnetic systems. 
For a spin-1/2 chain in a strong magnetic field, the resulting excitations can be mapped on  fermionic excitations in the Kitaev chain. We show that the response to turning on the modulation reveals dynamical bulk aspects of the nontrivial topology of the closed chain. In the topological regime, depending on the turn-on rate,  the system displays an absence of frequency dispersion of the time-averaged magnetization and an absence or a suppression of its  spatial correlations near resonance. The transition between the topological and trivial regimes is controlled by the modulation frequency. %The derivatives of the magnetization and the spatial spin correlators are discontinuous at the transition frequency. 
\end{abstract}

\maketitle

%\tableofcontents
\begin{figure}[t]
	\centering
	\includegraphics[scale=0.2]{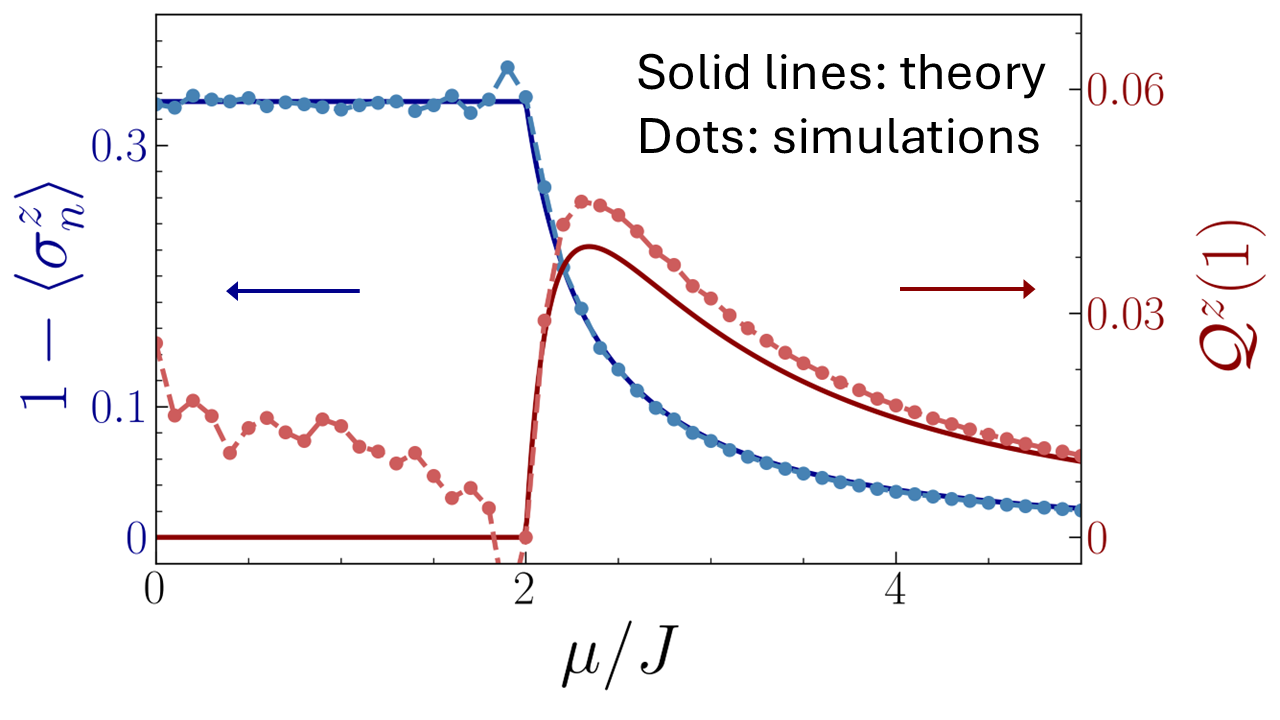}\\
	\includegraphics[scale=0.2]{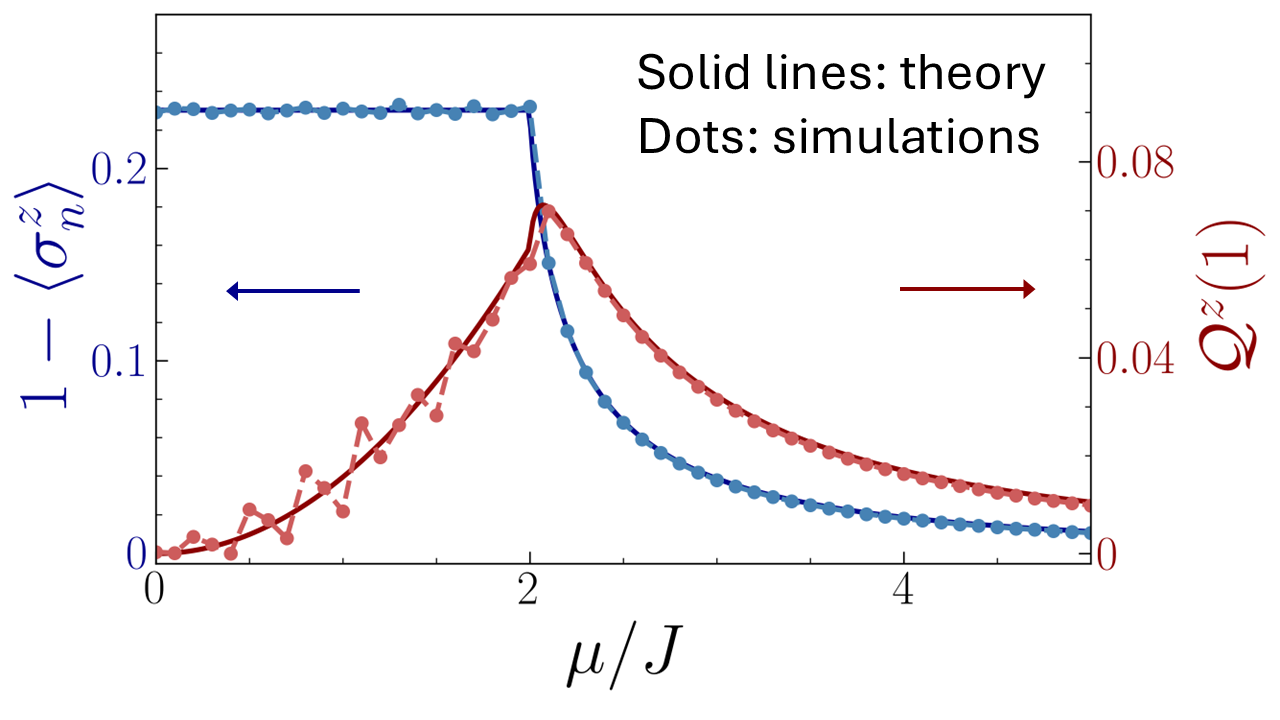}
	\caption{The time-averaged excitation density, $1-\braket{\sigma_n^z}$, and the nearest-neighbor correlator $\mathcal{Q}^z(1)=\braket{\sigma_n^z\sigma_{n+1}^z} - \braket{\sigma_n^z}^2$ in a modulated closed spin chain  as functions of the detuning of the modulation frequency $\mu$ scaled by the hopping integral $J$. The upper and lower panels refer to the sudden and quasi-adiabatic modulation turn-on. The matrix-product-state  simulations refer to the chain periods $N=20$ and $N=40$, respectively. The final modulation amplitude is $F(t_f)=J/2$. The slow turn-on regime in the simulations was implemented using $F(t) = F(t_f)t/\tau_\mathrm{sl}$ with $\tau_\mathrm{sl} = 10^3/J$. }
	\label{fig:z_quench}
\end{figure}

A sufficiently strong periodic modulation of magnetic systems can lead to parametric excitation of magnons \cite{Bloembergen1954,Suhl1957}. Various aspects of this  effect have been extensively studied.
Recent examples range from 
 dynamical demagnetization \cite{Shan2024} to a non-equilibrium topological magnon edge current \cite{Malz2019}, induced magnon-magnon coupling \cite{Arfini2025}, magnon excitation by amplitude-modulated light \cite{Kiselev2025},  Bose-Einstein condensation of magnons \cite{Demokritov2006,Lvov2023}, inverse spin-Hall effect \cite{Sandweg2011}, magnonic  turbulence \cite{Lvov2012}, magnon parametron \cite{Makiuchi2021,Elyasi2022}, and magnon-magnon entanglement and quantum magnonics \cite{Yuan2022}. Magnons are bosonic excitations; they are usually described by applying the Holstein–Primakoff or Villain transformations \cite{holstein1940,VILLAIN1974} to large-spin systems. 

Here we are interested in parametric  modulation  of spin-$1/2$ chains. Excitations in such chains  are mapped on spinless fermions \cite{Jordan1928,Lieb1961}. A remarkable feature of translationally symmetric chains of spinless fermions discovered by Kitaev \cite{Kitaev2001} is that, in the presence of pairing coupling, the system becomes topologically nontrivial.
Much of the previous work on the manifestations of this feature in  modulated spin-1/2 chains  was centered at the edge modes,  cf. \cite{Thakurathi2013,Dykman2019a,Yates2019,Mi2022, Schmid2024,vernier2024a,Tausendpfund2025,Schmid2025,Jin2025} and references therein. The analysis focused primarily on the modulation by periodically repeated short or rectangular  pulses. Such modulation is often  implemented in quantum simulations.

We consider {\it resonant} parametric excitation of spin-$1/2$ chains. Such excitation may lead to strong effects even where the modulating field is comparatively weak. Similar to magnons, if the modulation is long-wavelength, excitations in a translationally-symmetric system  should be created in pairs with opposite wave vectors $\kb$ and $-\kb$. The  modulation should be most efficient if its  frequency is  close  to the sum of the excitation frequencies. Mapping on the Kitaev chain in this case and the onset of nontrivial topology was considered in Ref.~\cite{Dykman2019a}. Resonant modulation has been implemented and some of its consequences explored in systems of up to six transmon qubits  that mimicked coupled spins \cite{Liang2024}. 

In contrast to studies focusing on topological edge-state effects, in this paper we  study how the Kitaev-chain topology affects  {\it bulk} properties of a modulated spin chain.
Respectively, we consider closed spin chains. 
We find profound bulk topological effects. The state of the system after the modulation is turned on is qualitatively different depending on  whether the modulation frequency lies in the topologically  trivial or nontrivial regime. The state also strongly depends on whether the modulation is turned on fast or slowly, as seen in Fig.~\ref{fig:z_quench}. The results are strongly different from what one would expect from tuning the system closer to resonance by reducing the detuning $\mu$ of the halved modulation frequency from the spin Larmor frequency $\omega_0$. In particular, the magnetization change is frequency-independent in the topological regime.

An insight into the origin of the dynamical manifestations of the topology can be gained for weak modulation. Without modulation, the excitation gap measured from $2\hbar\omega_0$  vanishes at two points in the Brillouin zone \cite{Lieb1961}. The modulation opens a gap and thus modifies the system dynamics, but primarily near these points, for weak modulation. This results in identical narrow peaks in the winding number density and, quite remarkably, in similar peaks in the excitation density \footnote{See Supplemental Material, which provides details of the calculations in the main text and contains Ref.~\cite{fishman2021}} %Ref.~\cite{fishman2021}}. 
Modulation frequency determines only the positions of the peaks in $k$-space, but not their shape. As a result, both the winding number (which is equal to $1$ or $-1$) and the magnetization are independent of $\mu$. With the increasing drive amplitude, the peaks broaden, but the winding number remains unchanged and the magnetization remains independent of  $\mu$ 
%\YYY{add a comment about hystresis?}

We consider a periodic $N$-spin chain, in a strong magnetic field, with nearest-neighbor $XY$ coupling. Resonant parametric excitation comes from sinusoidally modulating the coupling  at frequency $\omega_F$ close to twice the Larmor frequency $\omega_0$. The Hamiltonian of the  chain is 
\begin{align}
\label{eq:H_lab}
   & \hat{H}_0  = - \frac{1}{2}\omega_0\sum_{n=1}^N\sigma_{n}^{z}
   -J_{xx}(t)\sum_{n=1}^{N{}}\sigma_{n}^x\sigma_{n+1}^x
    -J_{yy}\sum_{n}^{N{}}\sigma_n^y\sigma_{n+1}^y, \nonumber\\
  & J_{xx}(t)=\bar J_{xx}{}+2F\cos(\omega_F t), \qquad (\hbar =1),
\end{align}
where $\sigma_n^i$ are the Pauli matrices ($i=\{x,y,z\}$) and $F$ is the modulation amplitude. We assume that the coupling parameters and the modulation amplitude are small compared to $\omega_0$. Then it is convenient to go to the rotating frame at frequency $\omega_F/2$ using the standard transformation $U=\exp[i(\omega_Ft/4)\sum_n\sigma_n^z]$. In the rotating wave approximation the Hamiltonian becomes
\begin{align}
\label{eq:H_RWA}
   & \hat{H}\RWA=\frac{1}{2}\mu\sum_{n=1}^N\sigma_n^z - J\sum_{n=1}^{N{}}(\sigma_{n}^+\sigma_{n+1}^-+ \sigma_{n+1}^+\sigma_{n}^-)    \nonumber\\
    &-F\sum_{n=1}^{N{}}(\sigma_{n}^+\sigma_{n+1}^+   + \sigma_{n+1}^-\sigma_{n}^-), 
    \quad \mu = \frac{1}{2}\omega_F - \omega_0,
\end{align}
where $J=\bar J_{xx}{}+J_{yy}$, $\mu$ is the detuning of the modulation frequency from the exact resonance, and $\sigma_n^\pm = (\sigma_n^x \pm i\sigma_n^y)/2$. The parameters $\mu, J$, and $F$ are generally of the same order of magnitude. It is seen from Eq.~(\ref{eq:H_RWA}) that the parametric modulation corresponds to flipping a pair of spins, reminiscent of creating a pair of magnons.

The spin chain (\ref{eq:H_RWA}) naturally maps by the Jordan-Wigner transformation \cite{Jordan1928,Lieb1961} on the Kitaev chain of spinless fermions. In particular, the parametric-modulation term maps on the term of the pairing interaction. This suggests that the modulation leads to a nontrivial topology of the spin chain. We will analyze it for periodic boundary conditions with even $N$. The  Hamiltonian of the Fourier-transformed fermions in the Nambu form  reads 
\begin{align} 
\label{eq:H_matrix1}
 &        \hat H =-\frac{1}{2} \sum_{k}
        \Vec{c}_k\!^{\dagger}
        \left(M_k\tau^z +F_k\,\tau^x \right)
        \Vec{c}_k ,\nonumber\\
   & M_k=  \mu - 2 J \cos k, \quad F_k = 2F\sin k    
 \end{align}
Here  $ c_k$ and $c_k^\dagger$ are the creation and annihilation operators of the fermions, $\vec{c}_k= \left(\begin{array}{c} c_k \\ c_{-k}^{\dagger}\end{array}\right)$, $\tau^{x,z}$ are the Pauli-matrices, and $-\pi<k\leq \pi$ \cite{Lieb1961,Kitaev2001}. It is well-known from the Kitaev chain theory that the range $|\mu/J|>2$ corresponds to a topologically trivial regime, whereas for $|\mu/J|<2$ the system of fermions, and  thus the resonantly modulated spin chain, are topologically nontrivial. 

The representation (\ref{eq:H_matrix1})  provides a convenient way to study the effect of turning on parametric modulation in the trivial and topological regimes.  We will consider this effect assuming that, in the initial state $\ket{\Psi_0}$, the spins are polarized along the strong magnetic field, $\bra{\Psi_0}\sigma_n^z\ket{\Psi_0}=1$.  In terms of the fermions, $\ket{\Psi_0}$ is the vacuum of the $c_k$-operators,  $c_k\ket{\Psi_0} = 0$; it is an eigenstate of $\hat H$ for $F=0$. 

Turning on modulation leads to a complicated time evolution of the wave function. We will be interested in the long-time behavior of physical observables of the spin chain, which is  determined by the time-averaged expectation values of the spin-chain operators and their spatial correlators, such as 
\begin{align}
\label{eq:averages_general}
&1-\braket{\sigma_n^z}  = \frac{2}{N}\sum_k\braket{c_k^\dagger c_k} ,\quad \mathcal{Q}^z(m)= \braket{\Delta\sigma_n^z \Delta\sigma_{n+m}^z}\nonumber\\
 &= \frac{4}{N^2}\sum_{k_1,...,k_4}\braket{c_{k_1}^\dagger c_{k_2}c_{k_3}^\dagger c_{k_4}} e^{i(k_1 - k_2)m}\delta_{k_1 - k_2, k_4 - k_3} \, .
\end{align}
Here $\Delta\sigma_n^z = \sigma_n^z -\braket{\sigma_n^z}$. We use $\braket{\cdot}$ to indicate the  time-averaged expectation value, 
\begin{align}
\label{eq:time_averaging}
\braket{A} = T^{-1}\int_{t_f}^{t_f+T}\bra{\Psi_0}A(t)\ket{\Psi_0},
\end{align} 
where $A(t)$ is an operator in the Heisenberg representation and $t_f$ is the time by which the modulation has reached its final value. Spin observables  appear to depend strongly  on how quickly the modulation is turned on.

\underline{Rapid turn-on of the modulation.} We will first assume that the modulation is sharply turned on at $t=0$, increasing from $F=0$ to a finite $F$ over time $\delta t\ll J^{-1}$ (yet $\delta t \gg \omega_0^{-1}$). The Hamiltonian $\hat H$ for $|F|>0$ is diagonalized by the fermionic operators $d_k, d_k^\dagger$ ,   
\begin{align}
\label{eq:ep_k}
&d_k =  c_k\cos(\theta_k/2) + c_{-k}^\dagger \sin(\theta_k/2),\quad \tan\theta_k=F_k/M_k.\nonumber\\
&\ep_k=(M_k^2+ F_k^2)^{1/2}.
\end{align}
Here $\ep_k$ is an eigenvalue of $\hat H$ in the Heisenberg representation $d_k(t)=d_k(0)\exp[-i\ep_k t\,\sgn(-M_k\cos\theta_k)]$. We set $\theta_k$ to be a continuous function of $k$; in particular, $M_k\cos\theta_k$ has the same sign for $M_k\to +0$ and $M_k\to -0$. For small $|F|$, in the topological regime $\ep_k$ has  characteristic narrow gaps, $\ep_k\approx |2J\sin k_\mu| [(|k|-|k_\mu|)^2 + F^2]^{1/2}$, where $k_\mu=\arccos (\mu/2J)$. They lead to the aforementioned narrow peaks of  the winding number density at $k=\pm k_\mu$. The peaks are identical by symmetry. Their areas are $\mu$-independent, since they add up to the $\mu$-independent winding number. The peaks of $\braket{c_k^\dagger c_k}$ are proportional to these peaks \footnotemark[1]. This shows an onset of unusual features of  parametric resonance in the spin-1/2 chain already for small $|F|$.

The time-averaged expectation values of spin correlators can be calculated in two steps. First, using Eq.~(\ref{eq:ep_k}) one finds the expectation values on the state $\ket{\Psi_0}$ of the  pairs $d_k^\dagger d_k, d_kd_{-k}$, and $d_k^\dagger d_{-k}^\dagger$ (pairs with nonzero total momentum are not excited). Only a contribution from the terms $\bra{\Psi_0}d_k^\dagger(t) d_k(t)\ket{\Psi_0}$ ``survives'' averaging over time for $T\gg [\mathrm{min}\,\ep_k]^{-1}$. We then  express, again from Eq.~(\ref{eq:ep_k}), the averages  $\braket{c_k^\dagger c_k}, \braket{ c_k c_{-k}}$, and $\braket{c_k^\dagger c_{-k}^\dagger}$ in terms  of  $\braket{d_k^\dagger d_k}$, which ultimately expresses the averages in terms of $\theta_k$. For example,  $\braket{c_k^\dagger c_k} = \frac{1}{2}\sin^2\theta_k$. This gives the spin correlators in terms of $\theta_k$. 

The nontrivial topology manifests when summation (integration) over $k$ is performed. The derivative $ d\theta_k/dk$ is the winding number density of the chain \cite{Chiu2016,Leumer2020}. From Eq.~(\ref{eq:ep_k}), the winding number is $\nu = (2\pi)^{-1}\int_{-\pi}^\pi dk(d\theta_k/dk) = -\Theta(1-|\mu/2J|)\,\sgn(FJ)$, where $\Theta(x)$ is the step function. As $k$ goes over the Brillouin zone, $\theta_k$ is incremented by $\pm 2\pi$ in the topologically nontrivial range $|\mu/2J| < 1$, whereas it returns to the initial value in the trivial regime. Since the sign of $F$ can be changed at the transition to the rotating frame by incrementing $t\to t+\pi/\omega_F$, of physical significance is $|\nu|$. The onset of nontrivial topology in a spin chain modulated at $\approx 2\omega_0$ has commonalities with the effects of resonant modulation in semiconductors with spin-orbit coupling, cf. \cite{Lindner2011,Lindner2013,Klinovaja2016,Thakurathi2017,Rudner2020a}. 
We note that different winding numbers emerge in static Ising chains with direct 3-spin coupling \cite{ Zhang2015c,Liu2025e}. 
%
%if $\tau^x$ and $\tau^z$ are thought of as components of the vector ${\bm\tau}$, $M_k\tau^z + F_k\tau^x$ describes an ellipse traced by this vector as $k$ goes from $-\pi$ to $\pi$. If this ellipse encircles zero, the system is topologically nontrivial.   

To illustrate how the topology affects spins we will first consider the correlator $\mathcal{Q}^z(m)$. The calculation is similar to the calculation $\int dk (d\theta_k/dk)$. Changing from integration over $k$ to integration over the unit circles $|w|=1$ with $w = \exp(ik)$ or $w=\exp(-ik)$ \footnotemark[1], after some algebra we obtain from Eqs.~(\ref{eq:averages_general}) and (\ref{eq:ep_k})
\begin{align}
\label{eq:z_correlator}
&\mathcal{Q}^z(m) =\sum_{\alpha=1,2}(-1)^\alpha |q_\alpha(m)|^2, \qquad q_\alpha(m) =(F/4\pi) %\mathrm{Im}\,\oint dw 
\nonumber\\
%q_\alpha(m) =\frac{F}{8\pi} 
&\times\mathrm{Im}\,\oint dw \,w^{|m|-1}(w^2-1)[P_1^{-1}(w) - (-1)^\alpha P_2^{-1}(w)],\nonumber\\
%&P_1(w) = (F-J)w^2 + \mu w -(F+J), \quad P_2(w) = (F+J)w^2  - \mu w - (F-J),\nonumber\\
&P_\beta(w) = F(w^2-1) -(-1)^\beta[\mu w - J(w^2+1)]\quad (\beta=1,2)
\end{align}
%
%($\beta=1,2$). Here, as well as in the calculation of $\nu$, we assumed that $M_k\cos\theta_k$ does not change sign with varying $k$; it means that, in the topological regime, $M_k\cos\theta_k$ has the same sign for $M_k\to \pm 0$. 

In the range $|\mu/2J|<1$,  where the system is topologically nontrivial, $P_1(w)$  has no roots with $|w|<1$ for $FJ>0$, whereas $P_2(w)$ has no such roots for $FJ<0$. Therefore,  from Eq.~(\ref{eq:z_correlator}), the time-averaged pair correlation function of the $z$-components of the spins is zero. Moreover,  higher-order correlators are zero, too \footnotemark[1].  Zero time-averaged correlations of $\sigma_n^z$ are a ``bulk'' manifestation of the nontrivial topology of the spin chain. In the trivial regime both $P_1(w)$ and $P_2(w)$ have one root with $|w|<1$, and then $\mathcal{Q}^z(m)$ is nonzero. From Eqs.~(\ref{eq:ep_k}) and (\ref{eq:z_correlator}), the structure of $P_{1,2}(w)$ determines also the value of the winding number.

Figure~\ref{fig:z_quench} shows the dependence of the correlator $\mathcal{Q}^z(1)$ on $\mu$, i.e., on the modulation frequency.  The correlator first sharply increases from zero with the increasing $|\mu|$ once $|\mu|$ reaches $2|J|$, and then falls off as $|\mu|$ further increases. The fall-off corresponds to the halved modulation frequency moving away  from the Larmor frequency, and thus from resonance. The pair correlators quickly fall-off with the increasing $|m|$, since the integrands in Eq.~(\ref{eq:z_correlator}) have a factor $w^{|m|-1}$ and $|w|<1$ at the poles inside the circle $|w|=1$. Higher-order spin correlators display a similar dependence on $\mu$ and the interspin distance \footnotemark[1]. 

Figure~\ref{fig:z_quench} also shows  the mean spin polarization  $\braket{\sigma^z_n}=1-2N^{-1}\sum_k\braket{c_k^\dagger c_k}$ as a function of $\mu$.  A change of $\braket{\sigma^z_n}$  is a conventional manifestation of parametric excitation. In the considered case it can be calculated in the same way as $\mathcal{Q}^z(m)$. Depending on the sign of $FJ$, it is conveniently expressed in term of a contour integral over $w=\exp(ik)$ or $w=\exp(-ik)$ that contains only $P_1^{-1}$ or $P_2^{-1}$, giving
\[1-\braket{\sigma^z_n} = (1+ |J/F|)^{-1},\quad |\mu/2J|<1.\]
Thus, the effect of parametric modulation is independent of the modulation frequency. Outside the topological regime,  the magnetization change falls off with the increasing $|\mu|$,  with $1-\braket{\sigma^z_n}\propto \mu^{-2}$ for large $|\mu|$ \footnotemark[1]. In the interesting paper \cite{Shi2022} 
it was found that, for the Kitaev chain, if the pairing coupling is suddenly turned on, $(i/N)\sum_{k>0}\braket{c_{-k} c_k - c_k^\dagger c_{-k}^\dagger}$ is independent of $\mu$ in the topological regime. This correlator does not contribute to $\braket{\sigma_n^z}$  and its relation to the nonzero winding number was not discussed.% to the other observables we study. } on in the Kitaev chain, there is a  combination of pairs of fermionic operators that has an independent \YYY{independence} of $\mu$ time-averaged expectation value in the topological regime; however, no spin-related observables were discussed and

\begin{figure}[h]
\includegraphics[scale=0.35]{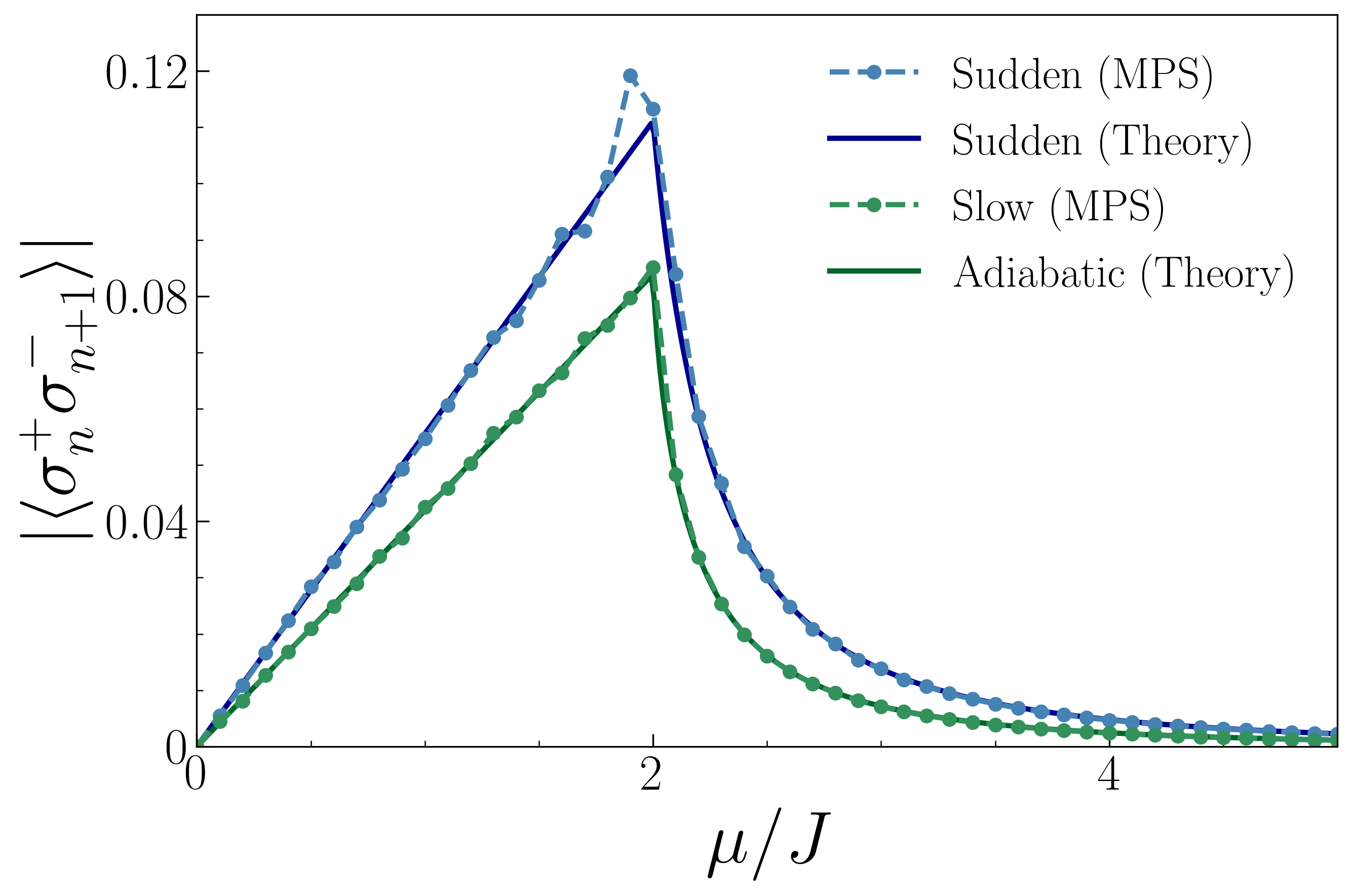}
	\caption{The time-averaged correlators $|\braket{\sigma_n^+\sigma_{n+1}^-}| = -\braket{\sigma_n^+\sigma_{n+1}^-} $ for slow and sudden turn-on of parametric modulation as functions of the scaled frequency detuning $\mu/J$. The results refer to the final modulation amplitude $ F(t_f)/J= 0.5$. The simulations of the sudden and slow turn-on refer to $N=20$ and $N=40$, respectively. The slow turn-on was implemented using $F(t) = F(t_f)t/\tau_\mathrm{sl}$ with $\tau_\mathrm{sl} = 10^3/J$. 
}
\label{fig:correlators}
\end{figure}

Other spin correlators, besides $\mathcal{Q}^z$, also display nontrivial behavior. In particular, $\braket{\sigma_n^+\sigma_{n+1}^-}= -N^{-1}\sum_k\braket{c_k^\dagger c_k}\cos k \to -[\mu\,\sgn (JF)/4F](1+|J/F|)^{-2}$  in the topological regime, i.e., the correlator linearly increases with $|\mu|$. For large $|\mu|$ it falls off as $1/|\mu|$. This behavior is shown in Fig.~\ref{fig:correlators}. The theory is in excellent agreement with the matrix product state (MPS) simulations already for $N=20$, except for $\mathcal{Q}^z(1)$. This is a consequence of the non-Wick contribution to the simulated  $\mathcal{Q}^z(1)$ from the terms, which are quartic in the fermionic operators; this contribution gives a correction $\propto 1/N$ for a finite chain \footnotemark[1].

We note that our results refer not only to spin, but also to qubit systems. For the both systems the time-averaged parameters can be directly measured in the experiment using conventional measurement techniques, cf.~\cite{Slichter1990,Blais2021}.

% One can also measure $\braket{\sigma_n^+\sigma_{n+1}^+}$ correlator. To do this it one shold allow for the phase of the drive, i.e., the drive-related parameter in $\hat H_0$ should be written  $\bar J_{xx} + 2F\co(\omega_Ft + \phi_F)$. Then the change to the fermionic operators should include the factor $\exp(i\phi_F/2)$. The measurement should be a lock-in type.  The resulting expression is...). 

\underline{Slow turn-on of the modulation.} The nontrivial topology also strongly affects the outcome of a slow turn-on of parametric excitation of a spin chain. We will assume that the modulation is increased from $F(t)=0$ for $t=0$ to $F(t_f)$ at time $t_f$ and is not changed afterwards. We will consider the state of the chain for $t>t_f$ and, again, will be interested in the time-averaged chain  parameters. 

Fermionic excitations of the chain are created pairwise with zero total momentum. The evolution of a fermion pair $(k,-k)$ is described by coupled Heisenberg equations for the operators $d_k, d_{-k}^\dagger$, which follow from Eqs.~(\ref{eq:H_matrix1}) - (\ref{eq:ep_k}),
\begin{align}
\label{eq:d_k_t}
&\dot d_k =- i\ep_k d_k + \frac{1}{2}\dot\theta_k d_{-k}^\dagger \quad (M_k\cos\theta_k<0).
\end{align}
Here $\ep_k\equiv \ep_k(t)$ is given by Eq.~(\ref{eq:ep_k}) with $F\equiv F(t)$. If $M_k\cos\theta_k>0$ one should replace $\ep_k\to -\ep_k$.

It is easy to solve  Eq.~(\ref{eq:d_k_t})  in the adiabatic approximation, where we disregard the term $\propto \dot\theta_k\propto \dot F_k$. This gives $d_k(t) = d_k(0)\exp[-i\int_0^t dt'\ep_k(t')]$.  % and similarly for $d_{-k}^\dagger$. 
Of interest for the time-averaged spin chain parameters are the expectation values $\braket{d_k^\dagger(t) d_k(t)}$. From  Eq.~(\ref{eq:ep_k}) we have 
\begin{align}
\label{eq:averaging_d_k}
\braket{d_k^\dagger(t) d_k(t)} =\bra{\Psi_0}d_k^\dagger(0) d_k(0)\ket{\Psi_0} =\sin^2[\theta_k(0)/2],
\end{align}
where we took into account that $c_k\ket{\Psi_0}=0$. Equation~(\ref{eq:ep_k}) also shows that, given the continuity of $\theta_k$, we have  $\theta_k(0)$ is  0 or $\pi$ for $F\to +0$ and $M_k<0$ or $M_k>0$, respectively. Then, as the modulation amplitude increases, $\theta_k(t)$ varies within the range $(-\pi/2,\pi/2)$ or $(\pi/2,3\pi/2)$, reaching the corresponding $\theta(t_f)$ once $F(t)$ reaches its stationary value $F(t_f)$. The value of $\theta(t_f)$ gives the relation between the operators $d_k(t)$ and $c_k(t)$ for $t>t_f$.

From Eq.~(\ref{eq:averaging_d_k}) we obtain $\braket{c_k^\dagger(t) c_k(t)} = [\ep_k(t_f)-|M_k|]/2\ep_k(t_f)$   and $\braket {c_{k}(t) c_{-k}(t)} = [F_k(t_f)/2\ep_k(t_f)]\sgn(M_k)$. These expressions allow us to find in the explicit form both $\braket{\sigma_n^z}$ and the spin correlators, as it was done earlier for a sudden turn-on. Again, the result depends on the singularities encountered in the contour integrals over $w=\exp(\pm ik)$, as in the case of the winding number, except that here the singularities are branching points rather than poles.

As seen from Fig.~\ref{fig:z_quench}, both in the slow and sudden turn-on regimes the magnetization  change $1-\braket{\sigma_n^z}$ is independent of $\mu$ in the topologically nontrivial range  $|\mu/2J|<1$ and falls off with the increasing $|\mu|$ for  $|\mu/2J|>1$. The results for the spin correlators $\braket{\sigma_n^+\sigma_{n+1}^-}$  are shown in Fig.~\ref{fig:correlators}. These correlators linearly depend on $\mu$ in the topological regime. However, for a slow turn-on  the correlators $\braket{\sigma_n^z \sigma_{n+1}^z}$ remain nonzero in the topological regime, in contrast to the sudden turn-on case.

The adiabaticity condition $|\dot\theta_k|\ll \ep_k$  breaks down in the topological regime $|\mu/2J|< 1$ when the drive is weak. This is because the gaps of $\ep_k$ at $k=\pm k_\mu$ go to zero as $F\to 0$. A different approach is required to study this regime \footnotemark[1]. Excitations are unavoidably generated as $F$ is turned on. The outcome depends on the relation between the final value of $F\equiv F_\mathrm{fin}$ and the rate at which it is reached. The independence of $\braket{\sigma_n^z}$ on $\mu$ persists even where $F_\mathrm{fin}$ is smaller than this rate, as it comes from the $\mu$-independent features of excitation dynamics in the small-gap regions. 

If $|\dot F|\ll F^2_\mathrm{fin}\ll J^2$, from Eq.~(\ref{eq:ep_k}), the adiabatic approximation still does not apply  for $\Bigl|\,|k|-|k_\mu|\Bigr| \lesssim |\dot F/G_\mu|^{1/2}$, where $G_\mu = J(4J^2 - \mu^2)^{1/2}$.
However, nonadiabatic corrections  to the spin correlators and $\braket{\sigma_n^z}$ are small, as they are proportional to the size of the corresponding  range of $k$ and thus scale as $|\dot F/G_\mu|^{1/2}$. They increase as  the system approaches the critical value of the modulation frequency where $|\mu/2J|=1$. Near this value the width of the nonadiabatic region scales as $(|\dot F|/J^2)^{1/3}$. 

The energy gap closing at $|\mu/2J|=1$ and $k=0$ or $k=\pi$ or all $F(t)$ also affects the time averaging.  In particular, in the sudden turn-on approximation, 
we disregarded the time-averaged matrix elements $[d_k(t)d_{-k}(t)+ \mathrm{H.c.}]$, which are  $\propto \sin(2\ep_kT)/2\ep_kT$. At $\mu=2J$, since $\ep_k \approx 2|F_\mathrm{fin}k|$ for $|k|\ll 1$,  the resulting correction is $\sim 1/F_\mathrm{fin}T$.   The smallness of this correction is determined by the restriction on $T$ imposed by spin relaxation due to the coupling to a thermal reservoir and the $\sigma^z_n\sigma^z_{m}$-coupling, which leads to pre-thermalization of excitations.

Both in the sudden and the adiabatic approximations, the derivatives of $\braket{\sigma_n^z}$ and the time-averaged spin correlators over $\mu$ are discontinuous at criticality, $\mu=\pm 2J$. This is seen in Figs.~\ref{fig:z_quench} and \ref{fig:correlators}. Such behavior is a consequence of the energy gap closing  in the thermodynamic limit and the winding number changing discontinuously. 

\begin{figure}[h]
\includegraphics[scale=0.25]{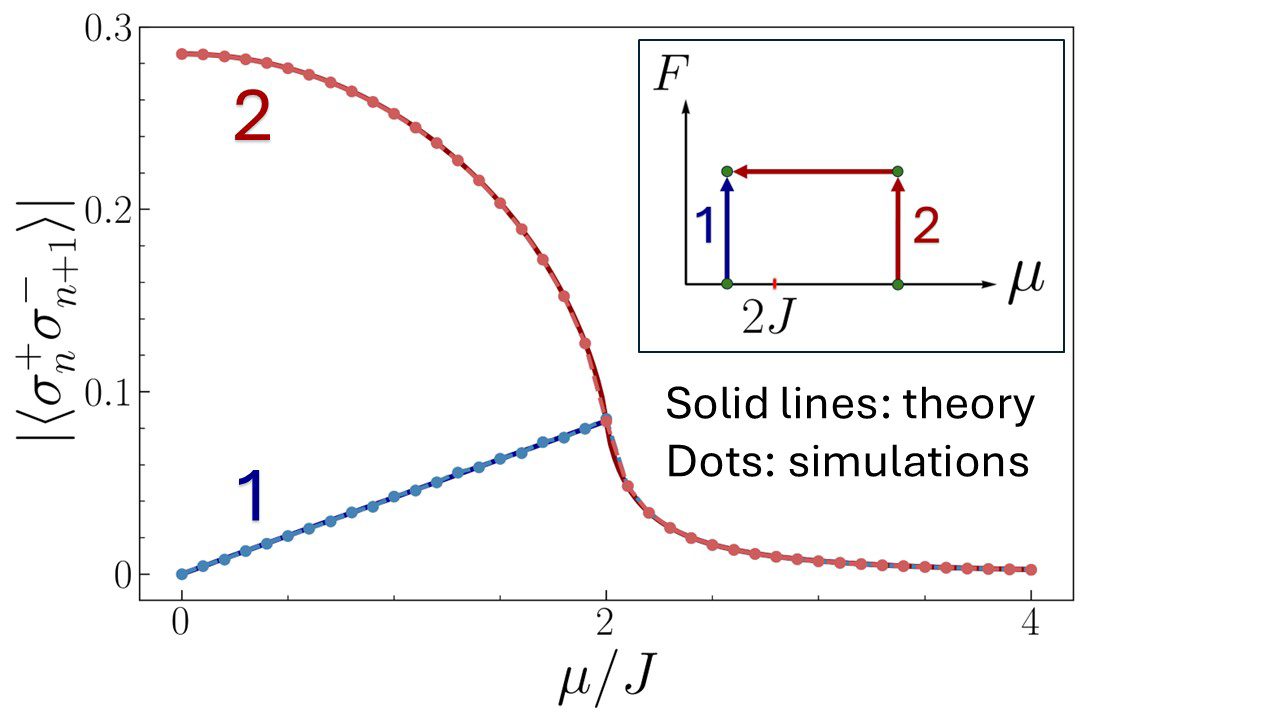}
	\caption{The time-averaged correlators $|\braket{\sigma_n^+\sigma_{n+1}^-}| = -\braket{\sigma_n^+\sigma_{n+1}^-} $ obtained by slowly increasing the drive $F(t)$ to $F_\mathrm{fin}=0.5J$ for a fixed $\mu$ ( line and circles 1) and by slowly turning on the drive to the same $F_\mathrm{fin}$ in the trivial regime and then slowly changing the modulation frequency to arrive at the same $\mu$  (line and circles 2). The simulations refer to $N=40$ and $F(t) = F_\mathrm{fin}t/\tau_\mathrm{sl}$ with $\tau_\mathrm{sl} = 10^3/J$. 
}
\label{fig:hysteresis}
\end{figure}

\underline{Hysteresis.} The different dynamics in the trivial and nontrivial regimes lead to a memory effect illustrated in Fig.~\ref{fig:hysteresis}. One can bring the system to a state with the same final value $F_\mathrm{fin}$ and $\mu_\mathrm{fin}$  (with $|\mu_\mathrm{fin}/2J|<1$) in different ways. First, one can slowly switch on the drive for a given $\mu=\mu_\mathrm{fin}$ in the topological regime. Second, one can switch on  the drive in the trivial regime and then slowly change $\mu$ to bring it to $\mu_\mathrm{fin}$. As $\mu$ crosses the critical value $\pm 2J$,   the adiabatic approximation breaks down via the Kibble-Zurek mechanism, cf.~\cite{Dziarmaga2005,*Dziarmaga2010,Polkovnikov2011}. However, for a slowly varying $\mu$ the resulting change of the time-averaged spin correlators is small. The large difference between the values of the correlators in Fig.~\ref{fig:hysteresis} (see also \footnotemark[1]) points to a strong memory effect: the system ``remembers'' how it was brought to a given state. 

Various regimes of tuning the transverse-field Ising chain (TFIC) across the quantum phase transition, which in our case corresponds to tuning the modulation frequency,  have been attracting much attention, including a sharp quench and linear, periodic, and nonlinear time evolution, cf. \cite{Barouch1970,*Barouch1971,*Barouch1971a, Sengupta2004,Patane2008, Calabrese2012,*Calabrese2012a,Mukherjee2009,Russomanno2012,Smelyanskiy2017,Heyl2013,
Dziarmaga2022,Paul2024,Grabarits2025} and references therein. Our paper refers to the paramagnetic phase in the  TFIC terms, as the Larmor frequency $\omega_0$ and the drive frequency ($\approx 2\omega_0$) are the largest frequencies. The unusual behavior of the magnetization and the spin correlators is a consequence of the excitation gap opening in the rotating frame by the high-frequency drive, which we relate to the behavior of the winding number density in the topological regime. The effects map on varying the anisotropy of the spin coupling in the TFIC.

We show that, for a slowly turned-on drive, $|\dot\ep_k|\ll \ep_k^2$, the Schr\"odinger equation  for the fermion wave function can be solved in the WKB approximation \footnotemark[1]. Furthermore, for a linear ramp of $F(t)$, the general solution is given by the parabolic cylinder functions for arbitrary $\dot\ep_k/\ep_k^2$; the WKB solution can be matched to it, providing a complete description of the dynamics.
We note also that, for $k=k_\mu$, the correlators $\braket{c_{k}^\dagger c_{k}}$ and $\braket{c_{k}c_{-k}}$  can be found for an arbitrary protocol of switching $F(t)$.

Small thermal population of excited states of the spin chain prior to parametric modulation leads to corrections $\propto \exp(-\hbar\omega_0/k_B T)\ll 1$ to the modulation effects. For $k_BT \gg \hbar |J|$ these corrections do not change the dependence of the magnetization and spin correlators on the modulation parameters \footnotemark[1].

This paper reveals dynamical  quantum effects of nontrivial topology and shows that they can be observed by studying parametric resonance in a closed  spin-1/2 chain. The topology  manifests in the response to turning on parametric modulation. It leads to an anomalous resonant behavior.  For a rapid turn-on, the magnetization is independent of the frequency detuning from resonance and displays no spatial correlations in the topologically nontrivial frequency range. For a slow turn-on, the dependence of the spin correlators on the modulation parameters is also qualitatively different from the conventional behavior where the response increases as the modulation approaches resonance \cite{Bloembergen1954}. Moreover,  spin correlators display hysteresis depending on whether the modulation is turned on in the trivial or nontrivial frequency range. The analytical results  describing the singular behavior of the spin system are in excellent agreement with the numerical results.

An experimental exploration of the predicted effects and their extensions is facilitated by the possibility of observing them  simply by tuning the modulation frequency. Appropriate platforms are provided by multi-qubit  simulators of spin chains, cf. \cite{Mi2022,Jin2025}, and by  various types of spin chains studied in condensed-matter systems \cite{Editorial2025}.

The authors acknowledge partial support
from the U.S. Defense Advanced Research Projects Agency
(Grant No. HR0011-23-2-004) and from the Gordon and Betty
Moore Foundation Award No. GBMF12214.

The source code and simulation data that support the findings of this study are openly available in the GitHub repository \cite{elewa_spin_chain_repo}.

\clearpage
\onecolumngrid 

\begin{center}
{\Large\bf SUPPLEMENTAL MATERIAL:}\\
\vspace{0.1in}
{\Large\bf Anomalous parametric resonance in a spin-1/2 chain: dynamical effects of nontrivial topology}

\vspace{0.1in}

Mahmoud T. Elewa
and M. I. Dykman

{\it Department of Physics and Astronomy, Michigan State University, East Lansing, Michigan 48824}%

%\date{\today}
\end{center}

% --- These commands reset all counters and add an "S" prefix ---
\setcounter{equation}{0}
\setcounter{figure}{0}
\setcounter{table}{0}
\setcounter{page}{1} % Resets page number to 1 for SM
\setcounter{section}{0} % Resets section counter
\makeatletter
\renewcommand{\theequation}{S\,\arabic{equation}}
\renewcommand{\thefigure}{S\,\arabic{figure}}
\renewcommand{\thetable}{S\,\arabic{table}}
\makeatother
\tableofcontents

\section{Mapping to the Kitaev chain}

%\subsection{Periodically modulated Spins}
We consider a chain  of $N$ spins in a strong magnetic field  with nearest-neighbor $XY$ coupling. The coupling along the $x$-axis is sinusoidally modulated at frequency $\omega_F$ close to twice the spin Larmor frequency. Equivalently, the formulation applies to a qubit chain, with $\omega_0$ being the transition frequency.  We assume that the chain is  closed, so that the site $N+1$ coincides with the site 1. The Hamiltonian of the chain is 
\begin{align}
\label{eq:H_lab}
   & \hat{H}_\mathrm{lab}  = - \frac{1}{2}\omega_0\sum_{n=1}^N\sigma_{n}^{z}
   -  J_{xx}(t)\sum_{n=1}^{N{}}\sigma_{n}^x\sigma_{n+1}^x- J_{yy}\sum_{n}^{N{}}\sigma_n^y\sigma_{n+1}^y, \nonumber\\
   &J_{xx}(t)=J_{xx}{}+2F\cos(\omega_F t), \qquad (\hbar =1),
\end{align}
where $\sigma_n^{x,y,z}$  are the Pauli matrices for an $n$th spin, $\omega_0$ is the Larmor (transition) frequency, $J_{xx}{}$ and $J_{yy}$ are the coupling parameters, and $F$ and $\omega_F$ are the modulation amplitude and frequency, respectively, $|\omega_F - 2\omega_0|\ll \omega_F$.  We consider the regime of weak coupling in the sense that $|J_{xx}|$, $|J_{yy}|$, and $|F|$ are  much smaller than $\omega_0$. The values of $F$ and $\omega_F$ can depend on time, generally, but they remain nearly constant on the time scale $\omega_F^{-1}$, that is $|(dF/dt)|, |d\omega_F/dt| \ll \omega_F^2$.

In contrast to the conventional problem of parametric resonance, where magnons are  bosonic excitations and it is necessary to take into account their interaction to prevent the runaway, i.e., to take into account the nonlinearity \cite{Suhl1957},  in the case of the spin chain the nonlinearity is already ``built in''. It is convenient to analyze resonant  parametric modulation by going to the rotating frame at frequency $\omega_F/2 \approx \omega_0$ using the transformation
\begin{align}
 \label{eq:Rotating_frame}
    \hat{U}(t) = \exp\left\{\frac{1}{4}i\int^t dt'\omega_F(t') \sum_{n=1}^N\sigma_n^z\right\}.
\end{align}
%
%We took into account here that $\omega_F$ can smoothly depend on time, $|(d/dt)\log\omega_F| \ll \omega_F$. We assume below that the drive amplitude can also smoothly depend on time, $F\equiv F(t)$, with $|F|\ll \omega_F$ and $|dF/dt|\ll \omega_F^2$.
%
In the rotating wave approximation (RWA), the transformed Hamiltonian of the system $U^\dagger H_\mathrm{lab}U - iU^\dagger \dot U$ has the form
\begin{align}
\label{eq:H_RWA}
%\begin{split}
   & \hat{H}\RWA=\frac{1}{2}\mu\sum_{n=1}^N\sigma_n^z - J\sum_{n=1}^{N{}}(\sigma_{n}^+\sigma_{n+1}^-+\sigma_{n+1}^+\sigma_{n}^-)-F\sum_{n=1}^{N{}}(\sigma_{n}^+\sigma_{n+1}^++\sigma_{n+1}^-\sigma_{n}^-),\nonumber\\
   &\mu = \frac{1}{2}\omega_F - \omega_0, \qquad J=J_{xx}+J_{yy},\qquad \sigma_m^\pm = \frac{1}{2}(\sigma_n^x \pm i\sigma_n^y).
%\end{split}
\end{align}
The Hamiltonian $\hat H\RWA$ clearly shows that resonant  parametric excitation leads to birth or annihilation of pairs of magnons or, equivalently, to upward or downward flips of qubits in pairs. 

%\subsection{Correspondence with the Kitaev Chain}

Next, we apply the standard Jordan-Wigner transformation
\begin{align}
    \label{eq:WJT}
 & \sigma_n^z = 1-2a_n^{\dagger}a_n , \qquad
    \sigma_n^+ = \left[\prod_{m=1}^{n-1}(2a_m^{\dagger}a_m-1)\right] a_n,\qquad
    a_n =\left[ \prod_{m=1}^{n-1}(-\sigma^z_m)\right]\sigma_n^+, \qquad n=1,\ldots, N
%\end{split}
\end{align}
where $a_n^{\dagger}$ and $a_n$ are fermionic creation and annihilation operators, respectively, which satisfy $\{a_n,a_m\}_+=\{a_n^{\dagger},a_m^{\dagger}\}_+=0$ and $\{a_n,a_m^{\dagger}\}_+=\delta_{n,m}$. cf. \cite{Lieb1961}. 

When changing to the fermionic operators in a chain of period $N$, care must be taken of the terms  $\sigma_N^\alpha \sigma_{N+1}^\beta
\equiv \sigma_N^\alpha \sigma_1^\beta$ in the Hamiltonian (\ref{eq:H_RWA}) with $\alpha,\beta = \pm$. These terms simplify since the original Hamiltonian and the RWA Hamiltonian preserve the parity of states, $P^\dagger\hat H_\mathrm{lab}P =\hat H_\mathrm{lab}$ and $P^\dagger\hat H\RWA P = \hat H\RWA$, where  $P=\exp[i(\pi/2)\sum_{n=1}^N\sigma_n^z]$. This is a consequence of spins flipped in pairs in resonant parametric excitation and, respectively, of the fermionic excitations  being created or annihilated in pairs. We assume that in the initial state $\ket{\Psi_0}$, i.e., in the state before the modulation is turned on, all spins are polarized along the magnetic field or, equivalently, all qubits are  in the ground state, $\ket{\Psi_0}=\ket{\uparrow_1...\uparrow_N}$. This state has even parity. For even $N$ this leads to the boundary condition $a_{N+1}  = -a_1$. We then change to the fermionic operators $c_k, c_k^\dagger$ using the Fourier transform
\begin{align} 
\label{eq:fourier}
    a_n = N^{-1/2}e^{-i\pi /4}\sum_k  c_k e^{ikn},\qquad k=\pi \frac{2m-1}{N} \qquad \Bigl( m = -(N/2)+1, -(N/2)+2,..., N/2\Bigr),
\end{align}
and write the RWA-Hamiltonian of the parametrically modulated chain in the form 
\begin{align} 
\label{eq:H_matrix1}
    \hat H =-\frac{1}{2} \sum_{k}
    \Vec{c}_k\!^{\dagger}
    \left(M_k\tau^z +F_k\,\tau^x \right)
    \Vec{c}_k , \quad M_k=  \mu - 2 J \cos k, \quad F_k = 2F\sin k.
\end{align}
Here $\Vec{c}_k=\left(\begin{array}{c} c_k \\ c_{-k}^{\dagger}\end{array}\right)$ are the Nambu vectors and $\tau^{x,y,z}$ are Pauli-matrices. 

It follows from the known results on the Kitaev chain \cite{Kitaev2001} that the range of relatively large detuning of the parametric drive from resonance, $|\mu/2J| \equiv |(\omega_F - 2\omega_0)/4J| >1$, is topologically trivial. The range  $|\mu/2J| \equiv |(\omega_F - 2\omega_0)/4J| <1$ is topologically nontrivial. We emphasize that we are considering a closed chain, there are no Majorana fermions at the chain edges. We are looking for the manifestation of topological features in the response to changing the modulation parameters.

In terms of the eigenstates $\ket{n_k{}}$ of the fermions described by the operators $c_k$, the initially occupied state is 
\[\ket{\Psi_0} = \prod_k\ket{0_k}, \qquad  c_k\ket{0_k{}}=0.
\]
While $\ket{\Psi_0}$ is an eigenstate of the Hamiltonian $\hat{H}$ in \eqref{eq:H_matrix1} for $F=0$, it is not the  ground state for $|F|>0$. This is not surprising, since the actual states of our  periodically driven system are Floquet states and there is no reason for the system to go to the state where $\hat H$ is minimal.

The Hamiltonian \eqref{eq:H_matrix1} can be diagonalized by the transformation $U(k) = \exp[-i(\theta_k/2)\tau^y]$, which relates $c_k, c_{k}^\dagger$ to new fermionic operators $d_k, d_k^\dagger$,
\begin{align}
\label{eq:canonical}
&d_k =  c_k\cos\theta_k/2 + c_{-k}^\dagger \sin\theta_k/2, \quad d_{-k}^\dagger =   -c_k\sin\theta_k/2 + c_{-k}^\dagger \cos\theta_k/2, \qquad \tan\theta_k=\frac{F_k}{M_k}.
\end{align}
In terms of these operators 
\begin{align}
\label{eq:H_diagonal}
&\hat H\to \hat{H}_0 =\frac{1}{2} \sum_{k} 
\ep_k [-\mathrm{sgn}(M_k\cos\theta_k)]
\vec{d}_k\!^\dagger \tau^z \vec{d}_k, \quad  \ep_k=\sqrt{M_k^2+F_k^2}.
\end{align}
%
%In what follows we call $c_k$-fermions and $d_k$ fermions the fermions described by the operators $c_k, c_k^\dagger$ and $d_k, d_k^\dagger$, respectively. \ME{\it We did not use this}
%
A standard convention is that  quasiparticles excited by  the operators $\{d_k^{\dagger}\}$ have  positive energies $\varepsilon_k/2$. This corresponds to the values $\theta_k$ such that $M_k\cos\theta_k <0$. As $k$ changes, function $\theta_k$ changes as well. However, the sign of  $M_k\cos\theta_k <0$ does not. As seen from Eq.(\ref{eq:canonical}), $\theta_{-k} = -\theta_k$.

We are interested in evaluating time-averaged correlation functions of the parametrically modulated spin/qubit chain. The averaging is done over time $T$ that exceeds the reciprocal gap $T\gg \max_k{\ep_k^{-1}}$. In the topologically trivial regime the gap $\min_k{\ep_k}$ is nonzero for all $F$ and the averaging is well-defined. The correction that emerges in  the nontrivial regime and close to its boundary is discussed in the main text.

%%%%%%%%%%%%%%%%%%%%%%%%%%%%%%%%%%%%%%%%
%%%%%%%%%%%%%%%%%%%%%%%%%%%%%%%%%%%%%%%%%%%%%%%%

\section{Analytical results for sudden turn-on}
\label{sec:sudden}

\subsection{Time-averaged  correlations of fermions}
\label{subsec:fermions_sudden}

We start with the dynamics where the drive amplitude is sharply turned on at time $t=0$ from $F=0$ to a final value $F$. The change of $F$ is slow on the time scale $\omega_F^{-1}$, but fast on the time scale $\ep_k^{-1}$. 
After the modulation has been turned on, the operators $d_k$ evolve in time as $d_k(t) = \exp(-i\ep_k t) d_k(+0)$, where we use $t=+0$ to indicate that this is the instant of time at which the field has been just turned on. The rotation angle $\theta_k$ is independent of time, $\theta_k(t) = \theta_k(+0)$. From Eq.~\eqref{eq:canonical} 

\begin{align}
\label{eq:d_k_eliminated}
    c_k^\dagger(t) c_k(t) = d_k^\dagger d_k \cos^2\theta_k/2 + d_{-k}d^\dagger_{-k}\sin^2\theta_k/2 - \frac{1}{2}\left(d_k^\dagger d_{-k}^\dagger e^{2i\ep_kt} + d_{-k}d_k e^{-2i\ep_k t}\right)\sin \theta_k,
\end{align}
where $d_k =d_k(+0)$. Only the terms proportional to $d_k^\dagger d_k$ and $d_{-k}d_{-k}^\dagger$ contribute to the time-averaged terms in this expression, 
\begin{align*}
    \overline{c_k^\dagger(t) c_k(t)} &= d_k^\dagger d_k \cos^2\theta_k/2 + d_{-k}d^\dagger_{-k}\sin^2\theta_k/2\\
    &= (\cos^4\theta_k/2 + \sin^4\theta_k/2) c_k^\dagger c_k +\frac{1}{4} \sin2\theta_k (c_k^\dagger c_{-k}^\dagger + c_{-k}c_k) + \frac{1}{2}\sin^2\theta_k c_{-k}c_{-k}^\dagger,
\end{align*}
where the overline indicates time-averaging and $c_k= c_k(+0)$, $c_k^\dagger= c_k^\dagger(+0)$.

For the sudden turn-on of the modulation, the expectation value of this average  has to be taken on the state $\ket{\Psi_0}$, which is the state of the system before the turn-on. It corresponds to the vacuum of the $c_k$ fermions, and we can disregard its change over the infinitesimal (on the scale $\max{\ep_k^{-1}}$) time it takes to turn on the modulation. Then only the term proportional to $c_{-k}(0)c_{-k}^\dagger(0)$ contributes to $\overline{c_k^\dagger(t) c_k(t)}$, so that 
\begin{equation}
    \label{eq:sudden_cdkck}
    \braket{c_k^\dagger(t)c_k(t)} = \frac{1}{2} \sin^2\theta_k,
\end{equation}
where, as in the main text,  $\braket{\cdot}$ denotes the time-averaged expectation value. 

Similarly, we have 
\[\overline{c_k^{\dagger}(t) c^{\dagger}_{-k}(t)} = - \frac{1}{2}\sin\theta_k \left[d_k(+0) d_k^\dagger(+0) - d_{-k}^\dagger(+0) d_{-k}(+0)\right].\] 
which gives for the expectation value
\begin{equation}
    \label{eq:sudden_ckcmk}
    \braket{c_k^{\dagger}(t)c^{\dagger}_{-k}(t)}=-\frac{1}{4}\sin2\theta_k. 
\end{equation}
We emphasize that, in Eqs.~(\ref{eq:sudden_cdkck}) and (\ref{eq:sudden_ckcmk}), the value of $\theta_k$ refers to the modulation having been turned on, $\theta_k\equiv \theta_k(+0)$. It is given by Eq.~(\ref{eq:canonical}).

%%%%%%%%%%%%%%%%%%%%%%%%%%%%%
%%%%%%%%%%%%%%%%%%%%%%%%%%%%%%

%We also note that $\braket{c_{-k}(t)c_{-k}^\dagger(t)}=1-(\sin^2\theta_k)/2$, and $\braket{c_{-k}(t)c_{k}(t)}=(\sin2\theta_k)/4$. 

\subsection{Magnetization and spatial correlations in the spin chain}
\label{subsec:spin_sudden}

The results on the correlators of the $c_k$ operators allow one to time-averaged parameters of the spin chain using Eqs.~(\ref{eq:WJT}) and (\ref{eq:fourier}). For the change of the scaled magnetization due to parametric excitation we have from Eqs.~(\ref{eq:canonical}) and (\ref{eq:sudden_cdkck}) in the thermodyunamic limit $N\to \infty$
\begin{align*}
%\label{eq:averages_density}
&1-\braket{\sigma_n^z}  = \frac{2}{N}\sum_k\braket{c_k^\dagger c_k}
= \frac{1}{2\pi}\int_{-\pi}^{\pi} \sin^2\theta_k dk 
= \frac{1}{2\pi}\int_{-\pi}^{\pi} \left(\frac{F_k}{\ep_k}\right)^2 dk\\
%\nonumber\\
& = -\frac{F}{\pi}\mathrm{Im}\int_{-\pi}^{\pi} \frac{2\sin k}{\mu-2J\cos k + 2 i F \sin k}dk 
 \end{align*}
 Changing to integration over $w=\exp(ik)$, we obtain
 \begin{align}
\label{eq:averages_density}
1-\braket{\sigma_n^z} = \frac{F}{2\pi}\mathrm{Im}\oint \frac{dw}{w}\frac{w^2-1}{P_1(w)}, \qquad P_1(w) =(F-J)w^2 + \mu w -(F+J), 
\end{align}
The contour integral is taken over the unit circle in the complex $w$-plane. The integrand has a pole at $w=0$ and two poles $w\1_\pm$ given by the zeros of $P_1(w)$, 
\begin{equation}
    w^{(1)}_{\pm} = \frac{-\mu\pm\sqrt{\mu^2-4(J^2-F^2)}}{2(F-J)}.
\end{equation} 

In the topological regime, both $w_{\pm}^{(1)}$ lie inside (outside) the unit circle for $FJ<0$ ($FJ>0$). In the trivial regime, for $\mu>0$ ($\mu<0$) the root  $w\1_+$ ($w_-^{(1)}$) 
lies inside the unit circle, whereas the other root lies outside. 
Then, from Eq.~\eqref{eq:averages_density},
\begin{align}
\label{eq:result_for_sudden}
 1-\braket{\sigma_n^z} =\left\{
 \begin{array}{cc}
 [F^2/(F^2-J^2)]\left\{1 -|\mu|\left[\mu^2 + 4(F^2-J^2)\right]^{-1/2}\right\},& |\mu|>2|J|\\
 (1+|J/F|)^{-1}, &|\mu|<2|J|.\\
 \end{array}
 \right.
 \end{align}

\subsubsection{The pair correlator}

A quantity of significant interest and importance is the spin correlator
\begin{align}
&    \mathcal{Q}^z(m) = \braket{\Delta\sigma_n^z \Delta\sigma_{n+m}^z} = \frac{4}{N^2}\sum_{k_1,...,k_4}\braket{c_{k_1}^\dagger c_{k_2}c_{k_3}^\dagger c_{k_4}} e^{i(k_1 - k_2)m}\delta_{k_1 - k_2, k_4 - k_3},\nonumber\\
    &\Delta\sigma^z_n = \sigma_n^z - \braket{\sigma_n^z},
\end{align} 
where $k_1\neq k_2$ and $k_3\neq k_4$. There are two different ways of pairing the values of $k_i$:
\[ k_1 = k_4\; \&\;  k_2 = k_3 \; \mathrm {or} \; k_1 = -k_3\; \& \; k_2 = -k_4.\] 
Applying the parings gives
\begin{align}
    \label{eq:two_correlations}
    \mathcal{Q}^z(m) &= \frac{4}{N^2}\sum_{k,q}\left[\braket{c_k^\dagger c_k}\braket{c_q c_q^\dagger}-\braket{c_k^\dagger c_{-k}^\dagger}\braket{c_q c_{-q}}\right]e^{i(k-q)m}
    %\nonumber\\
    %&= \frac{4}{N^2}\sum_{k,q}\left[-\braket{c_k^\dagger c_k}\braket{c_q^\dagger c_q}+\braket{c_k^\dagger c_{-k}^\dagger}\braket{c_{-q} c_{q}}\right]e^{i(k-q)m}\equiv 
    =-\left\vert q_1(m)\right\vert^2 +\left\vert q_2(m)\right\vert^2 \nonumber \\
& q_1(m)= \frac{2}{N}\sum_k \braket{c_k^\dagger c_k} e^{ikm}= \frac{1}{N}\sum_k \sin^2 \theta_k \cos(mk),\nonumber\\ 
& q_2(m) = \frac{-2i}{N}\sum_k\braket{c_k^\dagger c_{-k}^\dagger} e^{ikm}=\frac{-1}{2N}\sum_k \sin 2\theta_k \sin(mk),
\end{align}
where we used the results from equations \eqref{eq:sudden_cdkck} and \eqref{eq:sudden_ckcmk}. 
To calculate $q_{1,2}(m)$, we change from the sums to integrals over $k$ and then from integration over $k$ to integration over the unit circle $|w|=1$. We choose  $w=e^{ik}$ where the integrand contains $\exp(imk)$ and  $w=e^{-ik}$ when the integrand contains $\exp(-imk)$; in the latter case the direction of going around the contour $|w|=1$ is changed to the opposite. After simple algebra, we obtain, in the thermodynamic limit,
\begin{align}
\label{eq:Q1andQ2}
&q_\alpha(m) =\frac{F}{4\pi} \mathrm{Im}\,\oint dw \,w^{|m|-1}(w^2-1)[P_1^{-1}(w) - (-1)^\alpha P_2^{-1}(w)],\nonumber\\
%&P_1(w) = (F-J)w^2 + \mu w -(F+J), \quad P_2(w) = (F+J)w^2  - \mu w - (F-J),\nonumber\\
&P_\alpha (w) = F(w^2-1) -(-1)^\alpha [\mu w - J(w^2+1)]\quad (\alpha =1,2). 
\end{align}
The roots of the polynomial $P_2(w)$ are
\[  w^{(2)}_{\pm} = \frac{\mu \pm \sqrt{\mu^2-4(J^2-F^2)}}{2(F+J)}.
\]

In the topological regime, where $|\mu/2J|<1$, the polynomials $P_1(w)$ and $P_2(w)$ have no roots with $|w|<1$  for $FJ>0$ and $FJ<0$, respectively. Therefore,  from Eq.~(\ref{eq:Q1andQ2}), $ q_1(m)=- q_2(m)$ for $FJ>0$ and $ q_1(m)= q_2(m)$ for $FJ<0$. Thus, we see from Eq.~\eqref{eq:two_correlations} that the time-averaged pair correlation function of the $z$-components of the spins is zero for all values of $m$ for $|\mu/2J|<1$. This is an unexpected and  nontrivial feature noted in the main text.

In the trivial regime, where $|\mu/2J|>1$, both $P_1(w)$ and $P_2(w)$ have one root with $|w|<1$, and then $\mathcal{Q}^z(m)$ is nonzero. For instance, for the nearest-neighbor correlator, $m=1$, taking into account the contributions from the poles that are inside the unit circle, we obtain

\begin{align}
     &\mathcal{Q}^z(1)= \frac{2F^2(\mu^2-4J^2)}{r_\mu^2\left(\mu^2-2J^2+2F^2+|\mu|r_\mu\right)}, \quad |\mu/2J|>1,%\nonumber\\
  %  & \MD{\mathcal{Q}^z(1)= \frac{F^2}{4(F^2 - J^2)^2}
  %  \frac{(\mu^2 - |\mu| r_\mu -4J^2)^2 -16 F^2J^2}{r_\mu^2}, \quad r_\mu = [\mu^2 + 4(F^2 - J^2)]^{1/2}
   % }
\end{align}
where $r_\mu\equiv\sqrt{\mu^2-4J^2+4F^2}$. We see that  $\mathcal{Q}^z(1)$ is continuous at  $|\mu| \rightarrow 2|J|$, but its derivative over $\mu$ is discontinuous.  The function $\mathcal{Q}^z(1)$ first increases with the increasing $|\mu|$ beyond the critical value $|\mu| = 2|J|$, but then it falls off as $\mu^{-2}$ for large $|\mu|$. 

\subsubsection{The three-site correlator}

Calculating  the three-site correlator  $\braket{\Delta\sigma_{n_1}^z \Delta\sigma_{n_2}^z    \Delta\sigma_{n_3}^z}$ comes to integrating the expectation value of the time-averaged product $\overline{c_{k_1}^\dagger c_{k_2}c_{k_3}^\dagger c_{k_4} c_{k_5}^\dagger c_{k_6}}$ with the weight $\prod_j \exp[-in_j(k_j-k_{j+1})]$. In turn, this comes to calculating 8 combinations where we take $k_i$ so that the fermionic operators are paired into products with the total momentum transfer equal to zero, i.e., $c_{k_i}^\dagger c_{k_i}$,  $c_{k_i}^\dagger c_{-k_i}^\dagger$, and $c_{k_i}c_{-k_i}$. This allows us to express the three-site  correlator in terms of the functions $q_{1,2}(m)$ given by Eq.~\eqref{eq:Q1andQ2},

\begin{align}
    \label{eq:three_correlator}
    \braket{\Delta\sigma_{n}^z \Delta\sigma_{n+m}^z    \Delta\sigma_{n+l}^z} &
    = 2 q_1(m) q_1(l) q_1(m-l)
    -2 q_1(m) q_2(l) q_2(l-m)\nonumber\\
    &+2 q_1(l) q_2(m) q_2(l-m)
    -2 q_1(m-l) q_2(m) q_2(l),
\end{align}
where $l > m > 0$. 

As  discussed above, $ q_1(m)=-\sgn(FJ) q_2(m)$ in the topological regime. Therefore, it is seen from Eq.~\eqref{eq:three_correlator} that the three-site correlator is equal to zero in the topological regime. Higher order correlators can be shown to also vanish in the topological regime. 

\subsection{The correlator of the transverse spin components}

Another correlator  discussed in the main text is that of the transverse spin components  
\[\braket{\sigma_n^+\sigma_{n+1}^-}= -\braket{ a_{n+1}^{\dagger} a_n}=-N^{-1}\sum_k\braket{c_k^\dagger c_k}\cos k = - q_1(1)/2.\] 
This gives
\begin{align}
\label{eq:spsm_sudden}
 \braket{\sigma_n^+\sigma_{n+1}^-} =\left\{
 \begin{array}{cc}
 -\sgn (\mu) JF^2 (|\mu|-r_\mu)^2/[4r_\mu (J^2 - F^2)^2], &|\mu|>2|J|\\
 -\sgn(FJ)(\mu/4F)(1+|J/F|)^{-2}, &|\mu|<2|J|.\\
 \end{array}
 \right.
 \end{align}

In the main text, we plot the absolute value of  $\braket{\sigma_n^+\sigma_{n+1}^-}$. Given that $\braket{\sigma_n^+\sigma_{n+1}^-}$ is real and negative for $\mu J>0$, we plot the expressions \eqref{eq:spsm_sudden} taken with the opposite sign. %This concludes the results for the sudden switching discussed in the main text. 

%%%%%%%%%%%%%%%%%%%%%%%%%%%%%%%%%%%%

\subsection{Relation to the explicit calculation of the winding number}

As discussed in the main text, the winding number for the Kitaev chain is defined as $\nu=(2\pi)^{-1}\int_{-\pi}^{\pi}dk(d\theta_k/dk)$, where $\theta_k$ is defined in equation \eqref{eq:canonical}. From this equation, %Thus, we have
\begin{align}
    &\nu = \frac{1}{2\pi}\int_{-\pi}^{\pi}dk\left(\frac{M_k (dF_k/dk)-F_k (dM_k/dk)}{\ep_k^2}\right) = \frac{1}{\pi}\int_{-\pi}^{\pi}dk\left(\frac{F\cos k M_k - J\sin k F_k}{\ep_k^2}\right)\nonumber\\
    &= \frac{1}{\pi}\mathrm{Im}\left\{\int_{-\pi}^{\pi}dk \frac{iF\cos k+J\sin k}{\mu-2J\cos k + 2 i F \sin k}\right\}
\end{align}

As before, we can change to integration over $w=e^{ik}$, which gives 
\begin{align}
\label{eq:complex_winding}
    \nu = \frac{1}{2\pi}\mathrm{Im}\oint \frac{dw}{w}\frac{(F-J)w^2+(F+J)}{P_1(w)}, 
\end{align}
where $P_1(w)$ is defined in Eq.~\eqref{eq:averages_density}. We can also write $\nu$ in the same form, but with $F\to -F$ in the numerator and $P_1(w)\rightarrow P_2(w)$. Clearly, the value of the winding number is determined by the structure of the poles of $P_1(w)$ and $P_2(w)$, which also determine the structure of the time-averaged spin observables we discussed above. To further see this, we assume $FJ>0$. Then, we can write $\nu$ as
\begin{align}
 \label{eq:complex_winding}
    \nu = \frac{1}{2\pi}\mathrm{Im}\left\{(F-J)\oint dw\frac{w}{P_1(w)}+(F+J)\oint \frac{dw}{w}\frac{1}{P_1(w)}\right\}.
\end{align}
In the topological regime, the poles of $P_1(w)$ lie outside the unit circle, and therefore $\nu = (F+J)/P_1(0) = -1$. For $FJ<0$, writing the integral in terms of $P_2$ and taking into account that $P_2$ does not have poles with $|w|<1$ in the topological regime, we obtain $\nu=(J-F)/P_2(0) = 1$. A calculation for $|\mu/2J|>1$ shows that $\nu=0$ in this range. This is fully analogous to the calculations of the magnetization and the spin correlators above, revealing explicitly  the relation of these calculations to the winding number. 

%%%%%%%%%%%%%%%%%%%%%%%%%%%%%%%

\subsection{Weak-drive limit for fast turn-on}
\label{subsec:weak_drive}

\begin{figure}[h]
\includegraphics[scale=0.5]{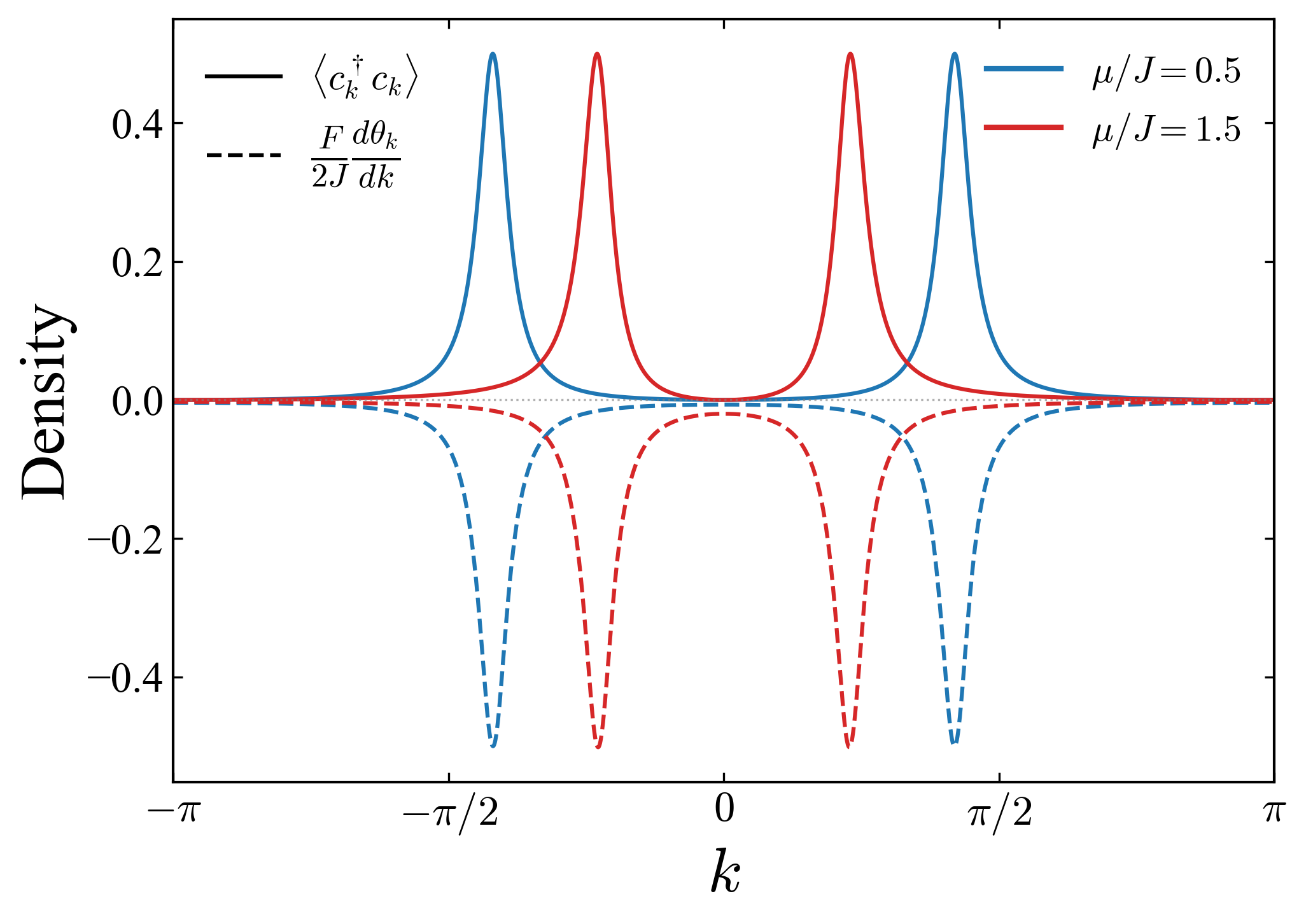}
	\caption{The densities $\braket{c_k^\dagger c_k}$ of the time-averaged magnetization change due to fast turn-on of the parametric drive (solid lines) and of the winding number $d\theta_k/dk$ (dashed lines).  The plots refer to $F/J=0.1$. The winding number density is multiplied by $F/J$ to bring the plots to the same scale. For small $|F/J|$, both densities are sets of two narrow Lorentzian peaks. The shapes of the peaks are independent of the scaled frequency detuning $\mu$, whereas the positions of their maxima $\pm k_\mu$ depend on $\mu$, with $k_\mu=\pm\arccos(\mu/2J)$.  
}
\label{fig:peaks_fast_turnon}
\end{figure}

An insight into the anomalous behavior of the average magnetization and the correlators in the topologically nontrivial regime can be gained from the analysis of the effect of weak drive. In this regime,  for $F=0$ the gap in the dispersion law $\ep_k$ closes for two values of $k$ in the Brillouin zone: $\ep_k = 0$ for $k=\pm k_\mu$, where $k_\mu = \arccos(\mu/2J)$. The drive  opens the gap. For a weak drive the dispersion near the gap opening has a characteristic form
\begin{align}
\label{eq:energy_near_gap}
\ep_k \approx |\sin k_\mu| [4J^2 (|k| -| k_\mu|)^2 + 4F^2]^{1/2}, \qquad |F|\ll |J|,\quad |k|- |k_\mu| \ll \pi.
\end{align}
When the gap is small, excitations are created just near the two gap openings, i.e., for  $k$ close to $k_\mu$ or $-k_\mu$. From Eq.~(\ref{eq:sudden_cdkck}), the number of excitations with a given $k$ is
\begin{align}
\label{eq:ck_near_gap}
\braket{c_k^\dagger c_k} = \frac{1}{2}\sin^2\theta_k =\frac{2F^2\sin^2k}{\ep_k^2}\approx 
\frac{2F^2 }{ 4J^2 (|k| -| k_\mu|)^2 + 4F^2}.
\end{align}
For small $F$, this expression as a function of $k$ describes two $\delta$-like peaks at $k= \pm k_\mu$ (more precisely, two narrow Lorentzian peaks). The area of each peak is $\pi |F/2J| $. It is independent of $\mu$. Therefore the magnetization change, which is given by $\int dk \braket{c_k^\dagger c_k}$, is also independent of $\mu$. This is in spite the positions of the peaks being $\mu$-dependent. The correlator $\braket{c_k^\dagger c_k}$ is shown in Fig.~\ref{fig:peaks_fast_turnon}.

As $|F|$ increases, the peaks shift and are broadened, but still they do not overlap, since no excitations with $k=0$ are created. As a result, the magnetization still remains independent of $\mu$. To see this a calculation is required, and such a calculation is presented in Sec.~\ref{subsec:spin_sudden}.  A formal insight is that the zeros of $\ep_k$ as a function of $k$ go further into complex-$k$ plane with the increasing $|F|$.

The correlator 
\[ \braket {c_k^\dagger c_{-k}^\dagger} = -\frac{1}{4}\sin 2\theta_k= -\frac{M_k F\sin k }{\ep_k^2} 
\]
goes to zero at $k= \pm k_\mu $. Respectively, the correlator of the magnetization on different sites $\mathcal{Q}^z(m)$ has no   terms linear in $F$ for $F\to 0$. The full calculation  in Sec.~\ref{subsec:spin_sudden} shows that it is equal to zero for any $F$ in the topological regime.

In contrast,  as seen from Eq.~(\ref{eq:ck_near_gap}), in the topologically nontrivial regime the correlator of the off-diagonal spin components  for small $F$ is
\[\braket{\sigma_n^+ \sigma_{n+1}^-} = -N^{-1}\sum_k\braket{c_k^\dagger c_k}\cos k \approx
-\frac{\mu}{4J} \,|F/J|.\]
This expression coincides with Eq.~(\ref{eq:spsm_sudden}) for $|F/J|\ll 1$ and $|\mu/2J|< 1$. It shows that the correlator linearly increases with $\mu$ in  the topological regime.

In the limit $F\to 0$ the  winding number density in the topologically nontrivial becomes a set of two very narrow Lorentzians,
\begin{align}
\label{eq:winding_weak_drive}
&\frac{1}{2\pi}\,\frac{d\theta_k}{dk} =\frac{1}{2\pi} \,\frac{2F (\mu-2J\cos k)\cos k -4JF\sin^2 k}{\ep_k^2} \nonumber\\
&\to - \frac{1}{2\pi} \,\frac{F/J}{(|k| - |k_\mu|)^2 + (F/J)^2}
\end{align}
The half-width of the Lorentzians is $|F/J|\ll 1$, their area (with the account taken of the factor $1/2\pi$) is $1/2$ independent of the scaled frequency detuning $\mu$. The similarity of Eqs.~(\ref{eq:ck_near_gap}) and (\ref{eq:winding_weak_drive}) is obvious,
\[ \braket{c_k^\dagger c_k} = -\frac{F}{2J} \,\frac{d\theta_k}{dk}.
\]
 The similarity of the densities $\braket{c_k^\dagger c_k}$ and $d\theta_k/dk$ is seen also from Fig.~\ref{fig:peaks_fast_turnon} that shows the full expressions for $\braket{c_k^\dagger c_k} $ and $d\theta_k/dk$.  
%%%%%%%%%%%%%%%%%%%%%%%%%%%%%%%%%%%%%%%%%%%%%%%%%%%%%%%%%%%%

\subsection{Thermal effect}
\label{subsec:thermal}

We now discuss the effect of a nonzero temperature. The previous analysis refers to the case where, prior to the turn-on of the modulation,  all spins are in the ground state, i.e., aligned along the strong magnetic field. For a nonzero temperature some spins are flipped. For a low temperature compared to the Zeeman energy, $\beta\omega_0 \ll 1$, where $\beta = \hbar/k_BT$, the portion of flipped spins is exponentially small, $\propto \exp(-\beta\omega_0)$. This leads to a small correction to the effect of the modulation. We will consider this correction in the thermodynamic limit where one can disregard corrections to the periodic boundary conditions; these corrections have an effect $\sim 1/N$ \cite{Lieb1961} (in fact, when switching to fermions, one may think of the system as a mix of two subsystems with even and odd numbers of fermions; because of the parity conservation, these subsystems do not interact with each other) 

The density matrix of the spin system in the absence of modulation is 
\begin{align}
    \label{eq:thermal_state}
    \rho_{\mathrm{th}} = Z^{-1}\exp(-\beta\hat{H}_T), \quad \hat{H}_T = - \frac{\omega_0}{2}\sum_{n=1}^N \sigma_n^z - J \sum_{n=1}^N (\sigma_n^+\sigma_{n+1}^- + \sigma_n^-\sigma_{n+1}^+),
\end{align}
We set $\hbar = k_B = 1$ and use $J_{xx}=J_{yy}=J/2$;  $Z=\mathrm{Tr}[\exp(-\beta\hat{H}_T)]$ is the partition function.

We can write the density matrix in terms of the Jordan-Wigner fermions in the momentum space, i.e., in terms of fermionic creation and annihilation operators $c_k^\dagger$ and $ c_k$,  
\begin{align}
    \label{eq:thermal_fermions}
    \rho_{\mathrm{th}} = \prod_{k} \exp\left[\beta\tilde M_k \left(c_k^\dagger c_k - \frac{1}{2}\right)\right]/2\cosh (\beta\tilde M_k/2), \quad \tilde M_k = -\omega_0 -2J\cos k,
    %
    %\frac{1}{2}\left(1-\tanh(\alpha_k)(c_k^\dagger c_k-c_k c_k^\dagger)\right).
\end{align}
cf. Eq.~(\ref{eq:H_matrix1}). To the first order in $\exp(-\beta\omega_0)$ this gives
\[\llangle c_k^\dagger c_k\rrangle \approx \exp(\beta\tilde M_k) 
\]
where $\llangle \hat A \rrangle = \mathrm{Tr}\rho_\mathrm{th} \hat A$. Substituting this expression into Eq.~\eqref{eq:d_k_eliminated}, we obtain that, for a fast turn-on of the drive, the change $\Delta\braket{c_k^\dagger c_k}$ of the  time-averaged  fermion density is  
\begin{align}
\label{eq:thermal_density_change_fast}
\Delta\braket{c_k^\dagger c_k} \equiv \braket{c_k^\dagger c_k} - \llangle c_k^\dagger c_k\rrangle= \frac{F_k^2}{2\ep_k^2} [1-2\exp(\beta\tilde M_k)],
\end{align}
The corresponding change  $\llangle \sigma_n^z\rrangle -\braket{\sigma_n^z}$ of the magnetization   is given by the integral of $\Delta\braket{c_k^\dagger c_k}$ over $k$. It is seen that the thermal correction is dependent on $\mu$ in the topological range $|\mu|<2|J|$; however, this dependence becomes very weak if, while $\exp(\omega_0/T)\gg 1$, we have $T\gg |J|$, so that the factor $\tilde M_k$ is essentially independent of $k$.  

A similar calculation gives for a fast turn-on
\[\braket{c_k^\dagger c_{-k}^\dagger} = -\frac{F_k M_k}{2\ep_k^2}
[1-2\exp(\beta\tilde M_k)].\]
It follows from these expressions and Eq.~\eqref{eq:two_correlations} that, for a fast turn-on and for $T\gg |J|$, the magnetization correlators $\mathcal{Q}^z(m)$ remain equal to zero even where the terms $\sim \exp(-\beta\omega_0)$ are taken into account. 

For a slow turn-on of the drive, in the adiabatic approximation the time-averaged fermion density is 
\begin{align}
\label{eq:thermal_density_change_slow}
\Delta\braket{c_k^\dagger c_k}= \frac{\ep_k - |M_k|}{2\ep_k}[1-2\exp(\beta \tilde M_k)].
\end{align}
In this case the thermal correction to the magnetization change is also independent of $\mu$ for $T\gg |J|$.

%%%%%%%%%%%%%%%%%%%%%%%%%%%%%%%%%%%%%%%%%
%%%%%%%%%%%%%%%%%%%%%%%%%%%%%%%%%%%%%%%

 \section{Analytical results for slow turn-on}
 \label{sec:quasi-adiabatic}
 \subsection{Adiabatic approximation}
 \label{subsec:adiabatic_approximation}

In this section, we consider the dynamics where the drive amplitude is slowly switched on, varying monotonously from $F(t=0)=0$ to $F(t_f)=F_\mathrm{fin}\equiv F$ (to maintain consistence with the notations in the previous section we use $F$ for the finite value of the drive amplitude). Expectation values of the spin operators are then time-averaged from $t>t_f$ till $t=t_f+T$. 

The Heisenberg  equations for the operators $d_k$ are
\[\dot{\vec{d}}_k = -i[\vec{d}_k, H_0] - U^\dagger(k)\dot U(k) \vec{d}_k, \quad\vec {d}_k=\left(\begin{array}{c} d_k \\ d_{-k}^{\dagger}\end{array}\right), \quad U(k) = \exp\left[-i\frac{\theta_k(t)}{2}\tau^y\right],
\]
where we take into account that the rotation angle $\theta_k(t)=\mathrm{Arctan}[F_k(t)/M_k]$ depends on time, since $F_k(t) = 2F(t)\sin k$ varies in time; we remind that we define $\theta_k$ as a continuous function of $k$. Therefore, in the topological regime, $\sgn\,\theta_k$ does not change where $M_k$ goes through zero; also, $\sgn\,\theta_k(t)$ does not change when $F_k(t)$ is a monotonous function of time.

For $M_k \cos\theta_k <  0$ the above equations take the form
\begin{align}
\label{eq:d_k_t}
\dot d_k =- i\ep_k d_k + \frac{1}{2}\dot\theta_k d_{-k}^\dagger, \quad
\dot d_{-k}^\dagger = i\ep_k d_{-k}^\dagger - \frac{1}{2}\dot\theta_k d_{k}.
\end{align}
where $\ep_k\equiv\ep_k(t)=\sqrt{M_k^2+F_k(t)^2}$.

The adiabatic approximation refers to the case where  $|\dot\theta_k|\ll\ep_k$. In this approximation, 
\[d_k(t)=d_k(0)\exp\left[-i\int_0^t dt'\ep_k(t')\right] \quad (M_k\cos\theta_k <0).\]
The value of $d_k(0)$ here is found from Eq.~\eqref{eq:canonical} by noting that $\theta_k(0)$ equals either $0$ or $\pi$ for $M_k\neq 0$ and $F\to 0$. In the considered case  $M_k \cos\theta_k<0$, we have $d_k(0)=c_k(0)$ for $M_k<0$ and $d_k(0)=c_{-k}^\dagger(0)$ for $M_k>0$. Then from equation $\vec {c}_k= U(k) \vec d_k$
\begin{align}
\label{eq:ck_quasi-adiabatic}
 c_k(t) =\left\{
 \begin{array}{cc}
 c_k(0)e^{-i\Phi(t)}\cos[\theta_k(t)/2]-c_{-k}^\dagger(0)e^{i\Phi(t)}\sin[\theta_k(t)/2],& \; M_k<0\;[\theta_k(0) = 0]\\
 c_{-k}^\dagger(0)e^{-i\Phi(t)}\cos[\theta_k(t)/2]-c_k(0)e^{i\Phi(t)}\sin[\theta_k(t)/2],& \; M_k>0\; [\theta_k(0)=\pi]\\
 \end{array}
 \right.
 \end{align}
where $\Phi(t)=\int_0^t dt'\ep_k(t')$. Thus, the fermionic expectation values are given by

\begin{align}
\label{eq:blocks_for_correlators}
&\braket{c_k^\dagger(t) c_k(t)}= \frac{1}{2}(1-|\cos\theta_k|) = \frac{\ep_k-|M_k|}{2\ep_k},
\nonumber\\
&\braket{c_k^\dagger(t) c_{-k}^\dagger(t)} = \frac{-1}{2}\sin\theta_k=-\frac{F_k}{2\ep_k}\sgn\,M_k. 
\end{align}
The last terms in the right-hand sides of Eq.~(\ref{eq:blocks_for_correlators}) apply also for $M_k\cos\theta_k>0$. Equations~(\ref{eq:blocks_for_correlators}) allow us to calculate the correlators $q_{1,2}(m)$, Eq.~\eqref{eq:two_correlations}, which determine the time-averaged expectation values and the correlation functions of the spins, 
\begin{align}
    \label{eq:corr_quasi-adiabatic}
    & q_1(m)= \frac{2}{N}\sum_k \braket{c_k^\dagger c_k} e^{ikm}= \delta_{m,0} - \frac{1}{N}\sum_k\sqrt{\frac{M_k^2}{M_k^2+F_k^2}}\cos(mk),\nonumber\\ 
    & q_2(m) = \frac{-2i}{N}\sum_k\braket{c_k^\dagger c_{-k}^\dagger} e^{ikm}=\frac{-1}{N}\sum_k \sin\theta_k \sin(mk).
\end{align}
One can change from  summation to integration over $k$ in these expressions, and then to integration over $w=\exp(ik)$ or $w=\exp(-ik)$. The resulting integrals have branching points inside the unit circle $|w|=1$ for  $P_1(w)P_2(w)=0$, where the polynomials $P_{1,2}(w)$ are defined in Eq.~(\ref{eq:Q1andQ2}). Therefore the structure of the integrals is different in the topologically trivial and topologically nontrivial regimes. This leads to a different dependence of the spin correlators on $\mu = \frac{1}{2}\omega_F-\omega_0$, and thus on the parametric drive frequency $\omega_F$. However, the explicit integration is cumbersome. In the main text we show the results obtained by calculating the integrals numerically.

%%%%%%%%%%%%%%%%%%%%%%%%%%%%%%%%%%%%%%%
%%%%%%%%%%%%%%%%%%%%%%%%%%%%%%%%%%%%%

\subsection{Scaling of non-adiabatic corrections}
\label{subsec:Nonquasi-adiabaticity}

The adiabaticity condition $|\dot\theta_k| \ll \ep_k$ breaks down where the energy gap is small.  Inside the topological regime, $|\mu/2J|<1$, the gap closes for $F(t=0)= 0$ and $k=\pm k_{\mu}$, where
\[k_{\mu}=\arccos(\mu/2J).\] 
Therefore, the adiabatic approximation does not apply for small $|F|$  and $\Bigl||k|-|k_\mu| \Bigr|$. The explicit applicability condition follows from Eq.~(\ref{eq:canonical}) for $\theta_k$, which shows that 
\[\dot\theta_k = 2\dot F \sin k\, M_k/\ep_k^2.\]
Thus, the  region of $k$, where $\ep_k \lesssim |\dot \theta_k|$ is limited to 
$\Bigl||k|-|k_\mu|\Bigr|\lesssim |\dot F/J^2\sin k_\mu|^{1/2}$.
The analysis of the contribution of this region and the corrections to the adiabatic approximation are discussed  in Sec.~\ref{sec:quasi_adiabatic_Schrodinger}. These corrections strongly depend on the relation between the size of the gap $|2F_k|$ after it has been opened and the bandwidth $J$. They also depend on the interrelation between $|2F_k|$ and the characteristic time $\tau_\mathrm{sl}$ it took the system to reach the stationary state or, equivalently, the rate at which the drive amplitude is increased.  The limit $|F_k|\tau_\mathrm{sl} \ll 1$ is discussed in Sec.~\ref{subsec:weak_drive_slow}. The above estimate of the smallness of the region of $k$ where the adiabaticity is broken refers to the opposite limit. See Sec.\ref{subsubsec:correction} for more details.

%\MD{\bf remove the next clause:} The resulting corrections to the spin correlators and to $\braket{\sigma_z}$ are proportional to the size of this region. As described in the main text, they scale as $|\dot F/G_\mu|^{1/2}$. We note that such quasi-adiabaticity breaking occurs only for small $|F(t)|$; for $|F(t)/J|\gtrsim |\dot F/G_\mu|^{1/2} $ the quasi-adiabaticity is restored.

The energy gap closes also at the boundary of the topological regime, for  $\mu/2J=1$ and $k=0$, and for $\mu/2J=-1$ and $k=\pi$. For $|\mu/2J|=1$, the range of $k$ where the quasi-adiabaticity is broken  scales as  $|\dot F/J^2|^{1/3}$.% for $|F(t)/J| <|\dot F/J^2|^{1/3}$.  

We note that the dynamical equations can be solved in the explicit form where the gap is closed, i.e.,  $M_k= 0$. In this case, from the Heisenberg equations for  the fermionic operators $c_k(t)$ we find
\begin{equation}
\label{eq:nonquasi-adiabatic_ck}
    c_k(t)= \cos\left(\int_0^tF_k(t')dt'\right)c_k(0)+i\sin\left(\int_0^tF_k(t')dt'\right)c_{-k}^\dagger(0), \quad k=\pm k_\mu.
\end{equation}
The time evolution of $c_k(t)$ as described by Eq.~(\ref{eq:nonquasi-adiabatic_ck}) is reminiscent of Rabi oscillations. Oscillations of $c_k(t)$ persist away from the points $\ep_k=0$. In the quasi-adiabatic regime they have frequency $\ep_k(t)$ and are superposed on the smooth quasi-adiabatic evolution; they lead to  oscillations of the spin variables. Of primary interest for the experiment are time-averaged spin correlators that we calculate. The simulation results shown in the main text also refer to time-averaged spin correlators, as we discuss in Sec.~\ref{sec:simulation}. Interestingly, it follows from Eq,~(\ref{eq:nonquasi-adiabatic_ck}) that $\braket{c_k^\dagger c_k} = 1/2$ for $k=\pm k_\mu$, which is the same value that follows form the adiabatic expression \eqref{eq:blocks_for_correlators}.

%%%%%%%%%%%%%%%%%%%%%%%%%%%%%%%%%%%%%%%%%%%%%%%%%%%%%
%%%%%%%%%%%%%%%%%%%%%%%%%%%%%%%%%%%%%%%%%

\subsection{Hysteresis}
\label{subsec:Hysteresis}

The difference between the topological and trivial regimes leads to a memory effect. Consider  a state with a given drive amplitude $F_\mathrm{fin}$ and the scaled detuning of the drive frequency $\mu_f$ in the topological regime $|\mu_f/2J|<1$. The system can be brought to this state in two different ways. First, one can slowly switch on the drive amplitude at $\mu=\mu_f$. The resulting average pair correlators of the fermions are given by Eq.~\eqref{eq:blocks_for_correlators}. Alternatively, one can quasi-adiabatically switch on the drive amplitude at a detuning $\mu=\mu_0$ that lies in the trivial regime, $|\mu_0/2J|>1$, and then slowly change $\mu$ to $\mu_f$. Since the initial value of $\sgn\,M_k=\sgn\,\mu_0$ is the same for all $k$ and the sign of $M_k\cos\theta_k$ is not changing with the slowly varying $\mu$, the resulting  expectation values of the correlators in the topological regime become 
\begin{align}
\label{eq:blocks_for_hyst}
&\braket{c_k^\dagger(t) c_k(t)} = \frac{\ep_k-M_k\,\sgn\,\mu_0}{2\ep_k},
\nonumber\\
&\braket{c_k^\dagger(t) c_{-k}^\dagger(t)} = -  \frac{F_k}{2\ep_k}\,\sgn\,\mu_0 
\end{align}
(we assume here that $k$ is away from the critical value $0$ or $\pi$, depending on $\sgn\,\mu_0$)

\begin{figure}[t]
\includegraphics[scale=0.5]{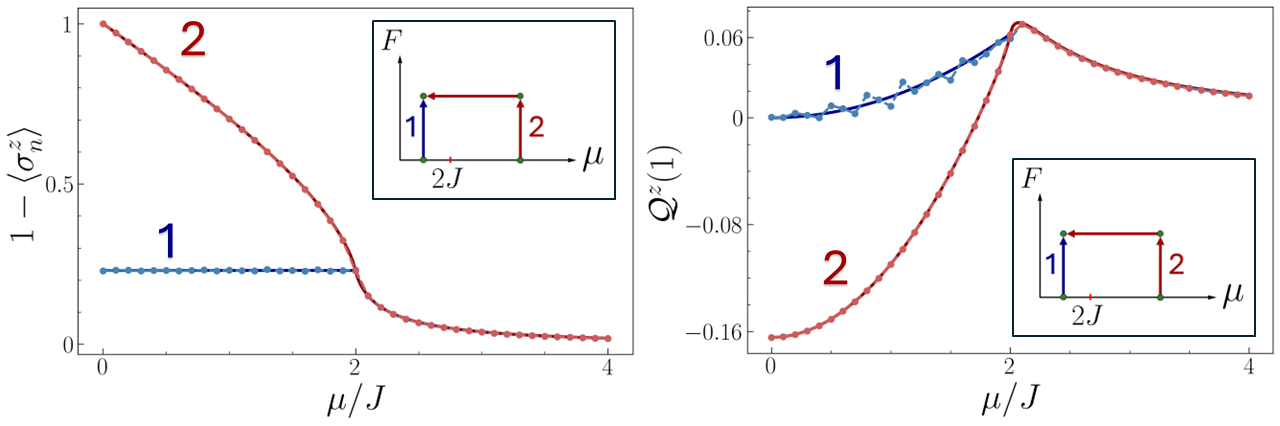}
	\caption{The time-averaged correlators $1-\braket{\sigma_z}$ (left panel) and $\mathcal{Q}^z(1)$ (right panel) obtained by slowly increasing the drive to $F_\mathrm{fin}/J=0.5$ for a fixed $\mu$ (line 1 and blue dots) and by slowly turning on the drive to the same $F_\mathrm{fin}/J$ in the trivial regime and then slowly changing the modulation frequency to arrive at the same $\mu$  (line 2 and red dots). The simulations refer to $N=40$ and $F(t) = F_\mathrm{fin} t/\tau_\mathrm{sl}$ with $\tau_\mathrm{sl} = 10^3/J$. 
}
\label{fig:hyst}
\end{figure}

As shown in the main text, this leads to a pronounced difference in the values of the spin correlators for the same drive amplitude and detuning depending on how these  parameters  have been reached, and thus to a memory effect. This effect can be observed using the observable $\braket{\sigma_n^+\sigma_{n+1}^-}$ shown in the main text as well as other observables, such as $1-\braket{\sigma_z}$ and $\mathcal{Q}^z(1)$. We present the results for the latter observables in Fig.~\ref{fig:hyst}. We note that, when the scaled detuning of the drive frequency $\mu$ is changed along the second path, i.e., starting with $|\mu(0)/2J|>1$, the phase transition at $|\mu(t)/2J|=1$ is crossed, leading to generation of  excitations near $k=0$ or $\pi$ via the Kibble-Zurek mechanism \cite{Dziarmaga2010,Polkovnikov2011}. The resulting corrections to the time-averaged expectation values of the spin correlators and $\braket{\sigma_z}$ remain small in the thermodynamic limit for a slowly varying $\mu$.

\section{Solving the Schr\"odinger equation for a slowly turned on drive}
\label{sec:quasi_adiabatic_Schrodinger}

The Hamiltonian \eqref{eq:H_matrix1} couples fermions of opposite momenta. Thus, the wave function of the system can be written as
\[\ket{\Psi(t)} = \prod_{k>0}\left[\sum_{m,n}A_{mn}(k,t)\ket{m_{-k},n_k}\right],\]
where $m_k,n_k =\{0,1\}$ are the occupation numbers of the states of $c_k$-fermions, $c_k\ket{n_k}=n_k\ket{n_k-1}$. The Schr\"{o}dinger equation for the system is reduced to a set of equations for $A_{mn}(k)$ for each $k$. Moreover, the equations for even and odd $m+n$ are uncoupled.

The initial condition $\ket{\Psi(0)}=\ket{\Psi_0}$ gives $A_{mn}(k,0)=\delta_{m,0}\delta_{n,0}$, $\forall k$. Therefore of interest are the equations that couple $A_{00}(k,t)$ and $A_{11}(k,t)$. In the Hamiltonian form, they read
\begin{align}
\label{eq:schrodinger_eq}
&i\dot A_{ii}(k,t) = \partial H_k/\partial A_{ii}^*(k,t) \quad (i=0,1), \nonumber\\ 
& H_k = M_k\left[|A_{00}(k,t)|^2 - |A_{11}(k,t)|^2\right] - F_k\left[A_{00}(k,t)A^*_{11}(k,t) + \mathrm{c.c.}\right],\nonumber\\
%&i\dot A_{00}(k,t)=M_k A_{00}(k,t)-F_k A_{11}(k,t),\nonumber\\
%&i \dot A_{11}(k,t)=-M_k A_{11}(k,t)-F_k A_{00}(k,t),\nonumber\\
&A_{00}(k,0) = 1,\quad A_{11}(k,0) = 0,
\end{align}
or, in the explicit form, 
\begin{align}
\label{eq:Schrodinger_explicit}
i\dot A_{00}=M_k A_{00}-F_k A_{11},\quad i\dot A_{11}=-M_k A_{11}-F_k A_{00} \quad [A_{ii} \equiv A_{ii}(k,t)],
\end{align}
whereas $A_{01}(k,t)=A_{10}(k,t)=0$. We note that the functions $A_{00}, A_{11}$ describe also the time evolution of the fermionic operators in the Heisenberg picture, with 
$c_{-k}^\dagger(t)=A_{00}(k,t)c_{-k}^\dagger(0) +A_{11}^*(-k,t)c_k(0)$. In particular, where we consider parity-conserving excitation,  $A_{01}=A_{10} = 0$. Once the parameters reach their stationary values and, for these values, averaging over time is performed, we have
\begin{align}
\label{eq:correlator_to A}
\braket{c_k^\dagger(t)c_k(t)} = \overline{|A_{11}(k,t)|^2},
\quad \braket{c_k^\dagger(t)c_{-k}^\dagger(t)} = \overline{A_{11}(k,t)^* A_{00}(k,t)},
\end{align}
where the overline indicates time averaging. These relations simplify  a comparison with the previous sections. However, in this section it is more natural to consider functions, rather than operators, and therefore we are using here the Schr\"odinger picture.

%%%%%%%%%%%%%%%%%%%%%%%%%%%%%%

\subsection{Weak-drive limit for slow turn-on}
\label{subsec:weak_drive_slow}

It is instructive to consider the solution of Eq.~\eqref{eq:schrodinger_eq} in the case of weak drive. Of particular interest is the  topological regime in the range of the gap closure for $F=0$. Here the dynamics is determined by the interplay of three small quantities: the gap $M_k$ for $F=0$, the time-dependent value of $F(t)$ and its final value $F\equiv F_\mathrm{fin}$ at the end of the switch-on,  and the rate at which $F(t)$ is varied (the results can be extended to the case where $M_k$ is time-dependent). Away from the gap closure, where $M_k$ largely exceeds the switching rate, one can use the adiabatic approximation of Sec.~\ref{subsec:adiabatic_approximation}. 

The analysis of the dynamics near the gap closure simplifies for small $F_\mathrm{fin}$. The smallness of $F_\mathrm{fin}$ includes not only that $|F_\mathrm{fin}|\ll |J|$ but also that $|F_\mathrm{fin}|$ is small compared to the characteristic rate at which $F(t)$ is turned one. Stand out in this respect are two simple models of the turn on: 
\begin{align}
\label{eq:exp_turnon}
F(t) = F_\mathrm{fin}\exp(\delta_{on}t), \quad -\infty < t\leq 0,   \quad 0<\delta_{on} \ll |J|,
\end{align}
and a linear ramp until $F(t) = F_\mathrm{fin}$, after which $F$ is independent of time,
\begin{align}
\label{eq:lin_turnon}
F(t) = \eta (t+\tau_\mathrm{sl}), \quad 0 \geq t\geq -\tau_\mathrm{sl}, \quad F_\mathrm{fin}=\eta\tau_\mathrm{sl}, \quad 0 < \eta \ll J^2.
\end{align}

It is tempting to think that, to the first order in $F$, one can solve Eq.~(\ref{eq:schrodinger_eq}) by setting $A_{00}(t) = \exp(-iM_k t)$. However, this approximation does not hold for all $M_k$ if $|F_\mathrm{fin}|\gtrsim \delta_{on}$, in the model (\ref{eq:exp_turnon}), or $|F_\mathrm{fin}|\gtrsim 1/\tau_\mathrm{sl}$, in the model \eqref{eq:lin_turnon}). 

\begin{figure}[h]
\includegraphics[scale=0.5]{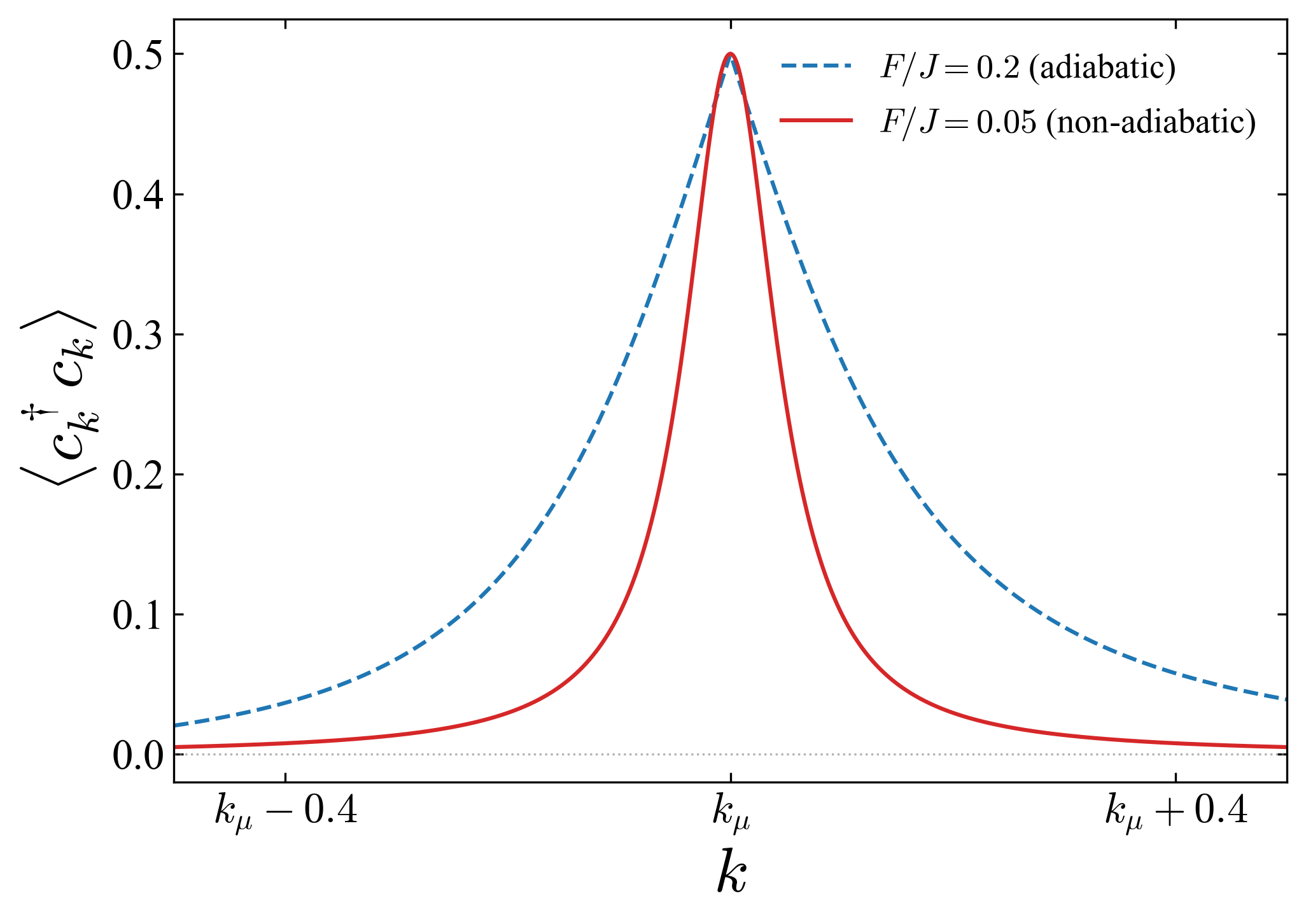}
	\caption{The densities $\braket{c_k^\dagger c_k}$ of the time-averaged magnetization change for a slow turn-on of the parametric drive.  The plots refer to $\mu/J = 1$, so that $k_\mu=\arccos(\mu/2J) =\pi/3$.  The solid line shows the Lorentzian peak at $k=k_\mu$  for a nonadiabatic regime near the gap closer, where the characteristic time to turn-on the drive is shorter than the reciprocal drive amplitude $F\equiv F_\mathrm{fin}$ (in frequency units), Eq.~\eqref{eq:corr_peak_exp_lin_turnon}. The dashed line shows the result of the adiabatic approximation for small $|F/J|$, but $F$ much larger than the reciprocal time of turning on the drive, Eq.~\eqref{eq:adiabatic_peak}.  In the both cases the shapes of the peaks are independent of the scaled drive frequency $\mu$.
}
\label{fig:non_adiabat}
\end{figure}

In the model \eqref{eq:exp_turnon} the perturbation theory in $F$ applies for $|F_\mathrm{fin}| \ll \delta_{on} \ll |J|$. In this case, for $t<0$, one can disregard the term $F_kA_{11}$ in the equation  \eqref{eq:Schrodinger_explicit}) for $A_{00}$, so that, in this time range, 
\begin{align}
\label{eq:pert_theory_A_11}
A_{11}(k,t) \approx \int_{-\infty}^t dt'\, \exp(iM_kt - 2iM_kt')F(t'), \quad t<0.
\end{align} 
This gives, for the time after the drive has reached a stationary value,
\[A_{11}(t) \approx -\frac{F_k}{2\ep_k(2M_k+i\delta_{on})}
\left[(\ep_k- M_k -i\delta_{on})e^{i\ep_k t}  
+ (\ep_k +M_k +i\delta_{on}) e^{-i\ep_k t}\right], \quad t\geq 0
\]
($F_k \equiv 2F_\mathrm{fin} \sin k$), 
so that $|A_{11}|\ll 1$ and
\begin{align}
\label{eq:exp_turn_answer}
\braket{c_k^\dagger(t)c_k(t)} =\overline{|A_{11}(t)|^2} \approx \frac{2F_\mathrm{fin}^2\sin^2 k}{\ep_k^2(4M_k^2 + \delta^2)}(\ep_k^2 + M_k^2 + \delta^2) 
\end{align}
Note that here $\ep_k = (M_k^2 + F_k^2)^{1/2}$, it is independent of time. The correlator $\braket{ c_k^\dagger(t) c_k(t)}$ displays two sharp Lorentzian peaks near $k=\pm k_\mu$  [$k_\mu = \arccos(\mu/2J)$], 
\begin{align}
\label{eq:corr_peak_exp_lin_turnon}
\braket {c_k^\dagger(t) c_k(t)}\approx \frac{|F_\mathrm{fin}|}{2|J|} \frac{|F_\mathrm{fin}/J|}{(|k|-|k_\mu|)^2 + |F_\mathrm{fin}/J|^2}
\end{align}
The areas of the peaks $\pi |F_\mathrm{fin}/J|$ are independent of $\mu$.

We also have
\begin{align}
\label{eq:dagger_exp_turnon}
\braket{c_k^\dagger c_{-k}^\dagger} = -\frac{F_\mathrm{fin}}{J}\,
\frac{k - |k_\mu|}{(k \pm k_\mu)^2 + (F_\mathrm{fin}/J)^2}, \quad |k|-|k_\mu|\ll 1.
\end{align}

In the model \eqref{eq:lin_turnon}, the perturbation theory in $F_\mathrm{fin}=\eta\tau_\mathrm{sl}$ applies for $|F_\mathrm{fin}| \ll |J|,\tau_\mathrm{sl}^{-1}$, and $ |\eta| \ll J^2$. Then
\begin{align*}
&A_{11}(t) \approx i \frac{F_k}{\ep_k} \sin\ep_k t +\frac{A_{11}(0)}{2 \ep_k}\left [(\ep_k+ M_k) e^{i\ep_k t} + (\ep_k - M_k)e^{-i\ep_kt}\right], \nonumber\\
&A_{11}(0) = -\frac{F_k}{2M_k} + i\frac{F_k}{4M_k^2\tau_\mathrm{sl}}\left(1 - e^{2iM_k\tau_\mathrm{sl}}\right)
\end{align*}
Here, too, of particular interest is the range $|M_k|\tau_\mathrm{sl}\ll 1$, where we have $A_{11}(0) \approx iF_k\tau_\mathrm{sl}/2$ and the correlator  $\braket {c_k^\dagger(t) c_k(t)}$ shows two Lorentzian peaks, which are given by Eq.~(\ref{eq:corr_peak_exp_lin_turnon}). For $|k|$ further away from $|k_\mu|$, the correlator is small. 

These results show that the linear in $F_\mathrm{fin}$ terms are not described by the adiabatic theory: one has to assume that the turn-on is faster than the final value of the drive amplitude. However, in the topologically nontrivial regime, the magnetization is independent of the scaled frequency detuning $\mu$. We note that the halfwidths of the peaks \eqref{eq:corr_peak_exp_lin_turnon} are $|F_\mathrm{fin}/J|$, they are much smaller than what one might expect from the finite duration of the turn-on, $\delta_{on}/J, J\tau_\mathrm{sl}^{-1}$. 

It is instructive to compare this result with Eq.~(\ref{eq:blocks_for_correlators}), which is obtained in the adiabatic approximation, i.e., disregarding the adiabaticity breaking near the gap closing. There, for small $F/J$, the correlator $\braket {c_k^\dagger(t) c_k(t)}$ also displays sharp peaks for small $|k| - |k_\mu|$, 
\begin{align}
\label{eq:adiabatic_peak}
\braket {c_k^\dagger(t) c_k(t)} \approx \frac{1}{2}  \,\frac{\left[(|k|-|k_\mu|)^2 + F^2/J^2\right]^{1/2} - \Bigl\vert |k|-|k_\mu|\Bigr\vert}{\left[(|k|-|k_\mu|)^2 + F^2/J^2\right]^{1/2}}
\qquad (F\equiv F_\mathrm{fin}).
\end{align}
The peaks are independent of $\mu$, and so is $\int dk\,\braket {c_k^\dagger(t) c_k(t)}\propto |F/J|$, which explains the independence of the magnetization of $\mu$ in the topologically nontrivial regime. Interestingly, they have the same height $1/2$ as the peaks \eqref{eq:corr_peak_exp_lin_turnon}, but have a different shape.

%%%%%%%%%%%%%%%%%%%%%
\subsubsection{Corrections to the adiabatic approximation for a slow turn-on rate}
\label{subsubsec:correction}

Adiabatic approximation breaks down in the regions of $k$ centered at $\pm k_\mu$. These regions are small for a small turn-on rate, as indicated in the main text. To confirm this, we solve numerically Eqs.~\eqref{eq:Schrodinger_explicit} with $F(t)$ linearly incremented to a certain value $F_\mathrm{fin}$, which is reached at $t=0$. We are interested in the parameter range $F_\mathrm{fin} =F(0)\gg \tau_\mathrm{sl}^{-1}$. For $t>0$ we have 
\begin{align*}
A_{mm}(k,t) = A_{mm}^+(k) \exp(i\ep_k t) + A_{mm}^- (k)\exp(-i\ep_k t) \quad (m=0,1)
\end{align*}
The parameters $A_{00}^\pm$ and $A_{11}^\pm$ are related by the equations of motion Eqs.~\eqref{eq:Schrodinger_explicit},
\[A_{11}^+(k) = F_k A_{00}^+(k)/(\ep_k-M_k), \quad A_{11}^-(k) = -F_k A_{00}^-(k)/(\ep_k+M_k),\]
$F_k\equiv 2F_\mathrm{fin}\sin k$. They also satisfy the conditions $A_{mm}^+(k) + A_{mm}^-(k) = A_{mm}(k,0)$ ($m=0,1$), where $A_{mm}(k,0)$ are found numerically by incrementing $F(t)$ from zero. This gives simple expressions for the time-averaged correlators 
\begin{align}
\label{eq:correlators_w_A_pm}
\braket{c_k^\dagger c_k} = |A_{11}^+(k)|^2 + |A_{11}^-(k)|^2, \quad
\braket{c_k^\dagger c_{-k}^\dagger} = \frac{\ep_k - M_k}{F_k}|A_{11}^+(k)|^2 -
\frac{\ep_k + M_k}{F_k}|A_{11}^-(k)|^2, 
\end{align}
in terms of the numerical solution of Eq.~(\ref{eq:Schrodinger_explicit}), with
\[ A_{11}^\pm(k) = [(\ep_k \pm M_k)A_{11}(k,0) \pm F_kA_{00}(k,0)]/2\ep_k.\]
\begin{figure}[h]
\includegraphics[scale=0.5]{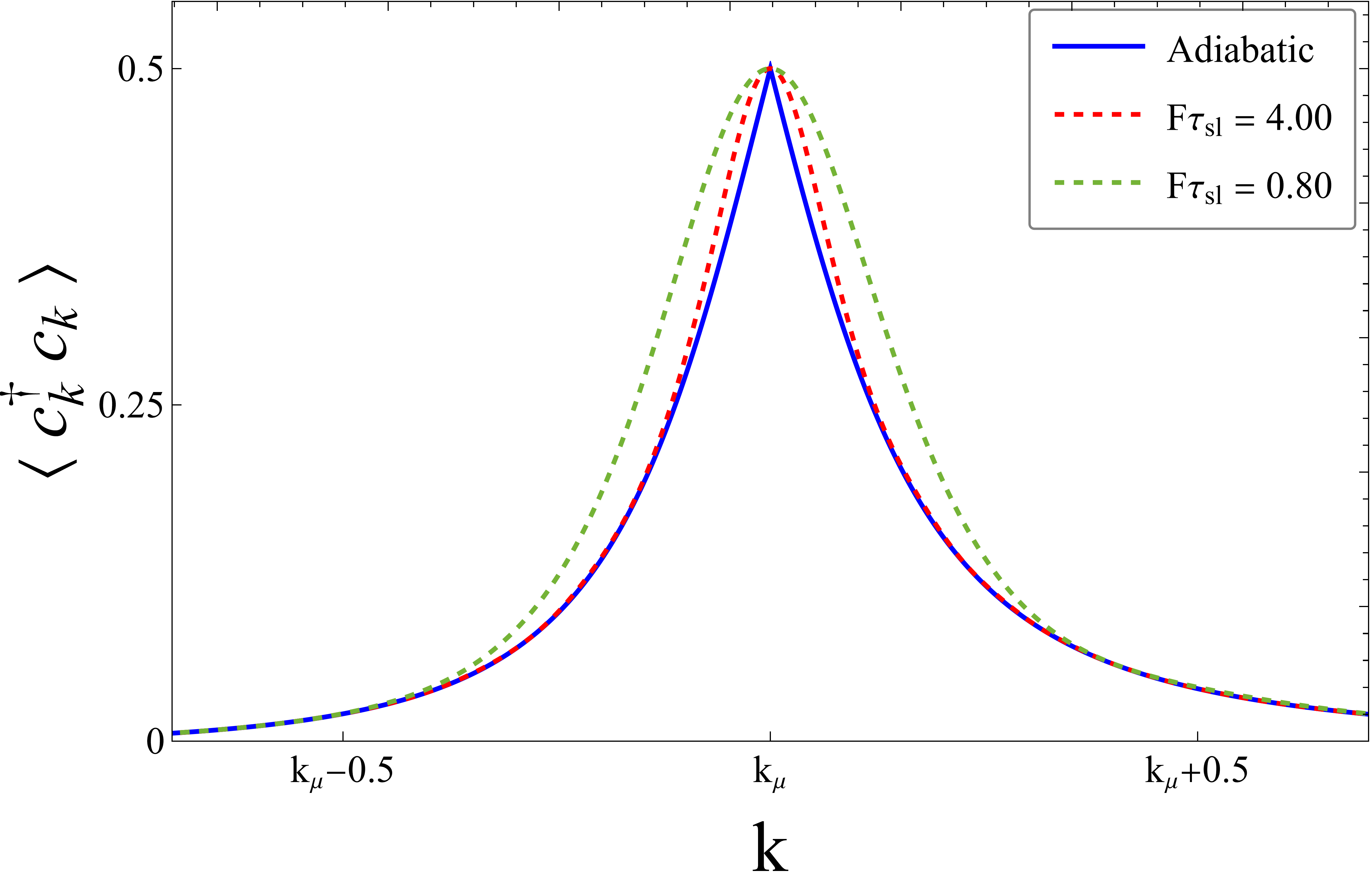}
	\caption{The density $\braket{c_k^\dagger c_k}$ of the time-averaged magnetization change for a slow turn-on of the parametric drive.  The plots refer to $\mu/J = 1$, so that $k_\mu=\arccos(\mu/2J) =\pi/3$ and to $F\equiv F_\mathrm{fin} = 0.2 J$.  The solid line shows the result of the adiabatic approximation.
The two dashed lines show $\braket{c_k^\dagger c_k}$ calculated from Eq.~\eqref{eq:correlators_w_A_pm} for a linearly increasing in time drive amplitude, $F(t) =F t/\tau_\mathrm{sl} $. The adiabatic approximation refers to the limit $J\tau_\mathrm{sl}\to \infty$, and moreover, $F\gg \tau_\mathrm{sl}^{-1}$. 
}
\label{fig:adiabatic_validity}
\end{figure}
These results are compared with the adiabatic expression in Fig.~\ref{fig:adiabatic_validity}. The agreement is very good already for $F\equiv F_\mathrm{fin} = 4\tau_\mathrm{sl}^{-1}$. Moreover even relatively far from the adiabatic limit, where  $F\equiv F_\mathrm{fin} = 0.8\tau_\mathrm{sl}^{-1}$ the results are close. This provides justification to using the adiabatic approximation to describe the magnetization. We checked that a close agreement holds also for spin correlators.

%%%%%%%%%%%%%%%%%%%%%%%%%%%%%%%%%%%
%%%%%%%%%%%%%%%%%%%%%%%%%%%%%%%%%%%%%

\subsection{The WKB approximation}
\label{subsec:WKB}

Equations~\eqref{eq:schrodinger_eq} for $A_{00}\equiv A_{00}(k,t), A_{11}\equiv A_{11}(k,t)$ can be written as%, we see that we can write $A_{11}$ as $$, where $A_{ii}\equiv A_{ii}(k,t)$. Then, this equation can be written as

\begin{equation}
\label{eq:second_order_ak}
    \ddot A_{00} = \frac{-i\dot F_k}{F_k}\left(i\dot A_{00} -M_k A_{00}\right)-\ep_k^2 A_{00}-i \dot M_k A_{00}, \qquad A_{11}=-(i\dot{A}_{00} -M_k A_{00})/F_k(t)
\end{equation}
We will seek the solution of these equations in the WKB approximation. This approximation relies on the slowness of the variation of $F_k$ and $M_k$. The small  parameter of the theory is $|\dot\ep_k|/\ep_k^2$. For simplicity, we present the results for the case of time-independent $M_k$, although they  immediately extend to a slowly varying in time $M_k$. 

The structure of Eq.~(\ref{eq:second_order_ak}) for $A_{00}$ is different from the more conventional structure of the equations solved in the WKB approximation, where there are no terms with first derivatives. These terms can be eliminated (see below), but it is more convenient to seek the solution as a sum of an eikonal and non-eikonal terms. To the first order in $|\dot\ep_k|/\ep_k^2 \ll 1$ the eikonal term  is 
\begin{align}
\label{eq:WKB_alpha_1_1}
&A_{00}(k,t) = A_1(k,t) + A_2(k,t); \quad A_1(k,t)=C_{k1}\exp[iS(t)], \quad S(t) \approx S\0(t) + S\1(t), \nonumber\\
&S\0(t) =-\int_0^t\ep_k(t')dt'\,(\sgn\, M_k), \qquad
S\1(t) = \frac{i}{2}\left(\log\frac{\ep_k(t)}{\ep_k(0)}-\log\frac{\ep_k(t)+ |M_k|}{\ep_k(0)+ |M_k|}\right),
\end{align}
where $C_{k1}$ is a constant to be determined by the initial conditions. The WKB approximation implies that $|\dot S\0|\gg |\dot S\1|$, which can be shown to be equivalent to the condition $|\dot\ep_k|/\ep_k^2 \ll 1$.% If  the drive is turned on in the trivial regime, $\sgn\, M_k = \sgn \,\mu_0$, and if $\mu_0$ is then slowly changed, $\sgn\,M_k$ does not change.

The second solution is given by the method of parameter variations,
\begin{align}
\label{eq:WKB_alpha_2}
&A_2(k,t)(t) = C_{k2}\exp[iS(t)] \int_0^t dt' F_k(t') \exp[-2iS(t')].
\end{align}
The constants $C_{k1}$ and  $C_{k2}$ can be found if the inequality $|\dot\ep_k|\ll \ep_k^2$ holds in the whole time interval during which the drive is turned on. For $F_k(t)\to 0$ for $t\to 0$, for the values of  we obtain from $A_{00}(k,t=0)=1$
\begin{align}
    \label{eq:fully_quasi-adiabatic_WKB}
    C_{k1} =1,\quad C_{k2} = 0,\quad A_{11}(k,t) \approx 
    -\sgn\,M_k\,\frac{F_k(t)}{\ep_k(t) + |M_k|} \exp[iS(t)].
\end{align}

It is easy to see that, for $\Bigl\vert |k| - |k_\mu|\Bigr| \gg |\dot F/J\sin k_\mu|^{1/2}$, Eqs.~(\ref{eq:WKB_alpha_1_1}) and (\ref{eq:fully_quasi-adiabatic_WKB})  give the quasi-adiabatic results of Eq.~\eqref{eq:ck_quasi-adiabatic}. In particular,
\begin{align}
\label{eq:relate_A_11_to_spins}
&    \braket{c_k^\dagger(t)c_k(t)} = |A_{11}(k,t)|^2 \to \frac{\ep_k -|M_k|}{2\ep_k},\\
& \braket{c_k^\dagger(t)c_{-k}^\dagger(t)} = A_{11}^*(k,t)A_{00}(k,t) \to -\frac{F_k(t)\,\sgn\,M_k}{2\ep_k(t)}.
\end{align}

%%%%%%%%%%%%%%%%%%%%%%%%%%%%%%%%%%
%%%%%%%%%%%%%%%%%%%%%%%%%%%%%%%%%%%%

%On the other hand, for $M_k=0$, where $\ep_k(t) = F_k(t)$,   to obtain the nonquasi-adiabatic solution, one needs to take the limit $M_k\rightarrow0$ before the limit $t\rightarrow0^+$. In this case, we get $C_{k2}=i$ which gives the results obtained in Eq.~\eqref{eq:nonquasi-adiabatic_ck}. This concludes that the solution developed here captures both the quasi-adiabatic and the fully nonquasi-adiabatic solutions discussed in earlier sections. 

\subsubsection{Vicinity of the gap closing for \texorpdfstring{$F_k=0$}{Fk=0}}

In the topological regime, the WKB condition $\dot \ep_k\ll \ep_k^2$ breaks down for small $|M_k|$, %in the range of small times $t$, 
i.e., for $|k|$ close to the  gap-closing value $|k_\mu|=|\arccos(\mu/2J)|$, which is determined by the condition  $\ep_{k_\mu}=0$  for $F_k=0$. Formally, for a nonzero $M_k$, we have $\dot\ep_k\propto F_k\dot F_k/M_k\to 0$ for $t\to 0$, since $F_k(t)\to 0$ for $t\to 0$. However, already for a small nonzero time, we can have $|M_k| \lesssim |F_k|$ for small $|M_k|$. Then the WKB appoximation requires that $|\dot\ep_k|\approx |\dot F_k|$ be small compared to $ \ep_k^2 \approx F_k^2$; this  condition breaks down for small $F_k$, and thus for a small nonzero time, even if $F_k$ is  varying slowly. 

For later times, as $F_k$ increases, the condition $|\dot F_k|\ll F_k^2$ can be met. Therefore, for small $|M_k|$, the WKB approximation becomes valid for not too small times, when
\begin{align}
\label{eq:restored_quasi-adiabaticity}
    F_k^2(t)\gg |\dot F_k(t)|.
\end{align}
The corresponding  WKB solution cannot be extended to $t=0$, and therefore the constants $C_{k1,2}$ in $A_{11}(k,t)$ are undetermined. 

The analysis is simplified if one takes into account that, in the  large-time limit, one can expand in Eq.~(\ref{eq:WKB_alpha_2}) $S(t') \approx S(t) + (t'-t)\dot S $, which gives
\begin{align}
    \label{eq:asymptotic_WKBA_2}
    A_2(k,t) \approx -i C_{k2} (\sgn\,M_k)\frac{F_k(t)}{2\ep_k(t)}\exp[-iS(t)], \quad
\end{align}
and
\begin{align}
    \label{eq:asymptotic_WKB_A11}
    A_{11}(k,t)\approx &-\frac{C_{k1} \exp[iS(t)]}{\ep_k+|M_k|}\left[F_k(t)\,\sgn\, M_k -i\frac{\dot F_k |M_k|}{2\ep_k^2} \right]\nonumber\\
   & -i\frac{C_{k2}\exp[-iS(t)]}{2\ep_k}\left[\ep_k+ |M_k|-\frac{F_k\dot F_k |M_k|}{\ep_k^2 (\ep_k+|M_k|)}\right]
\end{align}
In what follows we disregard the terms $\propto \dot F_k$ in $A_{11}(k,t)$. We remind that $A_{00}(k,t) = A_1(k,t) + A_2(k,t)$, with $A_{1,2}$ given by Eqs.~(\ref{eq:WKB_alpha_1_1}) and (\ref{eq:asymptotic_WKBA_2}).

The parameters $C_{k1,2}$ can be found by noticing that, quite generally, $F_k(t)\propto t$ for a time that is small, on the one hand but, on the other hand, is sufficiently long, so that $F_k(t)$
becomes large compared to $M_k$ and the above asymptotic  expressions for $A_{11}, A_{22}$ apply. We use this  to obtain the full WKB solution by matching these expressions to the explicit solution of the Schr\"odinger equation  (\ref{eq:schrodinger_eq}) for $F_k(t)\propto t$ in terms of the parabolic cylinder functions.

For $F_k(t) = \eta_k t$ and $|\eta_k|t\gg |M_k|$, the condition of the applicability of the WKB approximation  $\dot\ep_k \ll \ep_k^2$ is reduced to the inequality   $|\eta_k| \ll (\eta_k t)^2$.  The asymptotic expressions for $A_{00}(k,t)$, $A_{11}(k,t)$ strongly simplify in this case. We have in these expressions 
\begin{align}
    \label{eq:asymptotic_WKB_large_t}
    &S\0(t) \approx -\frac{1}{2}|\eta_k| (\sgn\,M_k)\left\{t^2 + \frac{M_k^2}{2\eta_k^2}\left[\log(4\eta_k^2 t^2/M_k^2)+1\right]\right\},\nonumber\\
    &S\1(t) \approx i\frac{1}{2}\log 2 - i\frac{|M_k|}{2|\eta_k|t},
    \quad \ep_k(t) \approx |\eta_k|                        t\left(1 + \frac{M_k^2}{2\eta_k^2 t^2}\right).
    \end{align}
%
%These expressions will be used to match the expansion of the parabolic cylinder functions for $|\eta_kt| \gg |M_k|$.
%
Respectively, the expressions for $A_{00}$ and $A_{11}$ take the form  
\begin{align}
    \label{eq:asymptotic_WKB_large_t_A}
    &A_{00}(k,t) \approx \frac{C_{k1}}{\sqrt{2}}\left(1+\frac{|M_k|}{2|\eta_k|t}\right)e^{iS^{(0)}(t)}-i\frac{C_{k2}}{\sqrt{2}}\left(\sgn(M_k\eta_k)-\frac{M_k}{2\eta_k t}\right)e^{-iS^{(0)}(t)},\nonumber\\
    &A_{11}(k,t) \approx -\frac{C_{k1}}{\sqrt{2}}\left(\sgn(M_k\eta_k)-\frac{M_k}{2\eta_kt}\right)e^{iS^{(0)}(t)}-i\frac{C_{k2}}{\sqrt{2}}\left(1+\frac{|M_k|}{2|\eta_k| t}\right)e^{-iS^{(0)}(t)}.
    \end{align}
These expressions will be used to match the expansion of the parabolic cylinder functions for $|\eta_kt| \gg |M_k|$. To this end, it is convenient
to introduce functions $A_{\pm}=(A_{00}\pm A_{11})/\sqrt{2}$. In particular,
\begin{align}
\label{eq:A_-_asymptot}
 A_{-}(k,t) =\left\{
 \begin{array}{cc}
  C_{k1} \exp(iS^{(0)})+iC_{k2} (M_k/2\eta_k t) \exp(-iS^{(0)}),& \; \sgn(M_k\eta_k)=+1\\
  -C_{k1} (M_k/2\eta_k t)\exp(iS^{(0)})+ iC_{k2} \exp(-iS^{(0)}) ,& \; \sgn(M_k\eta_k)=-1.\\
 \end{array}
 \right.
 \end{align}
whereas
\begin{align}
\label{eq:A_+_asymptot}
 A_+(k,t) =\left\{
 \begin{array}{cc}
  C_{k1} (M_k/2\eta_k t)\exp(iS^{(0)}) - iC_{k2} \exp(-iS^{(0)}),& \; \sgn(M_k\eta_k)=+1\\
  C_{k1} \exp(iS^{(0)})+ iC_{k2} (M_k/2\eta_k t)\exp(-iS^{(0)}) ,& \; \sgn(M_k\eta_k)=-1.\\
 \end{array}
 \right.
 \end{align}
Here $S\0\equiv S\0(t)$ is given by Eq.~(\ref{eq:asymptotic_WKB_large_t}).

The parameters $C_{k1,2}$ depend on the ratio $M_k^2/\eta_k$. They are found by matching to the parabolic cylinder functions. One can then find $A_{00},A_{11}$ in the stationary regime, i.e., where $F_k(t)$ has reached its stationary value and no longer varies in time, $\dot F_k=0$. To the first order in $M_k/F_k$ 
\begin{align}
\label{eq:stationary_with_C_{k1}_2}
&A_{00}(k,t) =\frac{ C_{k1}}{\sqrt{2}}e^{iS\0(t)}\left(1+\frac{|M_k|}{2|F_k|}\right) 
- i\frac{C_{k2}}{\sqrt{2}}\sgn(M_k\eta_k)e^{-iS\0(t)}\left(1 -\frac{|M_k|}{2|F_k|}\right),\nonumber\\
&A_{11}(k,t) = -\frac{ C_{k1}}{\sqrt{2}} \sgn(M_k\eta_k)e^{iS\0(t)}\left(1-\frac{|M_k|}{2|F_k|}\right) 
- i\frac{C_{k2}}{\sqrt{2}}e^{-iS\0(t)}\left(1 +\frac{|M_k|}{2|F_k|}\right),
 \end{align}
The factors $\exp[\pm iS\0(t)]$ in these expressions are fast oscillating, and therefore they drop out from the time-averaged parameters 
\begin{align}
\label{eq:averaged_nonquasi-adiabatic}
&\braket{|A_{11}(k,t)|^2}\approx \frac{1}{2}|C_{k1}|^2\left(1-\frac{|M_k|}{|F_k|}\right) +\frac{1}{2}| C_{k2}|^2\left(1 +\frac{|M_k|}{|F_k|}\right), \nonumber\\
&\braket{A^*_{11}(k,t) A_{00}(k,t)}
\approx \frac{1}{2}\left(|C_{k2}|^2 - |C_{k1}|^2\right)
\end{align}
These parameters  give the correlators we are interested in, cf. Eq.~(\ref{eq:relate_A_11_to_spins}).

%%%%%%%%%%%%%%%%%%%%%%%%%%%%%%%%%%%%%%%%

\subsection{Solution in terms of the parabolic cylinder functions}

For a linear turn-on of the drive, $F_k(t) = \eta_k t$, Eq.~(\ref{eq:schrodinger_eq}) can be solved using the parabolic cylinder functions. Indeed, from Eq.~(\ref{eq:schrodinger_eq}), we obtain the equation for $A_\pm(t)$, which has the form
\begin{align}
    \label{eq:weber_eq}
    &\frac{d^2A_\pm}{dz_k^2}+\left( s_k \mp\frac{1}{2}-\frac{z_k^2}{4}\right)A_\pm=0,\quad 
    z_k = e^{i(\pi/4) \sgn\,\eta_k}(2|\eta_k|)^{1/2}t,\quad s_k=- i\frac{M_k^2}{2\eta_k},
\end{align}
with 
\[A_+ = ie^{i(\pi/4) \sgn\,\eta_k}\sgn\,M_k\left(\frac{dA_-}{dz_k} +\frac{1}{2}z_k A_-\right)/|s_k|^{1/2}
%(i\dot A_- - \eta_k t A_-)/M_k
.\] Equation \eqref{eq:weber_eq} is the Weber equation. Its general solution can be expressed in terms of the parabolic cylinder functions $D_\nu(z)$ \cite{Gradshteyn2015}, 
\begin{align}
    \label{eq:parabolic_cylinder}
    &A_-=\alpha_k D_{-s_k-1}(-i z_k) + \beta_k D_{-s_k-1}(i z_k), \nonumber\\
    &A_+=\frac{e^{i(\pi/4) \sgn\,\eta_k}\sgn\,M_k}{\sqrt{|s_k|}}\left(-\alpha_k D_{-s_k}(-i z_k) + \beta_k D_{-s_k}(i z_k)\right),
\end{align}
we consider $\pm iz_k \equiv \exp(\pm i\pi/2)z_k$. The factors $\alpha_k$ and $\beta_k$ are determined by the initial conditions, which give
\begin{align}
\label{eq:alpha_beta}
    \alpha_k &= \frac{2^{s_k/2}}{\sqrt{4\pi}}
    \left[
        \Gamma\!\left(1+\frac{s_k}{2}\right)
        - e^{-i(\pi/4) \sgn\,\eta_k}\sgn\,M_k \sqrt{\frac{|s_k|}{2}}\Gamma\!\left(\frac{1}{2}+\frac{s_k}{2}\right)
    \right], \nonumber\\[6pt]
    \beta_k &= \frac{2^{s_k/2}}{\sqrt{4\pi}}
    \left[
        \Gamma\!\left(1+\frac{s_k}{2}\right)
        + e^{-i(\pi/4)\sgn\,\eta_k} \sgn\,M_k \sqrt{\frac{|s_k|}{2}}\Gamma\!\left(\frac{1}{2}+\frac{s_k}{2}\right)
    \right],
\end{align}
where $\Gamma(z)$ is the Gamma function. 

Equations (\ref{eq:parabolic_cylinder}) and (\ref{eq:alpha_beta}) provide a  general solution of the problem of turning on the drive linearly in time  at an arbitrary rate $\eta_k$ for an arbitrary $M_k$. The solution depends on $z\propto \eta_k t$ and the dimensionless parameter $s_k\propto M_k^2/\eta_k$. It allows us to find the coefficients $C_{k1}$ and $C_{k2}$, and thus to obtain the full WKB solution of the problem of the two-spin dynamics in the case where the drive is initially turned on linearly in time. This can be done  by matching the asymptotic expansion of the solution \eqref{eq:parabolic_cylinder} in the range $|\eta_k|^{1/2}t \gg 1$ with  Eqs.~\eqref{eq:A_-_asymptot} and (\ref{eq:A_+_asymptot}) in the same time range. For completeness, we provide the leading-order terms of the   expansion of the parabolic cylinder function $D_\nu(z)$ for $|z|\gg 1,|\nu|$ \cite{Gradshteyn2015}:
\begin{align}
\label{eq:D_expansions}
 D_{\nu}(z) \approx\left\{
 \begin{array}{cc}
  e^{-z^2/4}z^{\nu},& \; |\arg(z)|<3\pi/4\\
  &\\
  e^{-z^2/4} z^{\nu} -\frac{\sqrt{2\pi}}{\Gamma(-\nu)}e^{i\nu\pi} e^{z^2/4} z^{-\nu-1},& \; \pi/4<\arg(z)<5\pi/4\\
  &\\
  e^{-z^2/4} z^{\nu} -\frac{\sqrt{2\pi}}{\Gamma(-\nu)}e^{-i\nu\pi} e^{z^2/4} z^{-\nu-1},& \; -5\pi/4<\arg(z)<-\pi/4.\\
 \end{array}
 \right.
 \end{align}
We will use that, from Eqs.~(\ref{eq:asymptotic_WKB_large_t}) and (\ref{eq:weber_eq}), in the large-$t$ limit
\[e^{iS_0(t)} \approx e^{-(z_k^2/4)\sgn (M_k\eta_k)}z^{s_k\,\sgn (M_k\eta_k)}
\exp\left[\frac{1}{2}s_k\, \sgn(M_k\eta_k)\left(-i\frac{\pi}{2}\sgn\,\eta_k - \log|s_k|+1\right)\right]
\]
whereas 
\[\frac{M_k}{2\eta_k t} = \sgn (M_k\eta_k) \frac{\sqrt{|s_k|}}{z_k} e^{i(\pi/4)\sgn\,\eta_k}\]

To express the parameters of the WKB solution $C_{k1}, C_{k2}$ in terms of the coefficients $\alpha_k,\beta_k$ of the solution of the equations of motion in terms of the parabolic cylinder functions, it is convenient to introduce auxiliary parameters
\begin{align}
    \label{eq:auxiliary}
 % &J_k = \frac{\sqrt{2\pi}}{\Gamma(s_k+1)} |s_k|^{s_k/2}  \exp(-s_k/2),\nonumber\\
 % & L_k(\alpha_k,\beta_k) = |s_k|^{-(s_k +1)/2}\left\{ \alpha_k  \exp\left[\frac{s_k}{2}\left(1+ \frac{i\pi}{2}\right)\right] -\beta_k 
%    \exp\left[\frac{s_k}{2}\left(1 -\frac{3i\pi}{2}\right)\right]\right\}
%\end{align}
%
%\ME{
  &J_k = \frac{\sqrt{2\pi}}{\Gamma(s_k+1)} |s_k|^{s_k/2} 
    \exp\left(-\frac{s_k}{2}-\frac{\pi|s_k|}{4}\right),\nonumber\\
  & L_k = |s_k|^{-(s_k +1)/2} \exp\left(\frac{i\pi}{4}\sgn\,\eta_k+\frac{s_k}{2}+ \frac{\pi |s_k|}{4}\right).
\end{align}
%}

In terms of these parameters,
from Eqs.~(\ref{eq:A_-_asymptot}), (\ref{eq:A_+_asymptot}), and (\ref{eq:parabolic_cylinder}) we find that, for $M_k >0$ and $\eta_k >0$
\begin{align}
    \label{eq:C_k12_positive_eta_M}
% &C_{k1} = \beta_k J_k\exp(-i \pi s_k/4), \quad C_{k2} = L_k(\alpha_k,\beta_k)\exp(-i\pi/4)\\
 %\end{align}
  C_{k1}=\beta_{k}J_k, \quad C_{k2} =-i\left(\alpha_k-\beta_k e^{-\pi |s_k|}\right) L_k.
\end{align}
For $M_k <0$ and $\eta_k <0$, on the other hand,
\begin{align}
    \label{eq:C_k12_negative_eta_M}
 %   &C_{k1} = \alpha_k J_k \exp(i\pi s_k/4), \quad C_{k2} = L(\alpha_k,\beta_k) \exp[i\pi/4)(1+2s_k)]\\
  %\end{align}
%
%}
   C_{k1}=\alpha_{k}J_k, \quad C_{k2} =-i\left(\beta_k-\alpha_k e^{-\pi |s_k|}\right) L_k.
\end{align}
%}
%
For $M_k<0, \eta_k>0$ we find
\begin{align}
    \label{eq:C_k12_negative_M_positive_eta}
  %  &C_{k1} = L_k(\alpha_k,\beta_k)\exp(i\pi/4), \quad C_{k2} = -i\beta_k J_k\exp(-i\pi s_k/4)
   % \\
   %\end{align}
    C_{k1}=\left(\alpha_k-\beta_k e^{-\pi |s_k|}\right) L_k, \quad C_{k2}=-i\beta_{k}J_k .
\end{align}
%}
%
For $M_k>0, \eta_k<0$ we find
\begin{align}
    \label{eq:C_k12_positive_M_negative_eta}
 %   &C_{k1} = iL_k(\alpha_k,\beta_k)\exp[i(\pi/4)(1 +2s_k)], \quad
  %  C_{k2} = -i\alpha_k J_k\exp(i\pi s_k/4)
   % \\
   %\end{align}
%}
    C_{k1}=\left(\beta_k-\alpha_k e^{-\pi |s_k|}\right) L_k, \quad C_{k2}=-i\alpha_{k}J_k .
\end{align}

Equations (\ref{eq:auxiliary}) - (\ref{eq:C_k12_positive_M_negative_eta}), together with the expressions (\ref{eq:alpha_beta}) for the parameters $\alpha_k,\beta_k$ provide a complete solution of the problem of the slow  turn-on of the drive in the range of zero energy gap in the absence of the drive. For each $k$, the solution depends on a single parameter, the ratio of the squared gap $M_k^2$ to the rate of the increment of the drive $\eta_k$. We note that it is advantageous to express the coefficients $C_{k1}, C_{k2}$ in terms of $\alpha_k, \beta_k$, as this allows one to extend the WKB results to different initial conditions.

%%%%%%%%%%%%   Mahmoud %%%%%%%%%%%

%%%%%%%%%%%%%%%%%%%%%%%%%%%%%%%%%%%%%%%%%%%%%%%%%%%%%%%%%

\section{Numerical validation using matrix product states}
\label{sec:simulation}

To validate the analytical predictions derived in the thermodynamic limit ($N \rightarrow \infty$), we performed finite-size numerical simulations using the Matrix Product State (MPS) formalism. This approach is exceptionally well-suited for one-dimensional quantum systems with local interactions, such as the spin chain considered here.

\subsection{The matrix product state method}

The Hilbert space of a chain of $N$ spins has a dimension that grows exponentially as $2^N$. The MPS method resolves this dimensionality problem by efficiently representing a specific class of quantum states, characterized by a limited amount of entanglement. A quantum state of our spin chain can be written as:
\begin{equation}
    \ket{\Psi} = \sum_{s_1, \dots, s_N} C_{s_1 s_2 \dots s_N} \ket{s_1 s_2 \dots s_N}
\end{equation}
where $s_n =0,1$ enumerates the states of the spin at site $n$ and $C_{s_1 s_2 \dots s_N}$ is an element of a rank-$N$ tensor with $2^N$ elements. Following the standard prescription, we decompose this large tensor into a product of smaller rank-3 tensors, one for each site:
\begin{equation}
    \label{eq:mps}
    C_{s_1 s_2 \dots s_N} \approx A^{s_1}_{\alpha_0 \alpha_1} A^{s_2}_{\alpha_1 \alpha_2} \dots A^{s_N}_{\alpha_{N-1} \alpha_0}
\end{equation}
The maximal value of $\alpha_n$ is called the dimension of bond $n$. The maximal (over $n$) bond dimension is usually denoted $\chi$. It determines the maximum amount of entanglement the MPS takes into account, and thus it controls the accuracy of the approximation. The efficiency of MPS stems from the fact that many physically relevant states can be approximated with high fidelity using a bond dimension $\chi$ that is significantly smaller than what would be required to represent an arbitrary state in the Hilbert space.

We performed time  evolution of an MPS using the \textit{Time-Evolving Block Decimation (TEBD)} algorithm. For our nearest-neighbor Hamiltonian with an even number of sites, we write $\hat{H} = \hat{H}_{\text{even}} + \hat{H}_{\text{odd}}$, where $\hat{H}_{\text{even/odd}}=\sum_{j\in\text{even/odd}}\hat{h}_{j,j+1}$ with $\hat{h}_{j,j+1}$ being the  coupling of sites $j$ and $j+1$. A second-order Trotter decomposition is then:
$$
e^{-i\hat{H}\delta t} \approx e^{-i\hat{H}_{\text{odd}}\delta t/2} e^{-i\hat{H}_{\text{even}}\delta t} e^{-i\hat{H}_{\text{odd}}\delta t/2} + \mathcal{O}(\delta t^3).
$$
Applying a two-site gate $\exp(-ih_{j\,j+1}\delta t)$ increases the entanglement across the corresponding bond, which would lead to a growth in the required bond dimension. To keep the computation feasible, a truncation step is performed via singular value decomposition (SVD) across a bond, keeping only the $\chi$ largest singular values. This step controllably projects the state back into the manifold of states with a manageable bond dimension based on the desired accuracy.

After a quantum quench (sudden change in Hamiltonian parameters), entanglement typically grows linearly in time before it saturates. To maintain accuracy, the bond dimension must grow exponentially with time, which limits the maximum size of the simulated system. For an quasi-adiabatic evolution, the entanglement growth is much slower, allowing us to simulate larger systems.

Our simulations were implemented using the ITensor library \cite{fishman2021} with a TEBD time-step of $\delta t=0.1 J^{-1}$ ($J=1$ as the unit of energy, $\hbar = 1$). Reflecting the difference in entanglement growth, we used a chain of $N=40$ spins for the quasi-adiabatic protocols and a smaller chain of $N=20$ for the sudden turn-on protocol.

\subsection{Sudden turn-on simulations}
For the sudden turn-on, the system was initialized in the product state $\ket{\Psi_0}=\ket{\uparrow\dots\uparrow}$ and then evolved under a time-independent Hamiltonian for several values of detuning $\mu \in (0,5)$, separated by a step $\delta\mu = 0.1J$, for a fixed drive amplitude $F=0.5J$. The evolution was followed for $0<t\leq t_f = 50 J^{-1}$. To compare with the analytical predictions for the time-averaged expectation values, the expectation values of observables were averaged over the time interval from $t=20J^{-1}$ to $t=50J^{-1}$.

\begin{figure}[h]
\centering
\includegraphics[scale=0.15]{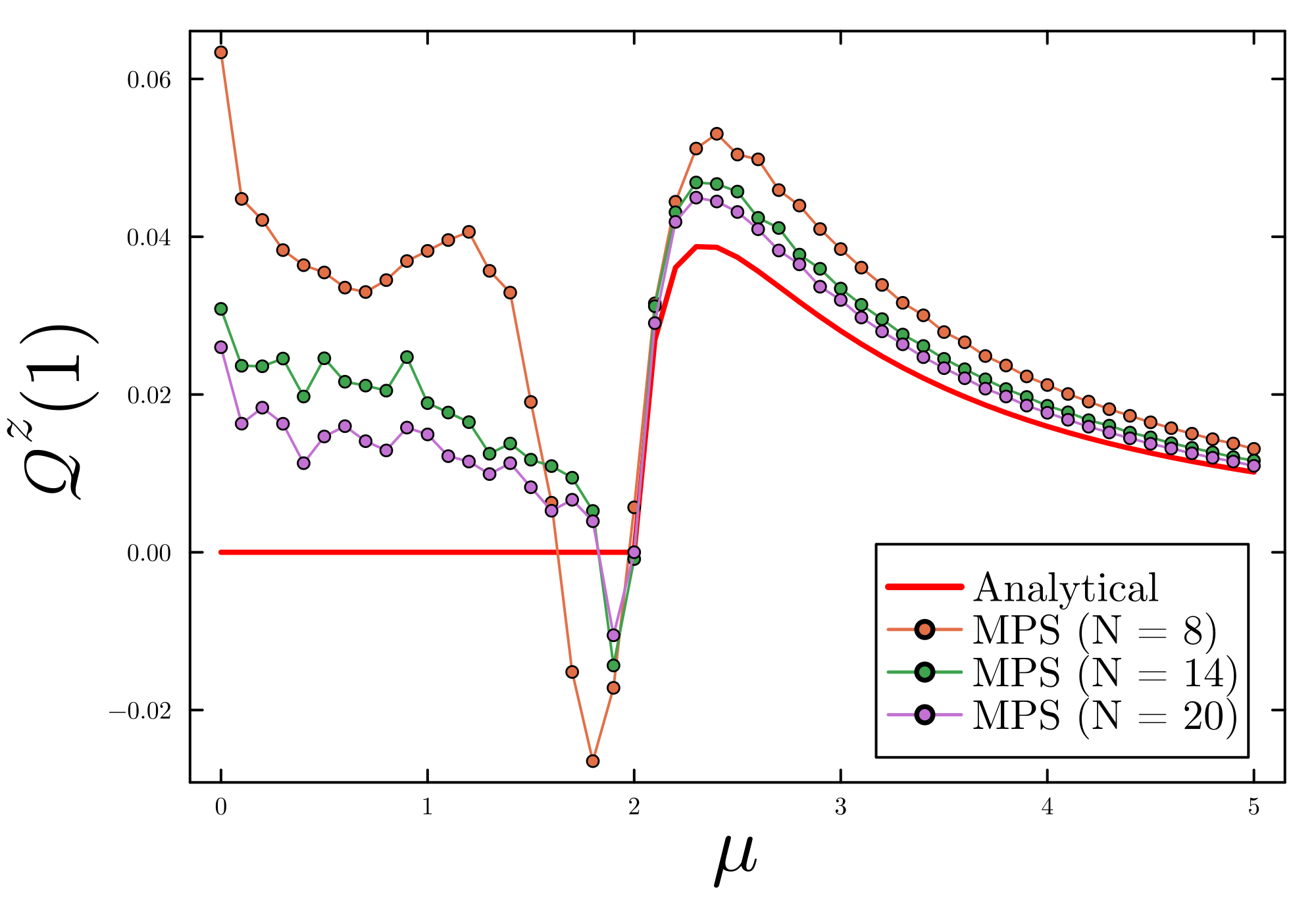}
	\caption{The time-averaged correlator $\mathcal{Q}^z(1)$ for a sudden turn-on of parametric modulation as a function of the scaled frequency detuning $\mu/J$. The final modulation amplitude is $F/J= 0.5$. The analytical result refers to the thermodynamic limit $N\to \infty$.}
\label{fig:Q_z_N}
\end{figure}

As discussed in the main text, the analytical results are derived in the thermodynamic limit, whereas our simulations are for finite $N$. We observe finite-size corrections in our results for the time-averaged nearest-neighbor $\sigma_z$ correlations. Figure~\ref{fig:Q_z_N} illustrates the deviation of the numerically computed nearest-neighbor correlator  $\mathcal{Q}^z(1)$ of the increments $\Delta\sigma^z_n$ from the analytical solution. The figure shows that the deviation quickly falls off with the increasing number of sites in the chain. 

\begin{figure}[h]
\centering
\includegraphics[scale=0.15]{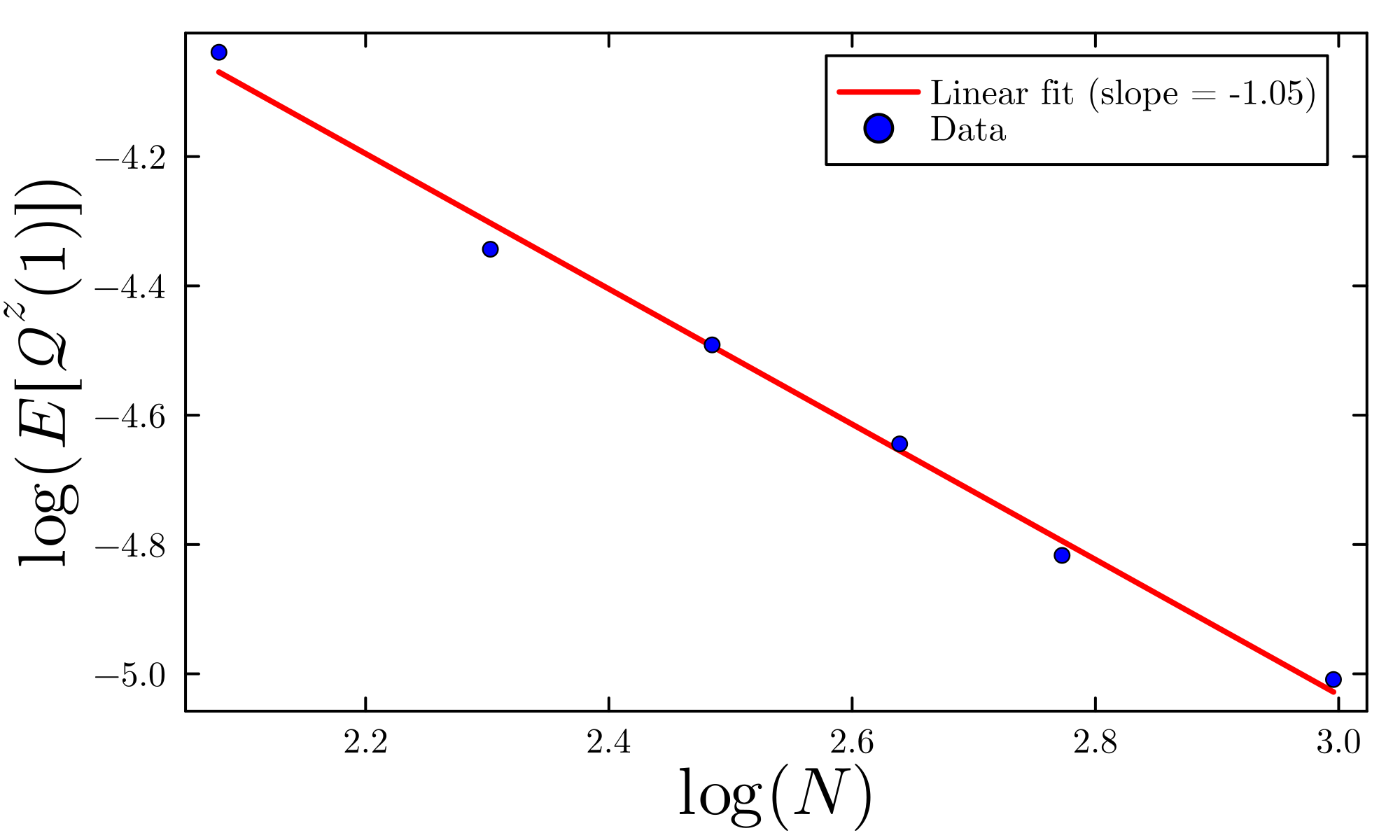}
	\caption{A log-log plot of the integrated error $E_N[\mathcal{Q}^z(1)]$ versus the number of sites $N$. The linear fit (dashed line) has a slope of approximately -1, confirming that the finite-size corrections to the correlator scale as $N^{-1}$.}
\label{fig:error_scaling}
\end{figure}

The scaling of the deviation wih the number of sites can be further quantified by an integrated error $E_N[\mathcal{Q}^z(1)]=\int_0^5d\mu|\mathcal{Q}^z(1)_{\infty}-\mathcal{Q}^z(1)_{N}|$, where $\mathcal{Q}^z(1)_{\infty}$ is the analytical result and $\mathcal{Q}^z(1)_{N}$ is the simulation result for $N$ sites. As shown in Fig.~\ref{fig:error_scaling}, this error scales as $N^{-1.05}$, confirming that our numerical results agree with the expected scaling of MPS simulations.

\subsection{Slow evolution and hysteresis}
The quasi-adiabatic switching protocol was simulated by linearly ramping the drive amplitude as $F(t)=F_\mathrm{fin}t/\tau_{sl}$ from $t=0$ to a ramp time of $t_f=\tau_{sl}=10^3J^{-1}$, reaching a final amplitude of $F_\mathrm{fin}=0.5J\gg 1/\tau_\mathrm{sl}$. This was independently performed for the same values of $\mu$ as in the sudden turn-on. After the ramp, the system was evolved with the constant final Hamiltonian for an additional time interval $\Delta t=30J^{-1}$, over which the time-averaging of observables was performed.

To demonstrate the hysteresis effect, we compare the results described above with the outcome of the following procedure. We switched on the drive amplitude quasi-adiabatically to a final value $F_\mathrm{fin}=0.5J$ at $t=t_{f_1}=10^{3}J^{-1}$ for $\mu=5J$ (i.e., in the trivial phase). Then $\mu$ was slowly varied as $\mu(t) =J[5-(t-t_{f_1})/\tau_\mathrm{sl}]$ from $t_{f_1}$ to $t_{f_1}+5 \tau_{sl}$ with the drive amplitude fixed at $F_\mathrm{fin}=0.5J$. This way we accumulated an array of wave functions for the values of $\mu \in (0,4)$ separated by $0.1 J$. For each of these wave functions the observables of interest were calculated. The system was then evolved for the corresponding $\mu$ for a duration $\Delta t = 30J^{-1}$. This allowed us to find the time-averaged values of the observables plotted in Fig.~\ref{fig:hyst} and in Fig.~3 of the main text.

%\bibliography{MyLibrary}

\begin{thebibliography}{58}%
\makeatletter
\providecommand \@ifxundefined [1]{%
 \@ifx{#1\undefined}
}%
\providecommand \@ifnum [1]{%
 \ifnum #1\expandafter \@firstoftwo
 \else \expandafter \@secondoftwo
 \fi
}%
\providecommand \@ifx [1]{%
 \ifx #1\expandafter \@firstoftwo
 \else \expandafter \@secondoftwo
 \fi
}%
\providecommand \natexlab [1]{#1}%
\providecommand \enquote  [1]{``#1''}%
\providecommand \bibnamefont  [1]{#1}%
\providecommand \bibfnamefont [1]{#1}%
\providecommand \citenamefont [1]{#1}%
\providecommand \href@noop [0]{\@secondoftwo}%
\providecommand \href [0]{\begingroup \@sanitize@url \@href}%
\providecommand \@href[1]{\@@startlink{#1}\@@href}%
\providecommand \@@href[1]{\endgroup#1\@@endlink}%
\providecommand \@sanitize@url [0]{\catcode `\\12\catcode `\$12\catcode
  `\&12\catcode `\#12\catcode `\^12\catcode `\_12\catcode `\%12\relax}%
\providecommand \@@startlink[1]{}%
\providecommand \@@endlink[0]{}%
\providecommand \url  [0]{\begingroup\@sanitize@url \@url }%
\providecommand \@url [1]{\endgroup\@href {#1}{\urlprefix }}%
\providecommand \urlprefix  [0]{URL }%
\providecommand \Eprint [0]{\href }%
\providecommand \doibase [0]{https://doi.org/}%
\providecommand \selectlanguage [0]{\@gobble}%
\providecommand \bibinfo  [0]{\@secondoftwo}%
\providecommand \bibfield  [0]{\@secondoftwo}%
\providecommand \translation [1]{[#1]}%
\providecommand \BibitemOpen [0]{}%
\providecommand \bibitemStop [0]{}%
\providecommand \bibitemNoStop [0]{.\EOS\space}%
\providecommand \EOS [0]{\spacefactor3000\relax}%
\providecommand \BibitemShut  [1]{\csname bibitem#1\endcsname}%
\let\auto@bib@innerbib\@empty
%</preamble>
\bibitem [{\citenamefont {Bloembergen}\ and\ \citenamefont
  {Wang}(1954)}]{Bloembergen1954}%
  \BibitemOpen
  \bibfield  {author} {\bibinfo {author} {\bibfnamefont {N.}~\bibnamefont
  {Bloembergen}}\ and\ \bibinfo {author} {\bibfnamefont {S.}~\bibnamefont
  {Wang}},\ }\bibfield  {title} {\bibinfo {title} {Relaxation {{Effects}} in
  {{Para-}} and {{Ferromagnetic Resonance}}},\ }\href
  {https://doi.org/10.1103/PhysRev.93.72} {\bibfield  {journal} {\bibinfo
  {journal} {Phys. Rev.}\ }\textbf {\bibinfo {volume} {93}},\ \bibinfo {pages}
  {72} (\bibinfo {year} {1954})}\BibitemShut {NoStop}%
\bibitem [{\citenamefont {Suhl}(1957)}]{Suhl1957}%
  \BibitemOpen
  \bibfield  {author} {\bibinfo {author} {\bibfnamefont {H.}~\bibnamefont
  {Suhl}},\ }\bibfield  {title} {\bibinfo {title} {The theory of ferromagnetic
  resonance at high signal powers},\ }\href@noop {} {\bibfield  {journal}
  {\bibinfo  {journal} {J. Phys. Chem. Solids}\ }\textbf {\bibinfo {volume}
  {1}},\ \bibinfo {pages} {209} (\bibinfo {year} {1957})}\BibitemShut {NoStop}%
\bibitem [{\citenamefont {Shan}\ \emph {et~al.}(2024)\citenamefont {Shan},
  \citenamefont {Curtis}, \citenamefont {Guo}, \citenamefont {Roh},
  \citenamefont {Rotundu}, \citenamefont {Lee}, \citenamefont {Narang},
  \citenamefont {Noh}, \citenamefont {Demler},\ and\ \citenamefont
  {Hsieh}}]{Shan2024}%
  \BibitemOpen
  \bibfield  {author} {\bibinfo {author} {\bibfnamefont {J.-Y.}\ \bibnamefont
  {Shan}}, \bibinfo {author} {\bibfnamefont {J.~B.}\ \bibnamefont {Curtis}},
  \bibinfo {author} {\bibfnamefont {M.}~\bibnamefont {Guo}}, \bibinfo {author}
  {\bibfnamefont {C.~J.}\ \bibnamefont {Roh}}, \bibinfo {author} {\bibfnamefont
  {C.~R.}\ \bibnamefont {Rotundu}}, \bibinfo {author} {\bibfnamefont {Y.~S.}\
  \bibnamefont {Lee}}, \bibinfo {author} {\bibfnamefont {P.}~\bibnamefont
  {Narang}}, \bibinfo {author} {\bibfnamefont {T.~W.}\ \bibnamefont {Noh}},
  \bibinfo {author} {\bibfnamefont {E.}~\bibnamefont {Demler}},\ and\ \bibinfo
  {author} {\bibfnamefont {D.}~\bibnamefont {Hsieh}},\ }\bibfield  {title}
  {\bibinfo {title} {Dynamic magnetic phase transition induced by parametric
  magnon pumping},\ }\href {https://doi.org/10.1103/PhysRevB.109.054302}
  {\bibfield  {journal} {\bibinfo  {journal} {Phys. Rev. B}\ }\textbf {\bibinfo
  {volume} {109}},\ \bibinfo {pages} {054302} (\bibinfo {year}
  {2024})}\BibitemShut {NoStop}%
\bibitem [{\citenamefont {Malz}\ \emph {et~al.}(2019)\citenamefont {Malz},
  \citenamefont {Knolle},\ and\ \citenamefont {Nunnenkamp}}]{Malz2019}%
  \BibitemOpen
  \bibfield  {author} {\bibinfo {author} {\bibfnamefont {D.}~\bibnamefont
  {Malz}}, \bibinfo {author} {\bibfnamefont {J.}~\bibnamefont {Knolle}},\ and\
  \bibinfo {author} {\bibfnamefont {A.}~\bibnamefont {Nunnenkamp}},\ }\bibfield
   {title} {\bibinfo {title} {Topological magnon amplification},\ }\href
  {https://doi.org/10.1038/s41467-019-11914-2} {\bibfield  {journal} {\bibinfo
  {journal} {Nat Commun}\ }\textbf {\bibinfo {volume} {10}},\ \bibinfo {pages}
  {3937} (\bibinfo {year} {2019})}\BibitemShut {NoStop}%
\bibitem [{\citenamefont {Arfini}\ \emph {et~al.}(2025)\citenamefont {Arfini},
  \citenamefont {{Bermejillo-Seco}}, \citenamefont {Bondarenko}, \citenamefont
  {Potts}, \citenamefont {Blanter}, \citenamefont {van~der Zant},\ and\
  \citenamefont {Steele}}]{Arfini2025}%
  \BibitemOpen
  \bibfield  {author} {\bibinfo {author} {\bibfnamefont {M.}~\bibnamefont
  {Arfini}}, \bibinfo {author} {\bibfnamefont {A.}~\bibnamefont
  {{Bermejillo-Seco}}}, \bibinfo {author} {\bibfnamefont {A.}~\bibnamefont
  {Bondarenko}}, \bibinfo {author} {\bibfnamefont {C.~A.}\ \bibnamefont
  {Potts}}, \bibinfo {author} {\bibfnamefont {Y.~M.}\ \bibnamefont {Blanter}},
  \bibinfo {author} {\bibfnamefont {H.~S.~J.}\ \bibnamefont {van~der Zant}},\
  and\ \bibinfo {author} {\bibfnamefont {G.~A.}\ \bibnamefont {Steele}},\
  }\href {https://doi.org/10.48550/arXiv.2506.11527} {\bibinfo {title}
  {Magnon-magnon interaction induced by nonlinear spin wave dynamics}}
  (\bibinfo {year} {2025}),\ \Eprint {https://arxiv.org/abs/2506.11527}
  {arXiv:2506.11527 [cond-mat]} \BibitemShut {NoStop}%
\bibitem [{\citenamefont {Kiselev}\ \emph {et~al.}(2025)\citenamefont
  {Kiselev}, \citenamefont {Karcher}, \citenamefont {Rudner}, \citenamefont
  {Duine},\ and\ \citenamefont {Lindner}}]{Kiselev2025}%
  \BibitemOpen
  \bibfield  {author} {\bibinfo {author} {\bibfnamefont {E.~I.}\ \bibnamefont
  {Kiselev}}, \bibinfo {author} {\bibfnamefont {J.~F.}\ \bibnamefont
  {Karcher}}, \bibinfo {author} {\bibfnamefont {M.~S.}\ \bibnamefont {Rudner}},
  \bibinfo {author} {\bibfnamefont {R.}~\bibnamefont {Duine}},\ and\ \bibinfo
  {author} {\bibfnamefont {N.~H.}\ \bibnamefont {Lindner}},\ }\href
  {https://doi.org/10.48550/arXiv.2507.08147} {\bibinfo {title} {Exciting
  terahertz magnons with amplitude modulated light: Spin pumping, squeezed
  states, symmetry breaking and pattern formation}} (\bibinfo {year} {2025}),\
  \Eprint {https://arxiv.org/abs/2507.08147} {arXiv:2507.08147 [cond-mat]}
  \BibitemShut {NoStop}%
\bibitem [{\citenamefont {Demokritov}\ \emph {et~al.}(2006)\citenamefont
  {Demokritov}, \citenamefont {Demidov}, \citenamefont {Dzyapko}, \citenamefont
  {Melkov}, \citenamefont {Serga}, \citenamefont {Hillebrands},\ and\
  \citenamefont {Slavin}}]{Demokritov2006}%
  \BibitemOpen
  \bibfield  {author} {\bibinfo {author} {\bibfnamefont {S.~O.}\ \bibnamefont
  {Demokritov}}, \bibinfo {author} {\bibfnamefont {V.~E.}\ \bibnamefont
  {Demidov}}, \bibinfo {author} {\bibfnamefont {O.}~\bibnamefont {Dzyapko}},
  \bibinfo {author} {\bibfnamefont {G.~A.}\ \bibnamefont {Melkov}}, \bibinfo
  {author} {\bibfnamefont {A.~A.}\ \bibnamefont {Serga}}, \bibinfo {author}
  {\bibfnamefont {B.}~\bibnamefont {Hillebrands}},\ and\ \bibinfo {author}
  {\bibfnamefont {A.~N.}\ \bibnamefont {Slavin}},\ }\bibfield  {title}
  {\bibinfo {title} {Bose--{{Einstein}} condensation of quasi-equilibrium
  magnons at room temperature under pumping},\ }\href
  {https://doi.org/10.1038/nature05117} {\bibfield  {journal} {\bibinfo
  {journal} {Nature}\ }\textbf {\bibinfo {volume} {443}},\ \bibinfo {pages}
  {430} (\bibinfo {year} {2006})}\BibitemShut {NoStop}%
\bibitem [{\citenamefont {L'vov}\ \emph {et~al.}(2023)\citenamefont {L'vov},
  \citenamefont {Pomyalov}, \citenamefont {Bozhko}, \citenamefont
  {Hillebrands},\ and\ \citenamefont {Serga}}]{Lvov2023}%
  \BibitemOpen
  \bibfield  {author} {\bibinfo {author} {\bibfnamefont {V.~S.}\ \bibnamefont
  {L'vov}}, \bibinfo {author} {\bibfnamefont {A.}~\bibnamefont {Pomyalov}},
  \bibinfo {author} {\bibfnamefont {D.~A.}\ \bibnamefont {Bozhko}}, \bibinfo
  {author} {\bibfnamefont {B.}~\bibnamefont {Hillebrands}},\ and\ \bibinfo
  {author} {\bibfnamefont {A.~A.}\ \bibnamefont {Serga}},\ }\bibfield  {title}
  {\bibinfo {title} {Correlation-{{Enhanced Interaction}} of a {{Bose-Einstein
  Condensate}} with {{Parametric Magnon Pairs}} and {{Virtual Magnons}}},\
  }\href {https://doi.org/10.1103/PhysRevLett.131.156705} {\bibfield  {journal}
  {\bibinfo  {journal} {Phys. Rev. Lett.}\ }\textbf {\bibinfo {volume} {131}},\
  \bibinfo {pages} {156705} (\bibinfo {year} {2023})}\BibitemShut {NoStop}%
\bibitem [{\citenamefont {Sandweg}\ \emph {et~al.}(2011)\citenamefont
  {Sandweg}, \citenamefont {Kajiwara}, \citenamefont {Chumak}, \citenamefont
  {Serga}, \citenamefont {Vasyuchka}, \citenamefont {Jungfleisch},
  \citenamefont {Saitoh},\ and\ \citenamefont {Hillebrands}}]{Sandweg2011}%
  \BibitemOpen
  \bibfield  {author} {\bibinfo {author} {\bibfnamefont {C.~W.}\ \bibnamefont
  {Sandweg}}, \bibinfo {author} {\bibfnamefont {Y.}~\bibnamefont {Kajiwara}},
  \bibinfo {author} {\bibfnamefont {A.~V.}\ \bibnamefont {Chumak}}, \bibinfo
  {author} {\bibfnamefont {A.~A.}\ \bibnamefont {Serga}}, \bibinfo {author}
  {\bibfnamefont {V.~I.}\ \bibnamefont {Vasyuchka}}, \bibinfo {author}
  {\bibfnamefont {M.~B.}\ \bibnamefont {Jungfleisch}}, \bibinfo {author}
  {\bibfnamefont {E.}~\bibnamefont {Saitoh}},\ and\ \bibinfo {author}
  {\bibfnamefont {B.}~\bibnamefont {Hillebrands}},\ }\bibfield  {title}
  {\bibinfo {title} {Spin {{Pumping}} by {{Parametrically Excited Exchange
  Magnons}}},\ }\href {https://doi.org/10.1103/PhysRevLett.106.216601}
  {\bibfield  {journal} {\bibinfo  {journal} {Phys. Rev. Lett.}\ }\textbf
  {\bibinfo {volume} {106}},\ \bibinfo {pages} {216601} (\bibinfo {year}
  {2011})}\BibitemShut {NoStop}%
\bibitem [{\citenamefont {L'vov}(2012)}]{Lvov2012}%
  \BibitemOpen
  \bibfield  {author} {\bibinfo {author} {\bibfnamefont {V.~S.}\ \bibnamefont
  {L'vov}},\ }\href@noop {} {\emph {\bibinfo {title} {Wave Turbulence under
  Parametric Excitation: Applications to Magnets}}},\ \bibinfo {edition} {1st}\
  ed.,\ Springer {{Series}} in {{Nonlinear Dynamics}}\ (\bibinfo  {publisher}
  {Springer Berlin, Heidelberg},\ \bibinfo {year} {2012})\BibitemShut {NoStop}%
\bibitem [{\citenamefont {Makiuchi}\ \emph {et~al.}(2021)\citenamefont
  {Makiuchi}, \citenamefont {Hioki}, \citenamefont {Shimazu}, \citenamefont
  {Oikawa}, \citenamefont {Yokoi}, \citenamefont {Daimon},\ and\ \citenamefont
  {Saitoh}}]{Makiuchi2021}%
  \BibitemOpen
  \bibfield  {author} {\bibinfo {author} {\bibfnamefont {T.}~\bibnamefont
  {Makiuchi}}, \bibinfo {author} {\bibfnamefont {T.}~\bibnamefont {Hioki}},
  \bibinfo {author} {\bibfnamefont {Y.}~\bibnamefont {Shimazu}}, \bibinfo
  {author} {\bibfnamefont {Y.}~\bibnamefont {Oikawa}}, \bibinfo {author}
  {\bibfnamefont {N.}~\bibnamefont {Yokoi}}, \bibinfo {author} {\bibfnamefont
  {S.}~\bibnamefont {Daimon}},\ and\ \bibinfo {author} {\bibfnamefont
  {E.}~\bibnamefont {Saitoh}},\ }\bibfield  {title} {\bibinfo {title}
  {Parametron on magnetic dot: {{Stable}} and stochastic operation},\ }\href
  {https://doi.org/10.1063/5.0038946} {\bibfield  {journal} {\bibinfo
  {journal} {Appl. Phys. Lett.}\ }\textbf {\bibinfo {volume} {118}},\ \bibinfo
  {pages} {022402} (\bibinfo {year} {2021})}\BibitemShut {NoStop}%
\bibitem [{\citenamefont {Elyasi}\ \emph {et~al.}(2022)\citenamefont {Elyasi},
  \citenamefont {Saitoh},\ and\ \citenamefont {Bauer}}]{Elyasi2022}%
  \BibitemOpen
  \bibfield  {author} {\bibinfo {author} {\bibfnamefont {M.}~\bibnamefont
  {Elyasi}}, \bibinfo {author} {\bibfnamefont {E.}~\bibnamefont {Saitoh}},\
  and\ \bibinfo {author} {\bibfnamefont {G.~E.~W.}\ \bibnamefont {Bauer}},\
  }\bibfield  {title} {\bibinfo {title} {Stochasticity of the magnon
  parametron},\ }\href {https://doi.org/10.1103/PhysRevB.105.054403} {\bibfield
   {journal} {\bibinfo  {journal} {Phys. Rev. B}\ }\textbf {\bibinfo {volume}
  {105}},\ \bibinfo {pages} {054403} (\bibinfo {year} {2022})}\BibitemShut
  {NoStop}%
\bibitem [{\citenamefont {Yuan}\ \emph {et~al.}(2022)\citenamefont {Yuan},
  \citenamefont {Cao}, \citenamefont {Kamra}, \citenamefont {Duine},\ and\
  \citenamefont {Yan}}]{Yuan2022}%
  \BibitemOpen
  \bibfield  {author} {\bibinfo {author} {\bibfnamefont {H.~Y.}\ \bibnamefont
  {Yuan}}, \bibinfo {author} {\bibfnamefont {Y.}~\bibnamefont {Cao}}, \bibinfo
  {author} {\bibfnamefont {A.}~\bibnamefont {Kamra}}, \bibinfo {author}
  {\bibfnamefont {R.~A.}\ \bibnamefont {Duine}},\ and\ \bibinfo {author}
  {\bibfnamefont {P.}~\bibnamefont {Yan}},\ }\bibfield  {title} {\bibinfo
  {title} {Quantum magnonics: {{When}} magnon spintronics meets quantum
  information science},\ }\href {https://doi.org/10.1016/j.physrep.2022.03.002}
  {\bibfield  {journal} {\bibinfo  {journal} {Physics Reports}\ }\textbf
  {\bibinfo {volume} {965}},\ \bibinfo {pages} {1} (\bibinfo {year}
  {2022})}\BibitemShut {NoStop}%
\bibitem [{\citenamefont {Holstein}\ and\ \citenamefont
  {Primakoff}(1940)}]{holstein1940}%
  \BibitemOpen
  \bibfield  {author} {\bibinfo {author} {\bibfnamefont {T.}~\bibnamefont
  {Holstein}}\ and\ \bibinfo {author} {\bibfnamefont {H.}~\bibnamefont
  {Primakoff}},\ }\bibfield  {title} {\bibinfo {title} {Field {{Dependence}} of
  the {{Intrinsic Domain Magnetization}} of a {{Ferromagnet}}},\ }\href
  {https://doi.org/10.1103/PhysRev.58.1098} {\bibfield  {journal} {\bibinfo
  {journal} {Phys. Rev.}\ }\textbf {\bibinfo {volume} {58}},\ \bibinfo {pages}
  {1098} (\bibinfo {year} {1940})}\BibitemShut {NoStop}%
\bibitem [{\citenamefont {Villain}(1974)}]{VILLAIN1974}%
  \BibitemOpen
  \bibfield  {author} {\bibinfo {author} {\bibfnamefont {J.}~\bibnamefont
  {Villain}},\ }\bibfield  {title} {\bibinfo {title} {Quantum-theory of 1- and
  2-dimensional ferromagnets and antiferromagnets {{W}} an easy magnetization
  plane. 1. {{Ideal}} 1-d or 2-d lattices without in-plane anisotropy},\ }\href
  {https://doi.org/10.1051/jphys:0197400350102700} {\bibfield  {journal}
  {\bibinfo  {journal} {Journal De Physique}\ }\textbf {\bibinfo {volume}
  {35}},\ \bibinfo {pages} {27} (\bibinfo {year} {1974})}\BibitemShut {NoStop}%
\bibitem [{\citenamefont {Jordan}\ and\ \citenamefont
  {Wigner}(1928)}]{Jordan1928}%
  \BibitemOpen
  \bibfield  {author} {\bibinfo {author} {\bibfnamefont {P.}~\bibnamefont
  {Jordan}}\ and\ \bibinfo {author} {\bibfnamefont {E.}~\bibnamefont
  {Wigner}},\ }\bibfield  {title} {\bibinfo {title} {{\"Uber das Paulische
  \"Aquivalenzverbot}},\ }\href {https://doi.org/10.1007/BF01331938} {\bibfield
   {journal} {\bibinfo  {journal} {Z. Physik}\ }\textbf {\bibinfo {volume}
  {47}},\ \bibinfo {pages} {631} (\bibinfo {year} {1928})}\BibitemShut
  {NoStop}%
\bibitem [{\citenamefont {Lieb}\ \emph {et~al.}(1961)\citenamefont {Lieb},
  \citenamefont {Schultz},\ and\ \citenamefont {Mattis}}]{Lieb1961}%
  \BibitemOpen
  \bibfield  {author} {\bibinfo {author} {\bibfnamefont {E.}~\bibnamefont
  {Lieb}}, \bibinfo {author} {\bibfnamefont {T.}~\bibnamefont {Schultz}},\ and\
  \bibinfo {author} {\bibfnamefont {D.}~\bibnamefont {Mattis}},\ }\bibfield
  {title} {\bibinfo {title} {Two soluble models of an antiferromagnetic
  chain},\ }\href {https://doi.org/10.1016/0003-4916(61)90115-4} {\bibfield
  {journal} {\bibinfo  {journal} {Annals of Physics}\ }\textbf {\bibinfo
  {volume} {16}},\ \bibinfo {pages} {407} (\bibinfo {year} {1961})}\BibitemShut
  {NoStop}%
\bibitem [{\citenamefont {Kitaev}(2001)}]{Kitaev2001}%
  \BibitemOpen
  \bibfield  {author} {\bibinfo {author} {\bibfnamefont {A.~Y.}\ \bibnamefont
  {Kitaev}},\ }\bibfield  {title} {\bibinfo {title} {Unpaired majorana fermions
  in quantum wires},\ }\href {https://doi.org/10.1070/1063-7869/44/10S/S29}
  {\bibfield  {journal} {\bibinfo  {journal} {Physics-Uspekhi}\ }\textbf
  {\bibinfo {volume} {44}},\ \bibinfo {pages} {131} (\bibinfo {year}
  {2001})}\BibitemShut {NoStop}%
\bibitem [{\citenamefont {Thakurathi}\ \emph {et~al.}(2013)\citenamefont
  {Thakurathi}, \citenamefont {Patel}, \citenamefont {Sen},\ and\ \citenamefont
  {Dutta}}]{Thakurathi2013}%
  \BibitemOpen
  \bibfield  {author} {\bibinfo {author} {\bibfnamefont {M.}~\bibnamefont
  {Thakurathi}}, \bibinfo {author} {\bibfnamefont {A.~A.}\ \bibnamefont
  {Patel}}, \bibinfo {author} {\bibfnamefont {D.}~\bibnamefont {Sen}},\ and\
  \bibinfo {author} {\bibfnamefont {A.}~\bibnamefont {Dutta}},\ }\bibfield
  {title} {\bibinfo {title} {Floquet generation of majorana end modes and
  topological invariants},\ }\href {https://doi.org/10.1103/PhysRevB.88.155133}
  {\bibfield  {journal} {\bibinfo  {journal} {Phys. Rev. B}\ }\textbf {\bibinfo
  {volume} {88}},\ \bibinfo {pages} {155133} (\bibinfo {year}
  {2013})}\BibitemShut {NoStop}%
\bibitem [{\citenamefont {Dykman}(2019)}]{Dykman2019a}%
  \BibitemOpen
  \bibfield  {author} {\bibinfo {author} {\bibfnamefont {M.~I.}\ \bibnamefont
  {Dykman}},\ }\bibfield  {title} {\bibinfo {title} {Coherent multiple-period
  states of periodically modulated qubits},\ }\href
  {https://doi.org/10.1103/PhysRevA.100.042101} {\bibfield  {journal} {\bibinfo
   {journal} {Phys. Rev. A}\ }\textbf {\bibinfo {volume} {100}},\ \bibinfo
  {pages} {042101} (\bibinfo {year} {2019})}\BibitemShut {NoStop}%
\bibitem [{\citenamefont {Yates}\ \emph {et~al.}(2019)\citenamefont {Yates},
  \citenamefont {Essler},\ and\ \citenamefont {Mitra}}]{Yates2019}%
  \BibitemOpen
  \bibfield  {author} {\bibinfo {author} {\bibfnamefont {D.~J.}\ \bibnamefont
  {Yates}}, \bibinfo {author} {\bibfnamefont {F.~H.~L.}\ \bibnamefont
  {Essler}},\ and\ \bibinfo {author} {\bibfnamefont {A.}~\bibnamefont
  {Mitra}},\ }\bibfield  {title} {\bibinfo {title} {Almost strong 0,
  \$\textbackslash pi\$ edge modes in clean, interacting {{1D}} floquet
  systems},\ }\href@noop {} {\bibfield  {journal} {\bibinfo  {journal} {Phys.
  Rev. B}\ }\textbf {\bibinfo {volume} {99}},\ \bibinfo {pages} {205419}
  (\bibinfo {year} {2019})}\BibitemShut {NoStop}%
\bibitem [{\citenamefont {Mi}\ \emph {et~al.}(2022)\citenamefont {Mi},
  \citenamefont {Sonner}, \citenamefont {Niu}, \citenamefont {Lee},
  \citenamefont {Foxen}, \citenamefont {Acharya}, \citenamefont {Aleiner},
  \citenamefont {Andersen}, \citenamefont {Arute}, \citenamefont {Arya},
  \citenamefont {Asfaw}, \citenamefont {Atalaya}, \citenamefont {Bardin},
  \citenamefont {Basso}, \citenamefont {Bengtsson}, \citenamefont {Bortoli},
  \citenamefont {Bourassa}, \citenamefont {Brill}, \citenamefont {Broughton},
  \citenamefont {Buckley}, \citenamefont {Buell}, \citenamefont {Burkett},
  \citenamefont {Bushnell}, \citenamefont {Chen}, \citenamefont {Chiaro},
  \citenamefont {Collins}, \citenamefont {Conner}, \citenamefont {Courtney},
  \citenamefont {Crook}, \citenamefont {Debroy}, \citenamefont {Demura},
  \citenamefont {Dunsworth}, \citenamefont {Eppens}, \citenamefont {Erickson},
  \citenamefont {Faoro}, \citenamefont {Farhi}, \citenamefont {Fatemi},
  \citenamefont {Flores}, \citenamefont {Forati}, \citenamefont {Fowler},
  \citenamefont {Giang}, \citenamefont {Gidney}, \citenamefont {Gilboa},
  \citenamefont {Giustina}, \citenamefont {Dau}, \citenamefont {Gross},
  \citenamefont {Habegger}, \citenamefont {Harrigan}, \citenamefont {Hoffmann},
  \citenamefont {Hong}, \citenamefont {Huang}, \citenamefont {Huff},
  \citenamefont {Huggins}, \citenamefont {Ioffe}, \citenamefont {Isakov},
  \citenamefont {Iveland}, \citenamefont {Jeffrey}, \citenamefont {Jiang},
  \citenamefont {Jones}, \citenamefont {Kafri}, \citenamefont {Kechedzhi},
  \citenamefont {Khattar}, \citenamefont {Kim}, \citenamefont {Kitaev},
  \citenamefont {Klimov}, \citenamefont {Klots}, \citenamefont {Korotkov},
  \citenamefont {Kostritsa}, \citenamefont {Kreikebaum}, \citenamefont
  {Landhuis}, \citenamefont {Laptev}, \citenamefont {Lau}, \citenamefont {Lee},
  \citenamefont {Laws}, \citenamefont {Liu}, \citenamefont {Locharla},
  \citenamefont {Martin}, \citenamefont {McClean}, \citenamefont {McEwen},
  \citenamefont {Meurer~Costa}, \citenamefont {Miao}, \citenamefont {Mohseni},
  \citenamefont {Montazeri}, \citenamefont {Morvan}, \citenamefont {Mount},
  \citenamefont {Mruczkiewicz}, \citenamefont {Naaman}, \citenamefont {Neeley},
  \citenamefont {Neill}, \citenamefont {Newman}, \citenamefont {O'Brien},
  \citenamefont {Opremcak}, \citenamefont {Petukhov}, \citenamefont {Potter},
  \citenamefont {Quintana}, \citenamefont {Rubin}, \citenamefont {Saei},
  \citenamefont {Sank}, \citenamefont {Sankaragomathi}, \citenamefont
  {Satzinger}, \citenamefont {Schuster}, \citenamefont {Shearn}, \citenamefont
  {Shvarts}, \citenamefont {Strain}, \citenamefont {Su}, \citenamefont
  {Szalay}, \citenamefont {Vidal}, \citenamefont {Villalonga}, \citenamefont
  {{Vollgraff-Heidweiller}}, \citenamefont {White}, \citenamefont {Yao},
  \citenamefont {Yeh}, \citenamefont {Yoo}, \citenamefont {Zalcman},
  \citenamefont {Zhang}, \citenamefont {Zhu}, \citenamefont {Neven},
  \citenamefont {Bacon}, \citenamefont {Hilton}, \citenamefont {Lucero},
  \citenamefont {Babbush}, \citenamefont {Boixo}, \citenamefont {Megrant},
  \citenamefont {Chen}, \citenamefont {Kelly}, \citenamefont {Smelyanskiy},
  \citenamefont {Abanin},\ and\ \citenamefont {Roushan}}]{Mi2022}%
  \BibitemOpen
  \bibfield  {author} {\bibinfo {author} {\bibfnamefont {X.}~\bibnamefont
  {Mi}} {\it et al.},\ }\bibfield
  {title} {\bibinfo {title} {Noise-resilient edge modes on a chain of
  superconducting qubits},\ }\href {https://doi.org/10.1126/science.abq5769}
  {\bibfield  {journal} {\bibinfo  {journal} {Science}\ }\textbf {\bibinfo
  {volume} {378}},\ \bibinfo {pages} {785} (\bibinfo {year}
  {2022})}\BibitemShut {NoStop}%
\bibitem [{\citenamefont {Schmid}\ \emph {et~al.}(2024)\citenamefont {Schmid},
  \citenamefont {Penner}, \citenamefont {Yang}, \citenamefont {Glazman},\ and\
  \citenamefont {{von Oppen}}}]{Schmid2024}%
  \BibitemOpen
  \bibfield  {author} {\bibinfo {author} {\bibfnamefont {H.}~\bibnamefont
  {Schmid}}, \bibinfo {author} {\bibfnamefont {A.-G.}\ \bibnamefont {Penner}},
  \bibinfo {author} {\bibfnamefont {K.}~\bibnamefont {Yang}}, \bibinfo {author}
  {\bibfnamefont {L.}~\bibnamefont {Glazman}},\ and\ \bibinfo {author}
  {\bibfnamefont {F.}~\bibnamefont {{von Oppen}}},\ }\bibfield  {title}
  {\bibinfo {title} {Robust {{Spectral}} \$\textbackslash
  ensuremath\textbraceleft\textbackslash pi\textbraceright\$ {{Pairing}} in the
  {{Random-Field Floquet Quantum Ising Model}}},\ }\href
  {https://doi.org/10.1103/PhysRevLett.132.210401} {\bibfield  {journal}
  {\bibinfo  {journal} {Phys. Rev. Lett.}\ }\textbf {\bibinfo {volume} {132}},\
  \bibinfo {pages} {210401} (\bibinfo {year} {2024})}\BibitemShut {NoStop}%
\bibitem [{\citenamefont {Vernier}\ \emph {et~al.}(2024)\citenamefont
  {Vernier}, \citenamefont {Yeh}, \citenamefont {Piroli},\ and\ \citenamefont
  {Mitra}}]{vernier2024a}%
  \BibitemOpen
  \bibfield  {author} {\bibinfo {author} {\bibfnamefont {E.}~\bibnamefont
  {Vernier}}, \bibinfo {author} {\bibfnamefont {H.-C.}\ \bibnamefont {Yeh}},
  \bibinfo {author} {\bibfnamefont {L.}~\bibnamefont {Piroli}},\ and\ \bibinfo
  {author} {\bibfnamefont {A.}~\bibnamefont {Mitra}},\ }\bibfield  {title}
  {\bibinfo {title} {Strong {{Zero Modes}} in {{Integrable Quantum
  Circuits}}},\ }\href {https://doi.org/10.1103/PhysRevLett.133.050606}
  {\bibfield  {journal} {\bibinfo  {journal} {Phys. Rev. Lett.}\ }\textbf
  {\bibinfo {volume} {133}},\ \bibinfo {pages} {050606} (\bibinfo {year}
  {2024})}\BibitemShut {NoStop}%
\bibitem [{\citenamefont {Tausendpfund}\ \emph {et~al.}(2025)\citenamefont
  {Tausendpfund}, \citenamefont {Mitra},\ and\ \citenamefont
  {Rizzi}}]{Tausendpfund2025}%
  \BibitemOpen
  \bibfield  {author} {\bibinfo {author} {\bibfnamefont {N.}~\bibnamefont
  {Tausendpfund}}, \bibinfo {author} {\bibfnamefont {A.}~\bibnamefont
  {Mitra}},\ and\ \bibinfo {author} {\bibfnamefont {M.}~\bibnamefont {Rizzi}},\
  }\bibfield  {title} {\bibinfo {title} {Almost strong zero modes at finite
  temperature},\ }\href {https://doi.org/10.1103/PhysRevResearch.7.023245}
  {\bibfield  {journal} {\bibinfo  {journal} {Phys. Rev. Res.}\ }\textbf
  {\bibinfo {volume} {7}},\ \bibinfo {pages} {023245} (\bibinfo {year}
  {2025})}\BibitemShut {NoStop}%
\bibitem [{\citenamefont {Schmid}\ \emph {et~al.}(2025)\citenamefont {Schmid},
  \citenamefont {Peng}, \citenamefont {Refael},\ and\ \citenamefont {{von
  Oppen}}}]{Schmid2025}%
  \BibitemOpen
  \bibfield  {author} {\bibinfo {author} {\bibfnamefont {H.}~\bibnamefont
  {Schmid}}, \bibinfo {author} {\bibfnamefont {Y.}~\bibnamefont {Peng}},
  \bibinfo {author} {\bibfnamefont {G.}~\bibnamefont {Refael}},\ and\ \bibinfo
  {author} {\bibfnamefont {F.}~\bibnamefont {{von Oppen}}},\ }\bibfield
  {title} {\bibinfo {title} {Self-{{Similar Phase Diagram}} of the
  {{Fibonacci-Driven Quantum Ising Model}}},\ }\href
  {https://doi.org/10.1103/hn66-j8pt} {\bibfield  {journal} {\bibinfo
  {journal} {Phys. Rev. Lett.}\ }\textbf {\bibinfo {volume} {134}},\ \bibinfo
  {pages} {240404} (\bibinfo {year} {2025})}\BibitemShut {NoStop}%
\bibitem [{\citenamefont {Jin}\ \emph {et~al.}(2025)\citenamefont {Jin},
  \citenamefont {Jiang}, \citenamefont {Zhu}, \citenamefont {Bao},
  \citenamefont {Shen}, \citenamefont {Wang}, \citenamefont {Zhu},
  \citenamefont {Xu}, \citenamefont {Song}, \citenamefont {Chen}, \citenamefont
  {Tan}, \citenamefont {Wu}, \citenamefont {Zhang}, \citenamefont {Gao},
  \citenamefont {Wang}, \citenamefont {Zou}, \citenamefont {Zhang},
  \citenamefont {Li}, \citenamefont {Zhong}, \citenamefont {Cui}, \citenamefont
  {Han}, \citenamefont {He}, \citenamefont {Wang}, \citenamefont {Yang},
  \citenamefont {Wang}, \citenamefont {Shen}, \citenamefont {Liu},
  \citenamefont {Deng}, \citenamefont {Dong}, \citenamefont {Zhang},
  \citenamefont {Li}, \citenamefont {Yuan}, \citenamefont {Lu}, \citenamefont
  {Sun}, \citenamefont {Li}, \citenamefont {Zhang}, \citenamefont {Song},
  \citenamefont {Wang}, \citenamefont {Guo}, \citenamefont {Machado},
  \citenamefont {Kemp}, \citenamefont {Iadecola}, \citenamefont {Yao},
  \citenamefont {Wang},\ and\ \citenamefont {Deng}}]{Jin2025}%
  \BibitemOpen
  \bibfield  {author} {\bibinfo {author} {\bibfnamefont {F.}~\bibnamefont
  {Jin}} {\it et al.}, \ }\bibfield  {title} {\bibinfo {title} {Topological prethermal
  strong zero modes on superconducting processors},\ }\href
  {https://doi.org/10.1038/s41586-025-09476-z} {\bibfield  {journal} {\bibinfo
  {journal} {Nature}\ }\textbf {\bibinfo {volume} {645}},\ \bibinfo {pages}
  {626} (\bibinfo {year} {2025})}\BibitemShut {NoStop}%
\bibitem [{\citenamefont {Liang}\ \emph {et~al.}(2024)\citenamefont {Liang},
  \citenamefont {Huang}, \citenamefont {Zhang}, \citenamefont {Tao},
  \citenamefont {Tang}, \citenamefont {Chu}, \citenamefont {Qiu}, \citenamefont
  {Sun}, \citenamefont {Zhou}, \citenamefont {Zhang}, \citenamefont {Zhang},
  \citenamefont {Guo}, \citenamefont {Liu}, \citenamefont {Chen}, \citenamefont
  {Liu}, \citenamefont {Zhong}, \citenamefont {Niu},\ and\ \citenamefont
  {Yu}}]{Liang2024}%
  \BibitemOpen
  \bibfield  {author} {\bibinfo {author} {\bibfnamefont {Y.}~\bibnamefont
  {Liang}} {\it et al.}, \ }\href {https://doi.org/10.48550/arXiv.2410.10208}
  {\bibinfo {title} {Floquet engineering of anisotropic transverse interactions
  in superconducting qubits}} (\bibinfo {year} {2024}),\ \Eprint
  {https://arxiv.org/abs/2410.10208} {arXiv:2410.10208} \BibitemShut {NoStop}%
\bibitem [{Note1()}]{Note1}%
  \BibitemOpen
  \bibinfo {note} {See Supplemental Material, which provides details of the
  calculations in the main text and contains Ref. \cite{fishman2021}.}\BibitemShut {Stop}%
   \bibitem [{\citenamefont {Fishman}\ \emph {et~al.}(2021)\citenamefont
  {Fishman}, \citenamefont {White},\ and\ \citenamefont
  {Stoudenmire}}]{fishman2021}%
   \BibitemOpen
  \bibfield  {author} {\bibinfo {author} {\bibfnamefont {M.}~\bibnamefont
  {Fishman}}, \bibinfo {author} {\bibfnamefont {S.~R.}\ \bibnamefont {White}},\
  and\ \bibinfo {author} {\bibfnamefont {E.~M.}\ \bibnamefont {Stoudenmire}},\
  }\href {https://doi.org/10.48550/arXiv.2007.14822} {\bibinfo {title} {The
  {{ITensor Software Library}} for {{Tensor Network Calculations}}}} (\bibinfo
  {year} {2021}),\ \Eprint {https://arxiv.org/abs/2007.14822} {arXiv:2007.14822
  [cs]} \BibitemShut {NoStop}%  
\bibitem [{\citenamefont {Chiu}\ \emph {et~al.}(2016)\citenamefont {Chiu},
  \citenamefont {Teo}, \citenamefont {Schnyder},\ and\ \citenamefont
  {Ryu}}]{Chiu2016}%
  \BibitemOpen
  \bibfield  {author} {\bibinfo {author} {\bibfnamefont {C.-K.}\ \bibnamefont
  {Chiu}}, \bibinfo {author} {\bibfnamefont {J.~C.~Y.}\ \bibnamefont {Teo}},
  \bibinfo {author} {\bibfnamefont {A.~P.}\ \bibnamefont {Schnyder}},\ and\
  \bibinfo {author} {\bibfnamefont {S.}~\bibnamefont {Ryu}},\ }\bibfield
  {title} {\bibinfo {title} {Classification of topological quantum matter with
  symmetries},\ }\href {https://doi.org/10.1103/RevModPhys.88.035005}
  {\bibfield  {journal} {\bibinfo  {journal} {Rev. Mod. Phys.}\ }\textbf
  {\bibinfo {volume} {88}},\ \bibinfo {pages} {035005} (\bibinfo {year}
  {2016})}\BibitemShut {NoStop}%
\bibitem [{\citenamefont {Leumer}\ \emph {et~al.}(2020)\citenamefont {Leumer},
  \citenamefont {Marganska}, \citenamefont {Muralidharan},\ and\ \citenamefont
  {Grifoni}}]{Leumer2020}%
  \BibitemOpen
  \bibfield  {author} {\bibinfo {author} {\bibfnamefont {N.}~\bibnamefont
  {Leumer}}, \bibinfo {author} {\bibfnamefont {M.}~\bibnamefont {Marganska}},
  \bibinfo {author} {\bibfnamefont {B.}~\bibnamefont {Muralidharan}},\ and\
  \bibinfo {author} {\bibfnamefont {M.}~\bibnamefont {Grifoni}},\ }\bibfield
  {title} {\bibinfo {title} {Exact eigenvectors and eigenvalues of the finite
  kitaev chain and its topological properties},\ }\href
  {https://doi.org/10.1088/1361-648X/ab8bf9} {\bibfield  {journal} {\bibinfo
  {journal} {J. Phys.: Condens. Matter}\ }\textbf {\bibinfo {volume} {32}},\
  \bibinfo {pages} {445502} (\bibinfo {year} {2020})}\BibitemShut {NoStop}%
\bibitem [{\citenamefont {Lindner}\ \emph {et~al.}(2011)\citenamefont
  {Lindner}, \citenamefont {Refael},\ and\ \citenamefont
  {Galitski}}]{Lindner2011}%
  \BibitemOpen
  \bibfield  {author} {\bibinfo {author} {\bibfnamefont {N.~H.}\ \bibnamefont
  {Lindner}}, \bibinfo {author} {\bibfnamefont {G.}~\bibnamefont {Refael}},\
  and\ \bibinfo {author} {\bibfnamefont {V.}~\bibnamefont {Galitski}},\
  }\bibfield  {title} {\bibinfo {title} {Floquet topological insulator in
  semiconductor quantum wells},\ }\href {https://doi.org/10.1038/nphys1926}
  {\bibfield  {journal} {\bibinfo  {journal} {Nature Phys}\ }\textbf {\bibinfo
  {volume} {7}},\ \bibinfo {pages} {490} (\bibinfo {year} {2011})}\BibitemShut
  {NoStop}%
\bibitem [{\citenamefont {Lindner}\ \emph {et~al.}(2013)\citenamefont
  {Lindner}, \citenamefont {Bergman}, \citenamefont {Refael},\ and\
  \citenamefont {Galitski}}]{Lindner2013}%
  \BibitemOpen
  \bibfield  {author} {\bibinfo {author} {\bibfnamefont {N.~H.}\ \bibnamefont
  {Lindner}}, \bibinfo {author} {\bibfnamefont {D.~L.}\ \bibnamefont
  {Bergman}}, \bibinfo {author} {\bibfnamefont {G.}~\bibnamefont {Refael}},\
  and\ \bibinfo {author} {\bibfnamefont {V.}~\bibnamefont {Galitski}},\
  }\bibfield  {title} {\bibinfo {title} {Topological {{Floquet}} spectrum in
  three dimensions via a two-photon resonance},\ }\href
  {https://doi.org/10.1103/PhysRevB.87.235131} {\bibfield  {journal} {\bibinfo
  {journal} {Phys. Rev. B}\ }\textbf {\bibinfo {volume} {87}},\ \bibinfo
  {pages} {235131} (\bibinfo {year} {2013})}\BibitemShut {NoStop}%
\bibitem [{\citenamefont {Klinovaja}\ \emph {et~al.}(2016)\citenamefont
  {Klinovaja}, \citenamefont {Stano},\ and\ \citenamefont
  {Loss}}]{Klinovaja2016}%
  \BibitemOpen
  \bibfield  {author} {\bibinfo {author} {\bibfnamefont {J.}~\bibnamefont
  {Klinovaja}}, \bibinfo {author} {\bibfnamefont {P.}~\bibnamefont {Stano}},\
  and\ \bibinfo {author} {\bibfnamefont {D.}~\bibnamefont {Loss}},\ }\bibfield
  {title} {\bibinfo {title} {Topological {{Floquet Phases}} in {{Driven Coupled
  Rashba Nanowires}}},\ }\href {https://doi.org/10.1103/PhysRevLett.116.176401}
  {\bibfield  {journal} {\bibinfo  {journal} {Phys. Rev. Lett.}\ }\textbf
  {\bibinfo {volume} {116}},\ \bibinfo {pages} {176401} (\bibinfo {year}
  {2016})}\BibitemShut {NoStop}%
\bibitem [{\citenamefont {Thakurathi}\ \emph {et~al.}(2017)\citenamefont
  {Thakurathi}, \citenamefont {Loss},\ and\ \citenamefont
  {Klinovaja}}]{Thakurathi2017}%
  \BibitemOpen
  \bibfield  {author} {\bibinfo {author} {\bibfnamefont {M.}~\bibnamefont
  {Thakurathi}}, \bibinfo {author} {\bibfnamefont {D.}~\bibnamefont {Loss}},\
  and\ \bibinfo {author} {\bibfnamefont {J.}~\bibnamefont {Klinovaja}},\
  }\bibfield  {title} {\bibinfo {title} {Floquet {{Majorana}} fermions and
  parafermions in driven {{Rashba}} nanowires},\ }\href
  {https://doi.org/10.1103/PhysRevB.95.155407} {\bibfield  {journal} {\bibinfo
  {journal} {Phys. Rev. B}\ }\textbf {\bibinfo {volume} {95}},\ \bibinfo
  {pages} {155407} (\bibinfo {year} {2017})}\BibitemShut {NoStop}%
\bibitem [{\citenamefont {Rudner}\ and\ \citenamefont
  {Lindner}(2020)}]{Rudner2020a}%
  \BibitemOpen
  \bibfield  {author} {\bibinfo {author} {\bibfnamefont {M.~S.}\ \bibnamefont
  {Rudner}}\ and\ \bibinfo {author} {\bibfnamefont {N.~H.}\ \bibnamefont
  {Lindner}},\ }\bibfield  {title} {\bibinfo {title} {Band structure
  engineering and non-equilibrium dynamics in {{Floquet}} topological
  insulators},\ }\href {https://doi.org/10.1038/s42254-020-0170-z} {\bibfield
  {journal} {\bibinfo  {journal} {Nat Rev Phys}\ }\textbf {\bibinfo {volume}
  {2}},\ \bibinfo {pages} {229} (\bibinfo {year} {2020})}\BibitemShut {NoStop}%
\bibitem [{\citenamefont {Zhang}\ and\ \citenamefont
  {Song}(2015)}]{Zhang2015c}%
  \BibitemOpen
  \bibfield  {author} {\bibinfo {author} {\bibfnamefont {G.}~\bibnamefont
  {Zhang}}\ and\ \bibinfo {author} {\bibfnamefont {Z.}~\bibnamefont {Song}},\
  }\bibfield  {title} {\bibinfo {title} {Topological {{Characterization}} of
  {{Extended Quantum Ising Models}}},\ }\href
  {https://doi.org/10.1103/PhysRevLett.115.177204} {\bibfield  {journal}
  {\bibinfo  {journal} {Phys. Rev. Lett.}\ }\textbf {\bibinfo {volume} {115}},\
  \bibinfo {pages} {177204} (\bibinfo {year} {2015})}\BibitemShut {NoStop}%
\bibitem [{\citenamefont {Liu}\ \emph {et~al.}(2025)\citenamefont {Liu},
  \citenamefont {Cheng}, \citenamefont {Qu}, \citenamefont {Guo},\ and\
  \citenamefont {Sun}}]{Liu2025e}%
  \BibitemOpen
  \bibfield  {author} {\bibinfo {author} {\bibfnamefont {Q.-G.}\ \bibnamefont
  {Liu}}, \bibinfo {author} {\bibfnamefont {H.-G.}\ \bibnamefont {Cheng}},
  \bibinfo {author} {\bibfnamefont {S.}~\bibnamefont {Qu}}, \bibinfo {author}
  {\bibfnamefont {B.}~\bibnamefont {Guo}},\ and\ \bibinfo {author}
  {\bibfnamefont {Z.-Y.}\ \bibnamefont {Sun}},\ }\bibfield  {title} {\bibinfo
  {title} {Multipartite nonlocality spectrum and topological quantum phase
  transitions in a one-dimensional extended quantum {{Ising}} model},\ }\href
  {https://doi.org/10.1016/j.physleta.2025.130858} {\bibfield  {journal}
  {\bibinfo  {journal} {Physics Letters A}\ }\textbf {\bibinfo {volume}
  {557}},\ \bibinfo {pages} {130858} (\bibinfo {year} {2025})}\BibitemShut
  {NoStop}%
\bibitem [{\citenamefont {Shi}\ \emph {et~al.}(2022)\citenamefont {Shi},
  \citenamefont {Zhang},\ and\ \citenamefont {Song}}]{Shi2022}%
  \BibitemOpen
  \bibfield  {author} {\bibinfo {author} {\bibfnamefont {Y.~B.}\ \bibnamefont
  {Shi}}, \bibinfo {author} {\bibfnamefont {K.~L.}\ \bibnamefont {Zhang}},\
  and\ \bibinfo {author} {\bibfnamefont {Z.}~\bibnamefont {Song}},\ }\bibfield
  {title} {\bibinfo {title} {Dynamic generation of nonequilibrium
  superconducting states in {{Kitaev}} chain},\ }\href
  {https://doi.org/10.1103/PhysRevB.106.184505} {\bibfield  {journal} {\bibinfo
   {journal} {Phys. Rev. B}\ }\textbf {\bibinfo {volume} {106}},\ \bibinfo
  {pages} {184505} (\bibinfo {year} {2022})}\BibitemShut {NoStop}%
\bibitem [{\citenamefont {Slichter}(1990)}]{Slichter1990}%
  \BibitemOpen
  \bibfield  {author} {\bibinfo {author} {\bibfnamefont {C.~P.}\ \bibnamefont
  {Slichter}},\ }\href@noop {} {\emph {\bibinfo {title} {Principles of Magnetic
  Resonance}}}\ (\bibinfo  {publisher} {Springer},\ \bibinfo {year} {Berlin,
  1990})\BibitemShut {NoStop}%
\bibitem [{\citenamefont {Blais}\ \emph {et~al.}(2021)\citenamefont {Blais},
  \citenamefont {Grimsmo}, \citenamefont {Girvin},\ and\ \citenamefont
  {Wallraff}}]{Blais2021}%
  \BibitemOpen
  \bibfield  {author} {\bibinfo {author} {\bibfnamefont {A.}~\bibnamefont
  {Blais}}, \bibinfo {author} {\bibfnamefont {A.~L.}\ \bibnamefont {Grimsmo}},
  \bibinfo {author} {\bibfnamefont {S.~M.}\ \bibnamefont {Girvin}},\ and\
  \bibinfo {author} {\bibfnamefont {A.}~\bibnamefont {Wallraff}},\ }\bibfield
  {title} {\bibinfo {title} {Circuit quantum electrodynamics},\ }\href
  {https://doi.org/10.1103/RevModPhys.93.025005} {\bibfield  {journal}
  {\bibinfo  {journal} {Rev. Mod. Phys.}\ }\textbf {\bibinfo {volume} {93}},\
  \bibinfo {pages} {025005} (\bibinfo {year} {2021})}\BibitemShut {NoStop}%
\bibitem [{\citenamefont {Dziarmaga}(2005)}]{Dziarmaga2005}%
  \BibitemOpen
  \bibfield  {author} {\bibinfo {author} {\bibfnamefont {J.}~\bibnamefont
  {Dziarmaga}},\ }\bibfield  {title} {\bibinfo {title} {Dynamics of a quantum
  phase transition: {{Exact}} solution of the quantum ising model},\
  }\href@noop {} {\bibfield  {journal} {\bibinfo  {journal} {Phys. Rev. Lett.}\
  }\textbf {\bibinfo {volume} {95}},\ \bibinfo {pages} {245701} (\bibinfo
  {year} {2005})}\BibitemShut {NoStop}%
\bibitem [{\citenamefont {Dziarmaga}(2010)}]{Dziarmaga2010}%
  \BibitemOpen
  \bibfield  {author} {\bibinfo {author} {\bibfnamefont {J.}~\bibnamefont
  {Dziarmaga}},\ }\bibfield  {title} {\bibinfo {title} {Dynamics of a quantum
  phase transition and relaxation to a steady state},\ }\href@noop {}
  {\bibfield  {journal} {\bibinfo  {journal} {Adv. Phys.}\ }\textbf {\bibinfo
  {volume} {59}},\ \bibinfo {pages} {1063} (\bibinfo {year}
  {2010})}\BibitemShut {NoStop}%
\bibitem [{\citenamefont {Polkovnikov}\ \emph {et~al.}(2011)\citenamefont
  {Polkovnikov}, \citenamefont{Sengupta}, \citenamefont {Silva},\ and\ \citenamefont
  {Vengalattore}}]{Polkovnikov2011}%
  \BibitemOpen
  \bibfield  {author} {\bibinfo {author} {\bibfnamefont {A.}~\bibnamefont
  {Polkovnikov}}, \bibinfo{author} {\bibfnamefont {K.}~\bibnamefont{Sengupta}}, \bibinfo {author}
  {\bibfnamefont {A.}~\bibnamefont {Silva}},\ and\ \bibinfo {author}
  {\bibfnamefont {M.}~\bibnamefont {Vengalattore}},\ }\bibfield  {title}
  {\bibinfo {title} {Colloquium: {{Nonequilibrium}} dynamics of closed
  interacting quantum systems},\ }\href
  {https://doi.org/10.1103/RevModPhys.83.863} {\bibfield  {journal} {\bibinfo
  {journal} {Rev. Mod. Phys.}\ }\textbf {\bibinfo {volume} {83}},\ \bibinfo
  {pages} {863} (\bibinfo {year} {2011})}\BibitemShut {NoStop}%
\bibitem [{\citenamefont {Barouch}\ \emph {et~al.}(1970)\citenamefont
  {Barouch}, \citenamefont {McCoy},\ and\ \citenamefont
  {Dresden}}]{Barouch1970}%
  \BibitemOpen
  \bibfield  {author} {\bibinfo {author} {\bibfnamefont {E.}~\bibnamefont
  {Barouch}}, \bibinfo {author} {\bibfnamefont {B.~M.}\ \bibnamefont {McCoy}},\
  and\ \bibinfo {author} {\bibfnamefont {M.}~\bibnamefont {Dresden}},\
  }\bibfield  {title} {\bibinfo {title} {Statistical {{Mechanics}} of the
  \$\textbackslash mathrm\textbraceleft{{XY}}\textbraceright\$ {{Model}}.
  {{I}}},\ }\href {https://doi.org/10.1103/PhysRevA.2.1075} {\bibfield
  {journal} {\bibinfo  {journal} {Phys. Rev. A}\ }\textbf {\bibinfo {volume}
  {2}},\ \bibinfo {pages} {1075} (\bibinfo {year} {1970})}\BibitemShut
  {NoStop}%
\bibitem [{\citenamefont {Barouch}\ and\ \citenamefont
  {McCoy}(1971{\natexlab{a}})}]{Barouch1971}%
  \BibitemOpen
  \bibfield  {author} {\bibinfo {author} {\bibfnamefont {E.}~\bibnamefont
  {Barouch}}\ and\ \bibinfo {author} {\bibfnamefont {B.~M.}\ \bibnamefont
  {McCoy}},\ }\bibfield  {title} {\bibinfo {title} {Statistical {{Mechanics}}
  of the \${{XY}}\$ {{Model}}. {{II}}. {{Spin-Correlation Functions}}},\ }\href
  {https://doi.org/10.1103/PhysRevA.3.786} {\bibfield  {journal} {\bibinfo
  {journal} {Phys. Rev. A}\ }\textbf {\bibinfo {volume} {3}},\ \bibinfo {pages}
  {786} (\bibinfo {year} {1971}{\natexlab{a}})}\BibitemShut {NoStop}%
\bibitem [{\citenamefont {Barouch}\ and\ \citenamefont
  {McCoy}(1971{\natexlab{b}})}]{Barouch1971a}%
  \BibitemOpen
  \bibfield  {author} {\bibinfo {author} {\bibfnamefont {E.}~\bibnamefont
  {Barouch}}\ and\ \bibinfo {author} {\bibfnamefont {B.~M.}\ \bibnamefont
  {McCoy}},\ }\bibfield  {title} {\bibinfo {title} {Statistical {{Mechanics}}
  of the \$\textbackslash mathrm\textbraceleft{{XY}}\textbraceright\$
  {{Model}}. {{III}}},\ }\href {https://doi.org/10.1103/PhysRevA.3.2137}
  {\bibfield  {journal} {\bibinfo  {journal} {Phys. Rev. A}\ }\textbf {\bibinfo
  {volume} {3}},\ \bibinfo {pages} {2137} (\bibinfo {year}
  {1971}{\natexlab{b}})}\BibitemShut {NoStop}%
\bibitem [{\citenamefont {Sengupta}\ \emph {et~al.}(2004)\citenamefont
  {Sengupta}, \citenamefont {Powell},\ and\ \citenamefont
  {Sachdev}}]{Sengupta2004}%
  \BibitemOpen
  \bibfield  {author} {\bibinfo {author} {\bibfnamefont {K.}~\bibnamefont
  {Sengupta}}, \bibinfo {author} {\bibfnamefont {S.}~\bibnamefont {Powell}},\
  and\ \bibinfo {author} {\bibfnamefont {S.}~\bibnamefont {Sachdev}},\
  }\bibfield  {title} {\bibinfo {title} {Quench dynamics across quantum
  critical points},\ }\href {https://doi.org/10.1103/PhysRevA.69.053616}
  {\bibfield  {journal} {\bibinfo  {journal} {Phys. Rev. A}\ }\textbf {\bibinfo
  {volume} {69}},\ \bibinfo {pages} {053616} (\bibinfo {year}
  {2004})}\BibitemShut {NoStop}%
\bibitem [{\citenamefont {Patan{\`e}}\ \emph {et~al.}(2008)\citenamefont
  {Patan{\`e}}, \citenamefont {Silva}, \citenamefont {Amico}, \citenamefont
  {Fazio},\ and\ \citenamefont {Santoro}}]{Patane2008}%
  \BibitemOpen
  \bibfield  {author} {\bibinfo {author} {\bibfnamefont {D.}~\bibnamefont
  {Patan{\`e}}}, \bibinfo {author} {\bibfnamefont {A.}~\bibnamefont {Silva}},
  \bibinfo {author} {\bibfnamefont {L.}~\bibnamefont {Amico}}, \bibinfo
  {author} {\bibfnamefont {R.}~\bibnamefont {Fazio}},\ and\ \bibinfo {author}
  {\bibfnamefont {G.~E.}\ \bibnamefont {Santoro}},\ }\bibfield  {title}
  {\bibinfo {title} {Adiabatic {{Dynamics}} in {{Open Quantum Critical
  Many-Body Systems}}},\ }\href
  {https://doi.org/10.1103/PhysRevLett.101.175701} {\bibfield  {journal}
  {\bibinfo  {journal} {Phys. Rev. Lett.}\ }\textbf {\bibinfo {volume} {101}},\
  \bibinfo {pages} {175701} (\bibinfo {year} {2008})}\BibitemShut {NoStop}%
\bibitem [{\citenamefont {Calabrese}\ \emph
  {et~al.}(2012{\natexlab{a}})\citenamefont {Calabrese}, \citenamefont
  {Essler},\ and\ \citenamefont {Fagotti}}]{Calabrese2012}%
  \BibitemOpen
  \bibfield  {author} {\bibinfo {author} {\bibfnamefont {P.}~\bibnamefont
  {Calabrese}}, \bibinfo {author} {\bibfnamefont {F.~H.~L.}\ \bibnamefont
  {Essler}},\ and\ \bibinfo {author} {\bibfnamefont {M.}~\bibnamefont
  {Fagotti}},\ }\bibfield  {title} {\bibinfo {title} {Quantum quench in the
  transverse field {{Ising}} chain: {{I}}. {{Time}} evolution of order
  parameter correlators},\ }\href
  {https://doi.org/10.1088/1742-5468/2012/07/P07016} {\bibfield  {journal}
  {\bibinfo  {journal} {J. Stat. Mech.}\ }\textbf {\bibinfo {volume} {2012}},\
  \bibinfo {pages} {P07016} (\bibinfo {year} {2012}{\natexlab{a}})}\BibitemShut
  {NoStop}%
\bibitem [{\citenamefont {Calabrese}\ \emph
  {et~al.}(2012{\natexlab{b}})\citenamefont {Calabrese}, \citenamefont
  {Essler},\ and\ \citenamefont {Fagotti}}]{Calabrese2012a}%
  \BibitemOpen
  \bibfield  {author} {\bibinfo {author} {\bibfnamefont {P.}~\bibnamefont
  {Calabrese}}, \bibinfo {author} {\bibfnamefont {F.~H.~L.}\ \bibnamefont
  {Essler}},\ and\ \bibinfo {author} {\bibfnamefont {M.}~\bibnamefont
  {Fagotti}},\ }\bibfield  {title} {\bibinfo {title} {Quantum quenches in the
  transverse field {{Ising}} chain: {{II}}. {{Stationary}} state properties},\
  }\href {https://doi.org/10.1088/1742-5468/2012/07/P07022} {\bibfield
  {journal} {\bibinfo  {journal} {J. Stat. Mech.}\ }\textbf {\bibinfo {volume}
  {2012}},\ \bibinfo {pages} {P07022} (\bibinfo {year}
  {2012}{\natexlab{b}})}\BibitemShut {NoStop}%
\bibitem [{\citenamefont {Mukherjee}\ and\ \citenamefont
  {Dutta}(2009)}]{Mukherjee2009}%
  \BibitemOpen
  \bibfield  {author} {\bibinfo {author} {\bibfnamefont {V.}~\bibnamefont
  {Mukherjee}}\ and\ \bibinfo {author} {\bibfnamefont {A.}~\bibnamefont
  {Dutta}},\ }\bibfield  {title} {\bibinfo {title} {Effects of interference in
  the dynamics of a spin- 1/2 transverse {{XY}} chain driven periodically
  through quantum critical points},\ }\href
  {https://doi.org/10.1088/1742-5468/2009/05/P05005} {\bibfield  {journal}
  {\bibinfo  {journal} {J. Stat. Mech.}\ }\textbf {\bibinfo {volume} {2009}},\
  \bibinfo {pages} {P05005} (\bibinfo {year} {2009})}\BibitemShut {NoStop}%
\bibitem [{\citenamefont {Russomanno}\ \emph {et~al.}(2012)\citenamefont
  {Russomanno}, \citenamefont {Silva},\ and\ \citenamefont
  {Santoro}}]{Russomanno2012}%
  \BibitemOpen
  \bibfield  {author} {\bibinfo {author} {\bibfnamefont {A.}~\bibnamefont
  {Russomanno}}, \bibinfo {author} {\bibfnamefont {A.}~\bibnamefont {Silva}},\
  and\ \bibinfo {author} {\bibfnamefont {G.~E.}\ \bibnamefont {Santoro}},\
  }\bibfield  {title} {\bibinfo {title} {Periodic {{Steady Regime}} and
  {{Interference}} in a {{Periodically Driven Quantum System}}},\ }\href
  {https://doi.org/10.1103/PhysRevLett.109.257201} {\bibfield  {journal}
  {\bibinfo  {journal} {Phys. Rev. Lett.}\ }\textbf {\bibinfo {volume} {109}},\
  \bibinfo {pages} {257201} (\bibinfo {year} {2012})}\BibitemShut {NoStop}%
\bibitem [{\citenamefont {Smelyanskiy}\ \emph {et~al.}(2017)\citenamefont
  {Smelyanskiy}, \citenamefont {Venturelli}, \citenamefont {{Perdomo-Ortiz}},
  \citenamefont {Knysh},\ and\ \citenamefont {Dykman}}]{Smelyanskiy2017}%
  \BibitemOpen
  \bibfield  {author} {\bibinfo {author} {\bibfnamefont {V.~N.}\ \bibnamefont
  {Smelyanskiy}}, \bibinfo {author} {\bibfnamefont {D.}~\bibnamefont
  {Venturelli}}, \bibinfo {author} {\bibfnamefont {A.}~\bibnamefont
  {{Perdomo-Ortiz}}}, \bibinfo {author} {\bibfnamefont {S.}~\bibnamefont
  {Knysh}},\ and\ \bibinfo {author} {\bibfnamefont {M.~I.}\ \bibnamefont
  {Dykman}},\ }\bibfield  {title} {\bibinfo {title} {Quantum annealing via
  environment-mediated quantum diffusion},\ }\href
  {https://doi.org/10.1103/PhysRevLett.118.066802} {\bibfield  {journal}
  {\bibinfo  {journal} {Phys. Rev. Lett.}\ }\textbf {\bibinfo {volume} {118}},\
  \bibinfo {pages} {066802} (\bibinfo {year} {2017})}\BibitemShut {NoStop}%
\bibitem [{\citenamefont {Heyl}(2013)}]{Heyl2013}%
  \BibitemOpen
  \bibfield  {author} {\bibinfo {author} {\bibfnamefont {M.}~\bibnamefont
  {Heyl}},\ }\bibfield  {title} {\bibinfo {title} {Dynamical {{Quantum Phase
  Transitions}} in the {{Transverse-Field Ising Model}}},\ }\href
  {https://doi.org/10.1103/PhysRevLett.110.135704} {\bibfield  {journal}
  {\bibinfo  {journal} {Phys. Rev. Lett.}\ }\textbf {\bibinfo {volume} {110}},\
  \bibinfo {pages} {135704} (\bibinfo {year} {2013})}\BibitemShut {NoStop}%
\bibitem [{\citenamefont {Dziarmaga}\ \emph {et~al.}(2022)\citenamefont
  {Dziarmaga}, \citenamefont {Rams},\ and\ \citenamefont
  {Zurek}}]{Dziarmaga2022}%
  \BibitemOpen
  \bibfield  {author} {\bibinfo {author} {\bibfnamefont {J.}~\bibnamefont
  {Dziarmaga}}, \bibinfo {author} {\bibfnamefont {M.~M.}\ \bibnamefont
  {Rams}},\ and\ \bibinfo {author} {\bibfnamefont {W.~H.}\ \bibnamefont
  {Zurek}},\ }\bibfield  {title} {\bibinfo {title} {Coherent {{Many-Body
  Oscillations Induced}} by a {{Superposition}} of {{Broken Symmetry States}}
  in the {{Wake}} of a {{Quantum Phase Transition}}},\ }\href
  {https://doi.org/10.1103/PhysRevLett.129.260407} {\bibfield  {journal}
  {\bibinfo  {journal} {Phys. Rev. Lett.}\ }\textbf {\bibinfo {volume} {129}},\
  \bibinfo {pages} {260407} (\bibinfo {year} {2022})}\BibitemShut {NoStop}%
  \bibitem [{\citenamefont {Paul}\ \emph {et~al.}(2024)\citenamefont {Paul},
  \citenamefont {Titum},\ and\ \citenamefont {Maghrebi}}]{Paul2024}%
  \BibitemOpen
  \bibfield  {author} {\bibinfo {author} {\bibfnamefont {S.}~\bibnamefont
  {Paul}}, \bibinfo {author} {\bibfnamefont {P.}~\bibnamefont {Titum}},\ and\
  \bibinfo {author} {\bibfnamefont {M.}~\bibnamefont {Maghrebi}},\ }\bibfield
  {title} {\bibinfo {title} {Hidden quantum criticality and entanglement in
  quench dynamics},\ }\href {https://doi.org/10.1103/PhysRevResearch.6.L032003}
  {\bibfield  {journal} {\bibinfo  {journal} {Phys. Rev. Res.}\ }\textbf
  {\bibinfo {volume} {6}},\ \bibinfo {pages} {L032003} (\bibinfo {year}
  {2024})}\BibitemShut {NoStop}%
\bibitem [{\citenamefont {Grabarits}\ \emph {et~al.}(2025)\citenamefont
  {Grabarits}, \citenamefont {Balducci},\ and\ \citenamefont {{del
  Campo}}}]{Grabarits2025}%
  \BibitemOpen
  \bibfield  {author} {\bibinfo {author} {\bibfnamefont {A.}~\bibnamefont
  {Grabarits}}, \bibinfo {author} {\bibfnamefont {F.}~\bibnamefont
  {Balducci}},\ and\ \bibinfo {author} {\bibfnamefont {A.}~\bibnamefont {{del
  Campo}}},\ }\bibfield  {title} {\bibinfo {title} {Driving a quantum phase
  transition at an arbitrary rate: {{Exact}} solution of the transverse-field
  {{Ising}} model},\ }\href {https://doi.org/10.1103/PhysRevA.111.042207}
  {\bibfield  {journal} {\bibinfo  {journal} {Phys. Rev. A}\ }\textbf {\bibinfo
  {volume} {111}},\ \bibinfo {pages} {042207} (\bibinfo {year}
  {2025})}\BibitemShut {NoStop}%
\bibitem [{\citenamefont {Editorial}(2025)}]{Editorial2025}%
  \BibitemOpen
  \bibfield  {author} {\bibinfo {author} {\bibnamefont {Editorial}},\
  }\bibfield  {title} {\bibinfo {title} {Spins in chains},\ }\href
  {https://doi.org/10.1038/s41563-025-02243-5} {\bibfield  {journal} {\bibinfo
  {journal} {Nature Materials}\ }\textbf {\bibinfo {volume} {24}},\ \bibinfo
  {pages} {651} (\bibinfo {year} {2025})}\BibitemShut {NoStop}%
  \bibitem [{\citenamefont {Gradshteyn}\ and\ \citenamefont
  {Ryzhik}(2015)}]{Gradshteyn2015}%
  \BibitemOpen
  \bibfield  {author} {\bibinfo {author} {\bibfnamefont {I.~S.}\ \bibnamefont
  {Gradshteyn}}\ and\ \bibinfo {author} {\bibfnamefont {I.~M.}\ \bibnamefont
  {Ryzhik}},\ }\href@noop {} {\emph {\bibinfo {title} {Tables of Integrals,
  Series and Products}}},\ \bibinfo {edition} {8th}\ ed.\ (\bibinfo
  {publisher} {Elsevier},\ \bibinfo {address} {Amsterdam},\ \bibinfo {year}
  {2015})\BibitemShut {NoStop}%
  \bibitem [{\citenamefont {Elewa}\ and\ \citenamefont
  {Dykman}(2025)}]{elewa_spin_chain_repo}%
  \BibitemOpen
 \bibfield  {author} {\bibinfo {author} {\bibfnamefont {M.~T.}\ \bibnamefont
  {Elewa}}\ and\ \bibinfo {author} {\bibfnamefont {M.~I.}\ \bibnamefont
  {Dykman}},\ }\bibinfo {title} {Source code and simulation data for ``Anomalous parametric resonance in a
  spin-1/2 chain: dynamical effects of nontrivial topology''} (\bibinfo {year}
  {2025}),\ \bibinfo {note} {\href{https://github.com/mah-elewa/modulated_spin_chain}{GitHub repository}}\BibitemShut {NoStop}%
\end{thebibliography}
%\end{document}

\end{document}